\documentclass{aa}
\usepackage[varg]{txfonts}
\usepackage{graphicx}

\begin{document}

\title{Water maser variability in a high-mass YSO outburst:}
\subtitle{VERA and ALMA observations of S255~NIRS~3}

\author{Tomoya Hirota\inst{\ref{inst1},\ref{inst2}}
\and
Riccardo Cesaroni\inst{\ref{inst3}}
\and 
Luca Moscadelli\inst{\ref{inst3}} 
\and 
Koichiro Sugiyama\inst{\ref{inst1},\ref{inst4}}
\and
Ross A. Burns\inst{\ref{inst1},\ref{inst5}}
\and
Jungha Kim\inst{\ref{inst1},\ref{inst2},\ref{inst5}}
\and
Kazuyoshi Sunada\inst{\ref{inst6},\ref{inst7}}
\and 
Yoshinori Yonekura\inst{\ref{inst8}}
}

\institute{Mizusawa VLBI Observatory, National Astronomical Observatory of Japan, 
       Osawa 2-21-1, Mitaka-shi, Tokyo 181-8588, Japan 
       \email{tomoya.hirota@nao.ac.jp}
       \label{inst1}
\and
Department of Astronomical Sciences, SOKENDAI (The Graduate University for Advanced Studies), 
       Osawa 2-21-1, Mitaka-shi, Tokyo 181-8588, Japan
       \label{inst2}
\and
INAF -- Osservatorio Astrofisico di Arcetri, Largo E. Fermi 5, I-50125 Firenze, Italy
       \label{inst3}
\and
National Astronomical Research Institute of Thailand (Public Organization), 260 Moo 4, T. Donkaew, Amphur Maerim, Chiang Mai, 50180, Thailand
       \label{inst4}
\and
Korea Astronomy and Space Science Institute, Hwaam-dong 61-1, Yuseong-gu, Daejeon, 305-348, Republic of Korea
       \label{inst5}
\and
Mizusawa VLBI Observatory, National Astronomical Observatory of Japan, Hoshigaoka 2-12, Mizusawa, Oshu-shi, Iwate 023-0861, Japan
       \label{inst6}
\and
Department of Astronomical Sciences, SOKENDAI (The Graduate University for Advanced Studies), Hoshigaoka 2-12, Mizusawa, Oshu-shi, Iwate 023-0861, Japan
       \label{inst7}
\and
Center for Astronomy, Ibaraki University, 2-1-1 Bunkyo, Mito, Ibaraki 310-8512, Japan
       \label{inst8}
}
\date{Received 02 11 2020/ Accepted 10 12 2020}

\abstract 
{Clarifying the relationship between mass accretion and ejection history is one of the key issues in understanding high-mass star formation processes. }
{We aim to investigate the possible relationship between the mass accretion burst event in mid-June 2015 and the jet ejection in the high-mass protostar S255~NIRS~3. } 
{The Very Long Baseline Interferometer (VLBI) monitoring observations of the 22~GHz H$_{2}$O masers were carried out using VLBI Exploration of Radio Astrometry (VERA) to reveal the 3D velocity and spatial structure of the outflow/jet traced by the H$_{2}$O masers in S255~NIRS~3. 
In addition, we conducted follow-up observations of the submillimeter continuum and the 321~GHz H$_{2}$O masers with the Atacama Large Millimeter/submillimeter Array (ALMA) at Band~7. } 
{We successfully measured the proper motions of the 22 GHz H$_{2}$O masers associated with a bipolar outflow. 
The structure is almost the same as was observed in 2005 and 2010. 
The expansion velocity of the blueshifted bow shock traced by the 22~GHz H$_{2}$O masers was measured to be 28~km~s$^{-1,}$ corresponding to a dynamical timescale of 60~years. 
The direction of the maser outflow is slightly tilted compared with the radio jet, which could suggest a more recent ejection episode during the accretion burst event. 
The total flux density of the 22~GHz H$_{2}$O masers gradually increases from the beginning of the VLBI monitoring in early 2017 and becomes almost constant in subsequent single-dish monitoring in 2018. 
The brightening of the H$_{2}$O masers is more prominent in the northeast outflow lobe. 
For the first time, we revealed extended H$_{2}$O maser emission at 22 GHz in a star-forming region, which is partly resolved out by VERA and even by the most extended Very Large Array (VLA) configurations. We find that the flux variation of such an extended component is similar to that of the unresolved maser emission.
The ALMA Band-7 continuum emission did not show significant variations compared with the previous observations performed five~months before. 
We mapped the 321~GHz H$_{2}$O masers in S255~NIRS~3 providing the fourth example, for this maser, of the spatial distribution in a high-mass star-forming region. }
{We conclude that the bow shock structure traced by the 22~GHz H$_{2}$O maser features is unlikely to originate at the interface between the radio jet powered by the recent accretion outburst and the surrounding medium.
The brightening of the 22~GHz H$_{2}$O masers could be due to radiative excitation by photons form the (declining) infrared (IR) outburst escaping along the cavity created by the newly ejected material. 
The lower ratio of the 22~GHz/321~GHz maser luminosity in the blueshifted bow shock suggests a temperature ($>$1000~K), higher than for the other maser features in this region. }


\titlerunning{H$_{2}$O masers in S255~NIRS~3}
\authorrunning{T. Hirota et al.}

\maketitle

\section{Introduction}
\label{sec-intro}

In the last few years, mass accretion events in high-mass young stellar objects (HMYSOs), which have masses and luminosities of $M_{\ast}>8M_{\odot}$ and $L_{\ast}>5\times10^{3}L_{\odot}$, respectively, have been discovered via the sudden increase of continuum emission from radio to infrared (IR) wavelengths. 
One of the first examples was an outburst from the 20$M_{\odot}$ HMYSO NIRS~3 in the high-mass star-forming region S255~IR \citep{CarattioGaratti2017} located at a distance of 1.78~kpc \citep{Burns2016}. 
At almost the same time, a similar accretion burst event was identified through the millimeter continuum emission from another HMYSO, NGC6334I-MM1 \citep{Hunter2017}. 
These discoveries suggest that mass accretion burst events are common in HMYSOs, as theoretically predicted for the formation of both low- and high-mass stars
\citep{AcostaPulido2007, CarattioGaratti2011, ContrerasPena2017} through disk-mediated accretion \citep{Krumholz2009, Kuiper2011}. 

These accretion bursts triggered a large increase in the flux density of the associated Class~II methanol (CH$_3$OH) masers at 6.7~GHz. 
For S255~IR, a CH$_{3}$OH maser flare was reported in November 2015 \citep{Fujisawa2015, Moscadelli2017, Szymczak2018}, through 
pumping by mid-infrared (MIR) radiation from dust grains. 
This maser flare could be attributed to the increase of the accretion luminosity, estimated to have begun in mid-June 2015 \citep{CarattioGaratti2017}. 
Follow-up near-infrared (NIR) observations showed a brightening of the source of $\Delta$K$\simeq$2.9~mag and $\Delta$H$\simeq$3.5~mag, and an increase in luminosity up to $1.3\times10^{5}L_\odot$, which corresponds to a mass accretion rate of $5\times10^{-3}~M_\odot$yr$^{-1}$ \citep{CarattioGaratti2017, Moscadelli2017, Cesaroni2018}. 
Subsequent submillimeter \citep{Liu2018} and NIR \citep{Uchiyama2020} observations also reported a significant increase of the corresponding flux densities. 
A similar 6.7~GHz CH$_{3}$OH maser flare was also detected in NGC6334I-MM1 \citep{Hunter2018, MacLeod2018}, along with a 22~GHz water (H$_{2}$O) maser burst \citep{Brogan2018}. 
Another HMYSO G358.93-0.03-MM1 \citep{Sugiyama2019, Burns2020} showed a flare of the 6.7~GHz CH$_{3}$OH masers and other new methanol transitions \citep{Breen2019, Brogan2019, MacLeod2019, Chen2020}, a result of the recent extensive observations by the global collaboration network, the Maser Monitoring Organization (M2O; MaserMonitoring.org). 
The outbursts from these HMYSOs represent a unique opportunity to investigate the accretion, and, in turn, ejection processes in these objects; and to derive useful parameters, such as the mass accretion and ejection rates, infall velocity, and duration of the events. 

According to subsequent monitoring observations of S255 NIRS~3, the outburst was also detected at centimeter wavelengths with the Karl Jansky Very Large Array (VLA), although with a delay of $\sim$1~yr \citep{Cesaroni2018}. 
The slope of the continuum spectrum ($\sim$0.78) is suggestive of emission from a thermal radio jet \citep{Reynolds1986}, of which the mass-loss rate has been boosted by the accretion burst. 
The delay between the IR and the radio burst \citep{Cesaroni2018, Uchiyama2020} 
is not surprising, as the former propagates at the speed of light and the latter at the typical speed of the ejecta (500-1000~km~s$^{-1}$). 
Indeed, \citet{Cesaroni2018} detected possible evidence of the thermal shock produced by the ejecta/wind as the jet recollimates. 
Assuming an escape velocity of 700~km~s$^{-1}$, the observed time delay corresponds to a traveling distance of $\sim$160-180~au from the source (or $\sim$100~mas).

Alongside the continuum emission observations with the VLA, we have been simultaneously monitoring the H$_{2}$O maser emission, which is also expected to undergo a variation due to the change of physical conditions in the shock ahead of the jet \citep[e.g., for NGC6334I-MM1; ][]{Brogan2018}, where the masers are located \citep{Goddi2007, Burns2016}. 
One can note that the richest H$_{2}$O maser clusters are separated from the continuum peak (the putative YSO position) by 100-200~mas, which is comparable to the expected distance traveled by the material ejected at the time of the IR burst. 
The ``new'' ejecta could now be reaching and interacting with the H$_{2}$O maser environment. 
 
In the present paper, we aim to discuss the relation between the accretion burst in S255~NIRS~3 and the H$_{2}$O masers associated with this HMYSO. 
We describe the observations and data analysis in Sect. \ref{sec-obs}. 
In particular, we devote Sects. \ref{subsec-vera}, \ref{subsec-alma}, and \ref{subsec-vla}, respectively, to the observations made with the Japanese very long baseline interferometer (VLBI) VLBI Exploration of Radio Astrometry (VERA), the Atacama Large Millimeter/submillimeter Array (ALMA), and the VLA. 
In Sect. \ref{sec-results}, we present the results from the observations with VERA single-dish and VLBI monitoring to investigate possible changes in flux density (Sect. \ref{subsec-spectra}) and structure (Sect. \ref{subsec-vlbi}) of the H$_{2}$O maser features, which are compared with the VLA data and previous VLBI observations \citep{Goddi2007, Burns2016}. 
The observations prior to the burst are compared with ours to verify a possible change in the H$_{2}$O maser spatial and velocity structure. 
At the same time, we also report maser absolute proper motions, and thus their 3D expansion velocities, for which technical information is provided in Appendix \ref{sec-appendix}. 
Furthermore, we give ALMA results for the submillimeter continuum and H$_{2}$O masers at Band~7 to study the physical properties very close to the central HMYSO (Sect. \ref{subsec-cont} and \ref{subsec-321ghz}). 
A possible origin of the maser variability and the maser physical properties are discussed in Sect. \ref{sec-discussion}. 
Finally, we summarize our conclusions in Sect. \ref{sec-summary}. 

\begin{table}[hbt]
\begin{center}
\caption{Summary of VERA observations}
\label{tab-obs}
\begin{tabular}{llll}
\hline
\hline
      & Observation           &          & Phase- \\
No & code & Date & referencing \\
\hline
1  & R17054A  & Feb. 23 2017  & Y \\
2  & R17056A  & Feb. 25 2017  & Y \\
3  & R17082A  & Mar. 23 2017  & \\
4  & R17109A  & Apr. 19 2017  & Y \\
5  & R17111A  & Apr. 21 2017  & \\
6  & R17134A  & May. 14 2017  & Y \\
7  & R17160A  & Jun. 09 2017  & Y \\
\hline
\end{tabular}
\end{center}
\end{table}

\section{Observations and data analysis}
\label{sec-obs}

\subsection{VERA}
\label{subsec-vera}

The VLBI monitoring observations of the H$_{2}$O ($6_{1,6}$-$5_{2,3}$) line at 22.235080~GHz were carried out with VERA in 2017, as summarized in Table \ref{tab-obs}. 
VERA consists of four 20-m radio telescopes located in Japan, with a maximum baseline length of 2270~km, providing a typical beam size of 1.2~mas. 
The pointing and phase-tracking center position of S255~NIRS~3 was 
\begin{eqnarray*}
\mbox{RA(J2000)} &=& 06\mbox{h}12\mbox{m}54.006\mbox{s} \\
\mbox{Decl(J2000)} &=& +17^{\circ}59\arcmin22.96\arcsec,
\end{eqnarray*}
which is the same as in \citet{Burns2016}. 
All observations lasted nine~hours from horizon to horizon, and the typical on-source time was $\sim$6.4~hours. 
Left-handed circular polarization data were recorded on hard disk devices at 1~Gbps. 
The total bandwidths were 16~MHz and 240~MHz for S255~NIRS~3 and the calibrator, J0613+1708, respectively. 
The spectral resolution for the maser line was chosen equal to 15.625~kHz, corresponding to a velocity resolution of 0.21~km~s$^{-1}$. 

We employed the standard dual beam observation mode \citep{VERA2020}, as in previous observations of S255~NIRS~3, in which the target and a 
nearby calibrator (J0613+1708) were observed simultaneously with a separation angle of 0.87~degrees \citep{Burns2016}. 
The latter was used as a phase or position reference \citep{Reid2014} to determine the absolute positions of the H$_{2}$O masers. 
During the observations, artificial wideband radio signals were injected into both dual-beam receivers to calibrate the phase difference between them \citep{Honma2008a}. 
Delay and bandpass calibration were done by observing a strong continuum source, DA193. 
Then, residual phase calibration was done using J0613+1708. 
This was detected at a sufficiently high signal-to-noise ratio, with a total flux density of 0.27-0.33~Jy for all epochs. 

Correlation processing was done using the software correlator developed at the National Astronomical Observatory of Japan (NAOJ) Mizusawa campus. 
After correlation, we applied the correction of an a priori delay-tracking model to reduce residual phase offsets during the following calibration procedure \citep{Honma2008b}. 
Standard calibration and synthesis imaging was performed using the National Radio Astronomy Observatory (NRAO) Astronomical Image Processing System (AIPS) software package \citep{Greisen2003}. 
We conducted a phase-referencing analysis to determine the absolute position of the reference maser spot\footnote{Throughout the paper, we define a spot as an individual maser emission at a single spectral channel, while a feature refers to a group of spots that are detected in more than three consecutive spectral channels at positions coincident within the beam size. Thus, a feature is regarded as a physical gas clump emitting the 22~GHz maser line. } for each epoch \citep[e.g.,][]{Burns2016, VERA2020}.
The resultant astrometric accuracy is estimated to be 0.5~mas in the worst case according to the absolute proper motion fitting, and this is discussed in the next section. 
The rms noise levels were 40-60~mJy~beam$^{-1}$ in the line-free channels. 

The H$_{2}$O masers in S255~NIRS~3 have been monitored with the single-dish radio antennas of VERA since October 20, 2016. 
The typical monitoring interval was two months, each time using in most cases one out of the four VERA antennas.
Observations were done in the conventional position switching mode, and amplitude calibrations were done by means of the chopper-wheel method. 
The spectral resolution of a digital spectrometer was 15.625~kHz or 31.25~kHz. 

\subsection{ALMA}
\label{subsec-alma}

We used ALMA cycle 5 data (ADS/JAO.ALMA \#2017.1.00178.S) to target the continuum emission in S255~NIR3 and identify the powering source of the 22~GHz H$_{2}$O masers. 
Observations were carried out at Band~7 (in the range of 320-340~GHz) on December 02, 2017, and the on-source time was ten minutes. 
The array consisted of 48 12-m antennas in the C43-7 configuration with baseline lengths of 41-6855~m. 
Amplitude and bandpass calibrations were done with J0510+1800, and we observed J0625+1440 for the phase calibration. 

We used the pipeline-calibrated visibility data delivered by the East-Asian ALMA Regional Center (EA-ARC) to carry out synthesis imaging using the Common Astronomy Software Applications (CASA) package \citep{McMullin2007}. 
We selected line-free channels from four spectral windows, three with a bandwidth of 1875~MHz and one 469~MHz wide, centered at 335.9~GHz, 334.0~GHz, 322.5~GHz (broader bands), and 321.2~GHz (narrower band), respectively. 
The effective bandwidth integrated over the line-free channels was 3593~MHz. 
Using these line-free channels, a continuum image was produced by means of the clean procedure with a Briggs robust weighting parameter of 0.5.
Self-calibration was done using the continuum image to solve only the phase terms. 
The flux density of the continuum emission increased by a factor of 1.5. 
The resultant rms noise level was 0.33~mJy~beam$^{-1}$. 
The synthesized beam size of the continuum emission was 0.073\arcsec$\times$0.059\arcsec, \ with a position angle of $-77$~degrees. 
Pure line data were obtained by subtracting the continuum from the original line and continuum data in the UV domain.
 
We assigned one of the spectral windows to the H$_{2}$O $10_{2,9}$-$9_{3,6}$ line at 321.225656~GHz \citep[][hereafter referred to as 321~GHz H$_{2}$O maser]{Chen2000}. 
The lower state energy level of the 321~GHz H$_{2}$O maser transition is 1846~K. 
The spectral resolution of the 321~GHz H$_{2}$O maser map was 0.244~MHz, which corresponds to a velocity resolution of 0.23~km~s$^{-1}$. 
We made a channel map with a velocity resolution of 0.25~km~s$^{-1}$ by regridding the spectral channels.
The rms noise level of the line-free channel data was 0.01~Jy~beam$^{-1}$. 
The synthesized beam size for the 321~GHz H$_{2}$O maser line was 0.082\arcsec$\times$0.070\arcsec, with a position angle of $-73$~degrees. 

\subsection{VLA}
\label{subsec-vla}

The object S255~NIRS~3 was also observed using the VLA of the NRAO during five epochs in 2016. 
Array configurations were in C-configuration (March 11, 2016), B-configuration (July 10 and August 01, 2016), and A-configuration (October 15 and November 24, 2016). 
All observations were done at frequencies of 6.0, 10.0, 22.2, and 45.5 GHz to image the continuum emission as described in \citet{Cesaroni2018}. 
In this paper, we analyze only the 22~GHz H$_{2}$O maser data. 
The WIDAR correlator was used to record dual circular polarization with a spectral resolution of 15.625~kHz for the C- and B-configurations, and 331.25~kHz for the A-configuration. 

\section{Results}
\label{sec-results}

\subsection{The 22~GHz H$_{2}$O maser spectra}
\label{subsec-spectra}

\begin{figure}[th]
\begin{center}
\includegraphics[width=7cm]{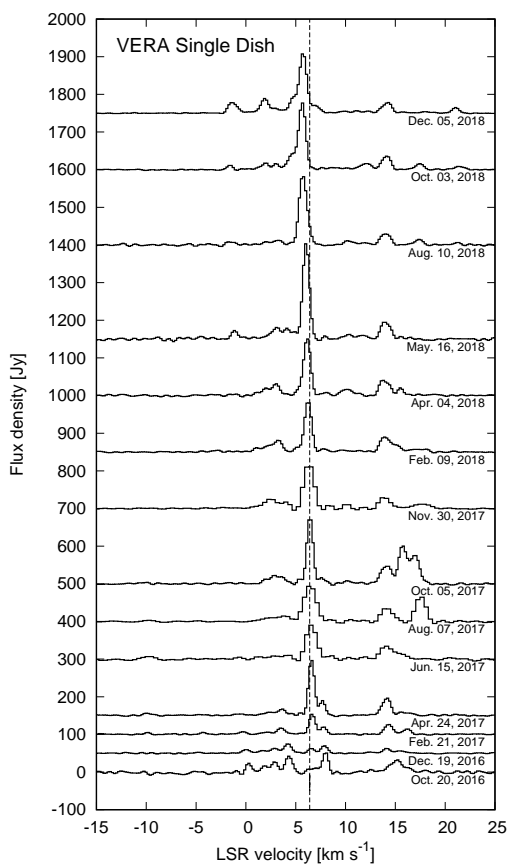}
\caption{Total-power spectra from the VERA 20-m single-dish monitoring observations. 
A vertical dashed line indicates the LSR velocity of 6.4~km~s$^{-1}$, corresponding to the brightest components during the VERA monitoring (Fig. \ref{fig-spVERA}). 
}
\label{fig-spSD}
\end{center}
\end{figure}

Figure \ref{fig-spSD} shows the total power spectra of the 22~GHz H$_{2}$O masers in S255~NIRS~3 observed with the VERA 20-m single-dish antennas. 
Some of the epochs have multiple independent data from different antennas/dates. 
In such cases, we plot the spectra that have higher frequency resolution or higher sensitivity in Fig. \ref{fig-spSD}. 
By comparing the data acquired with different VERA antennas, we conservatively estimate an uncertainty on the flux measurements of $\sim$30\%, due to calibration and/or pointing errors. 
The brightest spectral components can be seen at 6.4~km~s$^{-1,}$ as observed during VLBI monitoring (see below), which shows a slight offset from the systemic velocity of 5.2~km~s$^{-1}$ \citep{Wang2011, Zinchenko2015}. 
A significant flux increase is seen on April 24, 2017, which implies that such an increase began between February and April, 2017, namely about 21-23~months after the 6.7~GHz CH$_{3}$OH maser flare detected on July 07, 2015 \citep{Fujisawa2015, Moscadelli2017, Szymczak2018} and seven-to-nine~months after the flux increase of the radio continuum emission in July 2016 \citep{Cesaroni2018}. 
Other velocity components in the velocity range of 0 to 20~km~s$^{-1}$ are clearly detected and are highly variable. 

\begin{figure}[ht]
\begin{center}
\includegraphics[width=7cm]{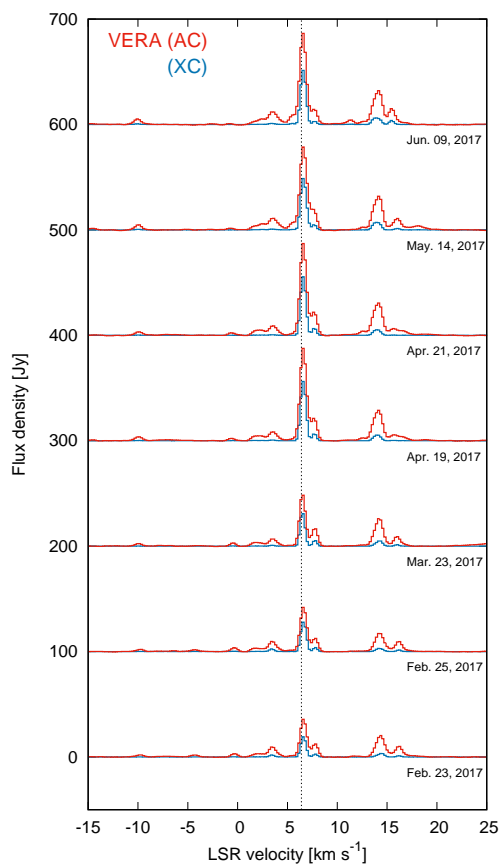}
\caption{
Auto-correlation total-power (solid red lines, AC) and scalar-averaged cross-power (blue solid lines, XC) spectra of the H$_{2}$O masers. 
The spectra were computed by averaging all the antennas/baselines and time ranges. 
Baselines were subtracted by fitting the line-free channels with a polynomial function. 
A vertical dashed line indicates the LSR velocity of 6.4~km~s$^{-1}$, corresponding to the brightest components. 
}
\label{fig-spVERA}
\end{center}
\end{figure}

\begin{figure}[ht]
\begin{center}
\includegraphics[width=7cm]{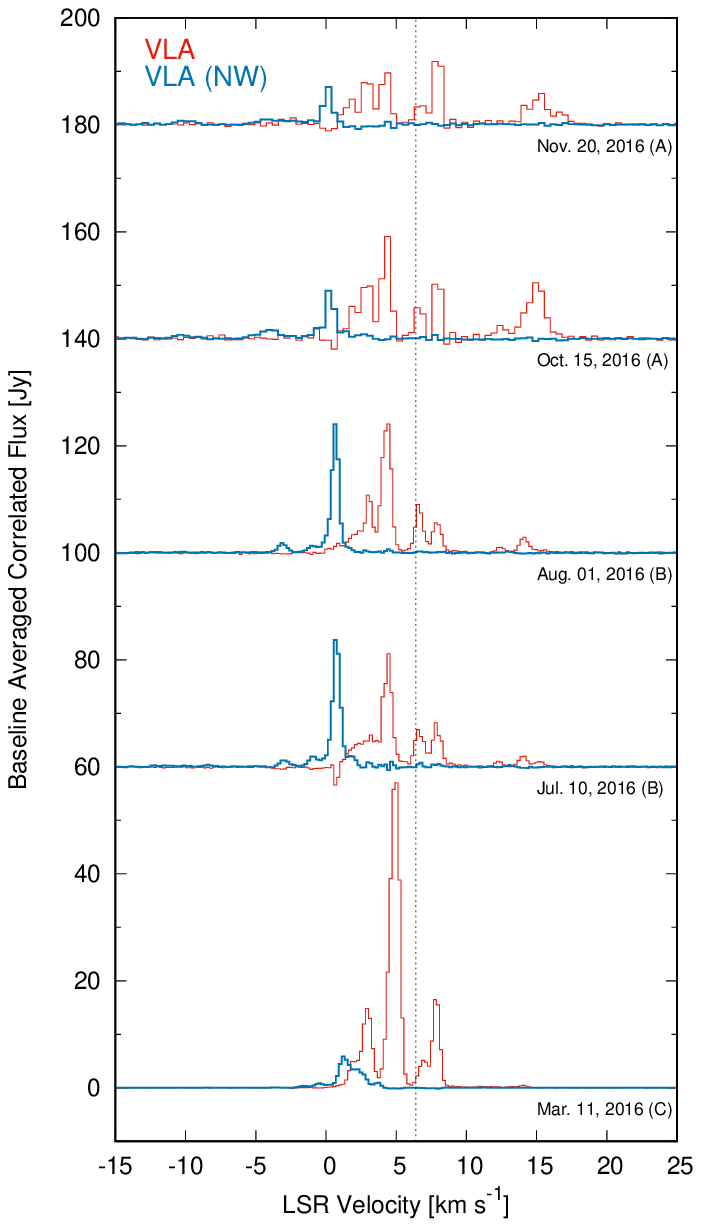}
\caption{Velocity-integrated flux densities of the 22~GHz H$_{2}$O masers observed with the VLA. 
A vertical dashed line indicates the LSR velocity of 6.4~km~s$^{-1}$, corresponding to the brightest components during the VERA monitoring (Fig. \ref{fig-spVERA}). 
The red and blue lines correspond to the spectrum integrated around the central source NIRS~3 and that of the newly found feature offset by 5\arcsec\ northwest (NW) of NIRS~3, respectively. }
\label{fig-spVLA}
\end{center}
\end{figure}

The H$_{2}$O maser spectra also exhibit significant changes in their peak velocities. 
In particular, the redshifted components at $\sim$15~km~s$^{-1}$ show drastic changes in their spectral profiles. 
Such variations were observed in a long-term monitoring of this source \citep{Felli2007}. 
In addition, the 6.4~km~s$^{-1}$ feature seems to drift in velocity in the interval between February 09, 2018 and December 05, 2018. 
Similar velocity drifts were found in the high-mass protostar IRAS~20126+4106, and in that case the motion was interpreted as deceleration \citep[Fig. 5 in ][]{Moscadelli2005}. 

The auto-correlation and cross-correlation spectra of the 22~GHz H$_{2}$O masers taken with VERA are shown in Fig. \ref{fig-spVERA}. 
By comparing the flux densities of two nearby epochs, February 23 and 25, 2017, we estimate a flux calibration error of $\sim$30\% for the cross-correlated spectrum. 
However, the spectra of two other epochs, April 19 and 21, 2017 show excellent agreement with each other, within a few \%. 
The cross-correlated flux density is increased from 19~Jy to 56~Jy at the spectral peak around 6.4~km~s$^{-1}$. 
The total power fluxes obtained from the auto-correlation data are 37-88~Jy at 6.4~km~s$^{-1}$. 
The ratio of the peak cross-correlation flux density to the total power flux is 60\%, and it is almost unchanged during the VLBI monitoring observations for this 6.4~km~s$^{-1}$ feature. 
On the other hand, the higher and lower velocity spectral features are more dominant in the total-power spectra, while the cross-correlated spectra show marginal detection in some of the velocity components. 
This implies that they are spatially more extended than the synthesized beam size, and hence resolved out with the VERA baselines.
We discuss this issue further in Sect.~\ref{subsec-extend}.

Figure \ref{fig-spVLA} shows the integrated spectra observed with the VLA in C-configuration (March 11, 2016), B-configuration (July 10 and August 01, 2016), and A-configuration (October 15 and November 20, 2016). 
The spectra shown in this plot were extracted from the VLA image cubes. 
The velocity of the strongest feature observed with the VLA is $\sim$5~km~s$^{-1}$, which is significantly different from those observed during the VERA monitoring in 2017-2018. 
The maser emission associated with NIRS~3 was brightest in the first epoch, on March 11, 2016, and it decreased after July 10, 2016. 
This decreasing trend in flux densities could be due either to variability of the maser or to the different array configurations used at the different epochs. 
In the latter case, there could be significant missing flux emitted from spatially extended components, in particular for the A-configuration data. 

The picture is further complicated by the presence of a secondary source of maser emission that we discovered about 5\arcsec\ northwest (NW) of NIRS~3 at the local-standard-of-rest (LSR) velocity of 0~km~s$^{-1}$. 
The position of this NW feature is RA(J2000)=06h12m53.7711s and Decl=+17$^{\circ}$59\arcmin26.208\arcsec. 
This feature increased in flux density during the monitoring and became brightest on August 01, 2016. 
At the maximum phase, the flux density was comparable to that of the main feature around NIRS~3. A more detailed discussion of the flux variability can be found in Sect. \ref{subsec-variability}. 

\subsection{VLBI mapping of the 22~GHz H$_{2}$O masers}
\label{subsec-vlbi}

\begin{table*}[hbt]
\begin{center}
\caption{Positions and proper motions of all features.}
\label{tab-feature}
\rotatebox{90}{
\begin{tabular}{rccrrrrr}
\hline
\hline
   & First         & Number of     &  RA offset at & Decl offset at & Proper motion in & Proper motion in &  LSR    \\
   & epoch of      & detected      & first epoch & first epoch &   RA                      & Decl             & velocity  \\
ID & detection     & epochs        & (mas)     & (mas)      & (mas~y$r^{-1}$) & (mas~yr$^{-1}$) & (km~s$^{-1}$) \\
\hline
 1 & Feb. 23, 2017  & 7 &   -7.56$\pm$0.17  & -20.73$\pm$0.19   & -1.36$\pm$0.38   & -0.88$\pm$0.58 &    1.62 \\
 2 & Feb. 23, 2017  & 7 &   -3.84$\pm$0.04  &  -1.15$\pm$0.01   & -0.09$\pm$0.32   &  0.47$\pm$0.21 &    2.41 \\
 3 & Mar. 25, 2017  & 5 &  -10.03$\pm$0.16  & -23.96$\pm$0.06   &  0.16$\pm$0.49   & -0.79$\pm$0.29 &    3.04 \\
 4 & Feb. 23, 2017  & 7 &  -10.85$\pm$0.02  & -30.90$\pm$0.01   & -3.23$\pm$0.34   & -1.61$\pm$0.20 &    3.22 \\
 5 & Feb. 25, 2017  & 3 &  -11.39$\pm$0.02  & -29.05$\pm$0.03   & -3.46$\pm$1.54   & -2.20$\pm$1.03 &    3.67 \\
 6 & May. 14, 2017  & 2 &  -11.71$\pm$0.05  & -27.08$\pm$0.01   &  ---             &  ---           &    3.90 \\
 7 & Feb. 23, 2017  & 7 &   19.63$\pm$0.03  & -39.59$\pm$0.01   &  2.61$\pm$0.50   & -0.67$\pm$0.46 &    3.96 \\
 8 & Mar. 25, 2017  & 2 &  230.30$\pm$0.14  & 299.38$\pm$0.08   &  ---             &  ---           &    5.04 \\
 9 & Feb. 23, 2017  & 4 &   14.06$\pm$0.18  & -45.21$\pm$0.08   &  0.45$\pm$0.37   & -4.01$\pm$0.33 &    5.08 \\
10 & Feb. 25, 2017  & 5 &  230.41$\pm$0.02  & 284.59$\pm$0.01   &  2.12$\pm$0.25   & -2.06$\pm$0.15 &    5.37 \\
11 & Feb. 23, 2017  & 1 &  223.22$\pm$0.08  & 281.29$\pm$0.21   &  ---             &  ---           &    6.42 \\
12 & Feb. 23, 2017  & 1 &  230.05$\pm$0.22  & 314.61$\pm$0.18   &  ---             &  ---           &    6.41 \\
13 & Feb. 23, 2017  & 7 &  226.79$\pm$0.06  & 299.80$\pm$0.02   &  0.89$\pm$0.16   &  0.04$\pm$0.14 &    6.40 \\
14 & Apr. 19, 2017  & 1 &  223.56$\pm$0.15  & 310.13$\pm$0.24   &  ---             &  ---           &    7.48 \\
15 & Feb. 23, 2017  & 7 &  222.00$\pm$0.06  & 301.14$\pm$0.03   & -0.57$\pm$0.52   &  0.40$\pm$0.56 &    7.59 \\
16 & May. 14, 2017  & 2 &  182.18$\pm$0.16  & 281.35$\pm$0.19   &  ---             &  ---           &   11.57 \\
17 & Apr. 21, 2017  & 3 &   71.39$\pm$0.16  & 127.44$\pm$0.53   &  0.20$\pm$0.55   &  1.33$\pm$2.37 &   12.45 \\
18 & Feb. 25, 2017  & 5 &  185.55$\pm$0.28  & 280.16$\pm$0.16   & -0.15$\pm$0.62   &  3.66$\pm$0.66 &   13.62 \\
19 & Feb. 23, 2017  & 7 &   78.37$\pm$0.17  & 143.48$\pm$0.11   & -1.06$\pm$0.38   & -0.13$\pm$0.27 &   13.84 \\
20 & Apr. 19, 2017  & 3 &  186.66$\pm$0.35  & 281.58$\pm$0.35   &  0.22$\pm$0.58   &  1.89$\pm$0.86 &   14.33 \\
21 & Feb. 23, 2017  & 2 &  179.20$\pm$0.07  & 279.93$\pm$0.09   &  ---             &  ---           &   14.88 \\
22 & Feb. 23, 2017  & 7 &  188.62$\pm$0.06  & 281.72$\pm$0.05   &  1.71$\pm$0.21   &  4.37$\pm$0.21 &   15.96 \\
23 & May. 14, 2017  & 1 &   46.76$\pm$0.02  &  66.49$\pm$0.04   &  ---             &  ---           &   17.79 \\
24 & May. 14, 2017  & 1 &   48.43$\pm$0.02  &  62.95$\pm$0.03   &  ---             &  ---           &   17.90 \\
25 & May. 14, 2017  & 1 &  103.95$\pm$0.10  & 124.09$\pm$0.13   &  ---             &  ---           &   18.66 \\
\hline
\multicolumn{8}{l}{Feature 13 is the reference feature for fringe fitting and proper motion measurements. } \\
\multicolumn{8}{l}{The absolute position is measured by phase-referencing analysis for feature 19. } \\
\multicolumn{8}{l}{Proper motion is measured with respect to that of the barycenter. A proper motion of 1~mas~yr$^{-1}$ corresponds to 8.44~km~s$^{-1}$. }
\end{tabular}
}
\end{center}
\end{table*}

\begin{figure*}[th]
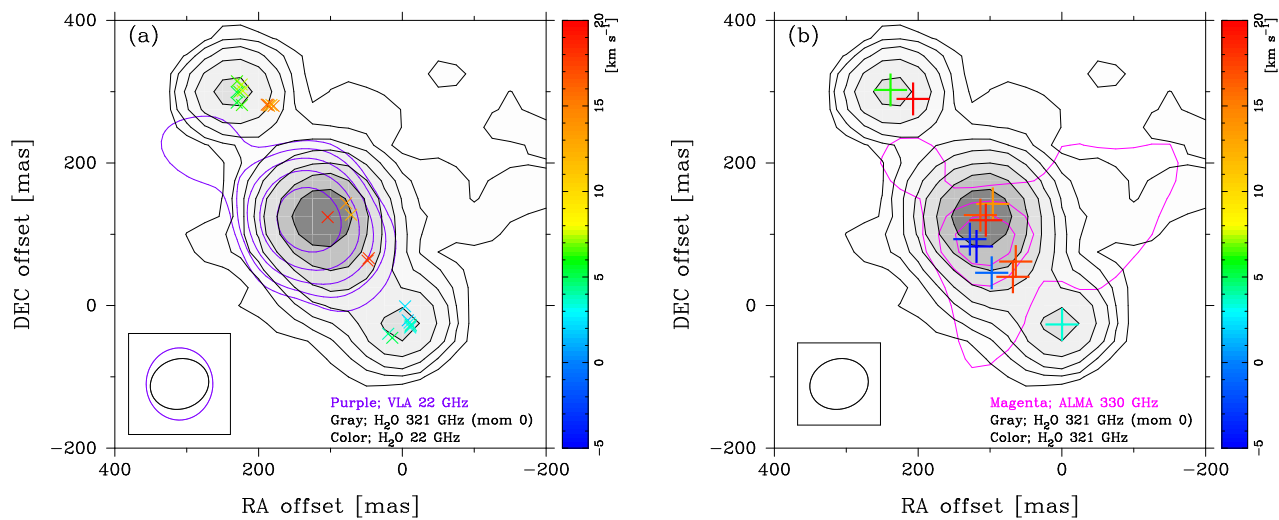

\begin{center}
\includegraphics[width=8cm]{fig04a.eps}
\hspace*{5mm}
\includegraphics[width=8cm]{fig04b.eps}
\caption{(a) Distribution of the 22~GHz H$_{2}$O maser features (colored symbols; Table \ref{tab-feature}) superposed on the VLA K-band continuum observed on December 27, 2016 \citep[purple contours; ][]{Cesaroni2018} and the moment~0 map of the 321~GHz H$_{2}$O maser emission (gray scale). 
Contour levels are 4, 8, 16, ... times the rms noise level of the 0.13~mJy~beam$^{-1}$ and the 32~mJy~beam$^{-1}$~km~s$^{-1}$ for the VLA K-band continuum and the moment 0 map of the 321~GHz H$_{2}$O, respectively. 
The (0, 0) position is the phase tracking center, RA(J2000)=06h12m54.006s and Decl=+17$^{\circ}$59\arcmin22.96\arcsec. 
The synthesized beam size of the ALMA Band-7 observation is shown in the bottom -eft corner of each panel. 
(b) Distribution of peak centroids of the 321~GHz H$_{2}$O maser features (colored symbols) superposed on the ALMA Band-7 continuum (magenta contours) and the moment~0 map of the 321~GHz H$_{2}$O maser emission (gray scale). 
Contour levels for the ALMA Band-7 continuum are 32, 64, and 128 times the rms noise level of 0.33~mJy~beam$^{-1}$, showing only the brightest part around the continuum peak. 
}
\label{fig-mapALMA}
\end{center}
\end{figure*}

The positions of the maser features were determined through VERA observations for seven epochs in 2017, as listed in Table \ref{tab-obs}. 
Spatial distributions and radial velocities of the H$_{2}$O masers detected in at least one of the observed epochs are plotted in the left panel of Fig. \ref{fig-mapALMA}. 
Details of the identified maser features are given in Table \ref{tab-feature} and Appendix \ref{sec-appendix} (see discussion below). 
The H$_{2}$O maser features are aligned along the northeast-southwest (NE-SW) direction. 
The most blueshifted masers tend to be located at the SW part, while the systemic and redshifted components are distributed on the opposite side, toward NE. 
The brightest feature at 6.4~km~s$^{-1}$, identified as feature ID13, is located in the NE lobe. 
Around the midway, or slightly south of the midpoint between these two main maser clusters, there are two smaller clusters including the most redshifted components. 
The position angle of the H$_{2}$O maser distribution is slightly different from that of the VLA K-band continuum emission \citep{Cesaroni2018}, which traces the radio jet driven by the central protostar. 

\begin{figure}[th]
\begin{center}
\includegraphics[width=9cm]{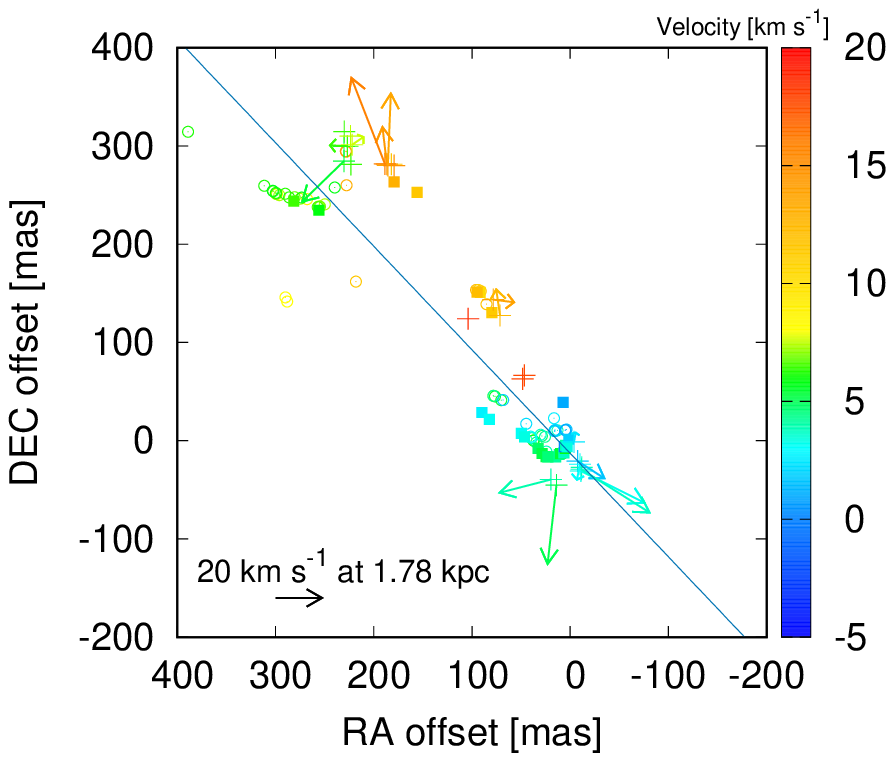}
\caption{Distributions of the 22~GHz H$_{2}$O masers indicated by the plus symbols with the proper motion vectors measured with respect to the barycenter. 
The color code indicates the LSR velocities of the maser features. 
The 22~GHz H$_{2}$O maser features observed in previous observations by \cite{Goddi2007} in 2005 and \citet{Burns2016} in 2010 are also plotted with open circles and filled squares, respectively. 
The solid line indicates the outflow axis derived from a linear fit to the positions of all maser features, including previous results. }
\label{fig-mapVLBI}
\end{center}
\end{figure}

The proper motions of all maser features were measured by fitting their position offsets as a function of time. 
In the present study, the monitoring period of about three~months was too short to determine a reliable value of the annual parallax. 
Furthermore, a detailed analysis of absolute motions and annual parallax measurements for the H$_{2}$O masers was already made by \citet{Burns2016}. 
Thus, we only focus on the internal proper motions of the H$_{2}$O masers, which trace outflow motions, as we discuss later. 
For this purpose, we followed the method described in \citet{Burns2016}. 
Details of the adopted procedure are given in Appendix \ref{sec-appendix}. 

We successfully produced phase-referenced images for five epochs: February 23, February 25, April 19, May 14, and June 09, 2017. 
The absolute coordinates of the reference maser spot ID19 derived from the Gaussian fitting to the spot map are
\begin{eqnarray*}
\mbox{RA(J2000)} &=& 06\mbox{h}12\mbox{m}54.0114935\pm 0.0000007\mbox{s} \\
\mbox{Decl(J2000)} &=& +17^{\circ}59\arcmin23.103476\pm 0.000011\arcsec
\end{eqnarray*}
for the first epoch, that is, February 23, 2017. 
The proper motions of all maser features with respect to the barycenter of S255~NIRS~3 are listed in Table \ref{tab-feature}. 
Due to the short monitoring period of three~months, some proper motions have large uncertainties. 
However, we can reveal systematic proper motions in the outflow, as shown in Fig. \ref{fig-mapVLBI}. 
The less ordered proper motions in the NE lobe could be due to interaction with the burst-regenerated radio jet (see Fig. \ref{fig-mapALMA}(a)).

Figure \ref{fig-mapVLBI} shows the maser features detected in all previous works \citep{Goddi2007, Burns2016}, and in the present study. 
These maps are registered with respect to the absolute coordinate frame with uncertainties of $\sim$0.5~mas (see Sect. \ref{subsec-vera}). 
The spatial and velocity structures traced by the H$_{2}$O masers are in good agreement with those of previous observations in 2005 made with the Very Long Baseline Array \citep[VLBA; ][]{Goddi2007} and conducted with VERA in 2010 \citep{Burns2016}, as mentioned above. 
The features are aligned along the NE-SW direction with a position angle of 43.5~degrees obtained by fitting a line to their positions (solid line in Fig. \ref{fig-mapVLBI}), as explained
in Sect.~\ref{subsec-distribution}. 

\begin{figure}[th]
\begin{center}
\includegraphics[width=7cm]{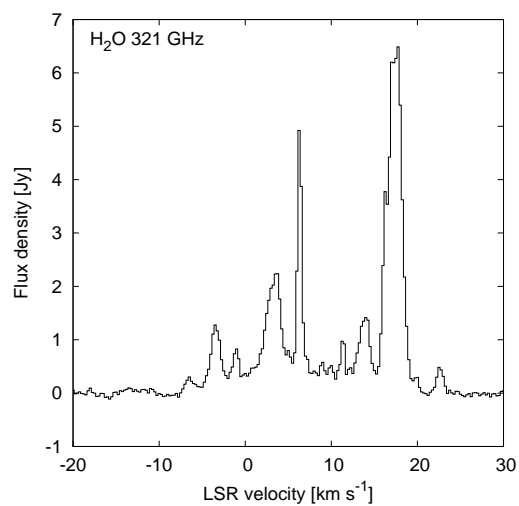}
\caption{Cross-power spectrum of the 321~GHz H$_{2}$O maser integrated over the 0.5\arcsec$\times$0.5\arcsec \ region. }
\label{fig-321GHz}
\end{center}
\end{figure}

\begin{table}[hbt]
\begin{center}
\caption{Summary of ALMA Band-7 continuum data.}
\label{tab-cont}
\begin{tabular}{lc}
\hline
\hline
Parameter & Gaussian fitting results \\
\hline
Date & December 02, 2017 \\ 
Peak intensity & 77.2$\pm$0.2~mJy~beam$^{-1}$ \\
                        & 127.4$\pm$0.8~mJy~beam$^{-1}$$^{a}$ \\
Flux density & 132.5$\pm$0.5~mJy \\
                    & 230.4$\pm$2.1~mJy$^{a}$ \\
Peak RA(J2000) & 06h12m54.01348s \\ 
Peak Decl(J2000) & 17d59\arcmin23.0525\arcsec \\ 
Measured size &   (93.3$\pm$0.3)~mas~$\times$~(80.2$\pm$0.2)~mas \\
Position angle & 100.6$\pm$0.7~degrees  \\
Deconvolved size &   (57.9$\pm$0.7)~mas~$\times$~(53.6$\pm$0.6)~mas \\
Position angle      & 88.8$^{+7.5}_{-5.9}$~degrees  \\
\hline
\multicolumn{2}{l}{$a$: Results from the deconvolved beam size of 0.14\arcsec. }
\end{tabular}
\end{center}
\end{table}

\subsection{The ALMA Band-7 continuum emission}
\label{subsec-cont}

As shown in the right panel of Fig. \ref{fig-mapALMA}, the peak of the submillimeter continuum emission is located on the outflow axis and is close to the central redshifted maser features. 
The submillimeter continuum peak is consistent with that observed previously with the Submillimeter Array (SMA) and ALMA, identified as SMA1 \citep{Wang2011, Zinchenko2015, Liu2018}. 
It is also coincident with the radio continuum emission peak observed with the VLA at various wavelengths \citep[see the left panel of Fig. \ref{fig-mapALMA}; ][]{Moscadelli2017, Cesaroni2018}. 
The intensity, flux density, peak position, and size of the submillimeter emission are determined by 2D Gaussian fitting on the image, as listed in Table \ref{tab-cont}. 
The size is slightly larger than the synthesized beam, and the deconvolved size is 57.9~mas$\times$53.6~mas with a position angle of 89~degrees. 
The deconvolved size is not elongated in the direction of the outflow, unlike the radio continuum emission tracing the newly ejected jet \citep{Cesaroni2018}. 
Thus, the submillimeter continuum source could trace the dust emission from an envelope and/or disk associated with the driving source of the NE-SW outflow. 

The flux density of the submillimeter source at $\sim$330~GHz was measured on December 02, 2017, after previous observations with the VLA, SMA, and ALMA \citep{Zinchenko2015, Moscadelli2017, Cesaroni2018, Liu2018}. 
The peak intensity and integrated flux density of the previous observations at 335.4~GHz were 0.14~Jy~beam$^{-1}$ and 0.41~Jy, respectively, on July 20, 2017 \citep{Liu2018}. 
The peak intensity of our measurement at almost the same frequency is $\sim$2 times lower than that of \citet{Liu2018}. 
It is most likely that the twice higher resolution ($\sim$0.066\arcsec) with respect to the previous ALMA observation (0.14\arcsec) can justify the different flux values. 
In fact, the integrated flux density over the 0.5\arcsec \ aperture is measured to be 440~mJy for our data, consistent with that from the previous observations \citep{Liu2018}. 
In order to compare these results, we also produced a continuum image with the same deconvolved beam size of 0.14\arcsec \ as the previous observations by \cite{Liu2018}. 
In this case, the peak intensity and integrated flux density are 127.4$\pm$0.8~mJy~beam$^{-1}$ and 230.4$\pm$2.1~mJy, respectively. The integrated flux density within the 0.5\arcsec aperture (420~mJy) does not depend on the resolution. 
The larger value of the integrated flux density than that of the Gaussian fitting (Table \ref{tab-cont}) is attributed to the diffuse extended emission around the compact continuum peak. 
Given the flux calibration accuracy and different UV coverage, our results suggest that the submillimeter continuum emission does not show any significant variation between July 20 \citep{Liu2018} and December 02, 2017 (present study). 

\begin{table*}[t]
\begin{center}
\caption{Summary of the 321~GHz and 22~GHz H$_{2}$O maser features.}
\label{tab-321GHz}
\rotatebox{90}{
\begin{tabular}{crrrrrccrrrrr}
\hline
\hline
\multicolumn{6}{c}{321~GHz H$_{2}$O maser} & & \multicolumn{5}{c}{22~GHz H$_{2}$O maser} & \\
\cline{1-6}  \cline{8-12} 
   &                     &                    &                       & Peak         & Brightness      & &    &                      &                       &                    & Peak  & Emissivity  \\

   & RA offset      & Decl offset     & Peak flux    &  velocity  &  temperature     & &    & RA offset      & Decl offset     & Peak flux    &  velocity &  ratio $R$ \\
   
ID & (mas)          &  (mas)         & (Jy)         & (km~s$^{-1}$)  &  (K) & & ID & (mas)          &  (mas)         & (Jy)         & (km~s$^{-1}$) & $L_{22}/L_{321}$ \\
\hline
A  &  118.7$\pm$1.3  &   82.8$\pm$1.2  &  0.189$\pm$0.007  &  -6.50   &   388 & & --- &             ---  &              ---  &     ---  &    ---  &   --- \\
B  &  128.0$\pm$0.3  &   93.1$\pm$0.3  &  0.872$\pm$0.007  &  -3.50   &  1790 & & --- &             ---  &              ---  &     ---  &    ---  &   --- \\
C  &   97.6$\pm$0.5  &   46.0$\pm$0.4  &  0.541$\pm$0.007  &  -1.00   &  1110 & & --- &             ---  &              ---  &     ---  &    ---  &   --- \\
D  &   -0.2$\pm$0.2  &  -26.5$\pm$0.2  &  1.431$\pm$0.007  &   3.50   &  2937 & &  4  & -10.85$\pm$0.02  &  -30.90$\pm$0.01  &   3.778$\pm$0.042  &   3.22  &   2.6 \\
E  &  238.4$\pm$0.1  &  302.5$\pm$0.1  &  3.812$\pm$0.007  &   6.25   &  7825 & & 13  & 226.79$\pm$0.06  &  299.80$\pm$0.02  &  53.994$\pm$0.208  &   6.40  &  14.2 \\
F  &   96.1$\pm$0.2  &  142.7$\pm$0.2  &  1.056$\pm$0.007  &   9.00   &  2168 & & 19  &  78.37$\pm$0.17  &  143.48$\pm$0.11  &  14.909$\pm$0.072  &  13.84  &  14.1 \\
G  &  113.4$\pm$0.1  &  126.8$\pm$0.1  &  4.796$\pm$0.007  &  14.00   &  9845 & & --- &             ---  &              ---  &     ---  &    ---  &   --- \\
H  &   64.2$\pm$0.2  &   61.8$\pm$0.1  &  0.697$\pm$0.007  &  17.00   &  1431 & & 23  &  46.76$\pm$0.02  &   66.49$\pm$0.04  &   1.427$\pm$0.038  &  17.79  &   2.0 \\
I  &   68.2$\pm$0.3  &   40.3$\pm$0.1  &  0.616$\pm$0.007  &  17.25   &  1264 & & --- &             ---  &              ---  &     ---  &    ---  &   --- \\
J  &  105.9$\pm$0.3  &  119.7$\pm$0.3  &  0.877$\pm$0.006  &  17.75   &  1800 & & 25  & 103.95$\pm$0.10  &  124.09$\pm$0.13  &   0.473$\pm$0.038  &  18.66  &   0.5 \\
K  &  207.0$\pm$0.6  &  289.8$\pm$0.5  &  0.426$\pm$0.007  &  18.75   &   874 & & --- &             ---  &              ---  &     ---  &    ---  &   --- \\
\hline
\multicolumn{13}{l}{The 22~GHz H$_{2}$O maser features are taken from epoch May 14, 2017. }
\end{tabular}
}
\end{center}
\end{table*}

\subsection{The 321~GHz H$_{2}$O masers}
\label{subsec-321ghz}

We also observed the 321~GHz H$_{2}$O masers with ALMA. 
Although this strong H$_{2}$O maser emission was identified in single-dish radio observations 30 years ago \citep{Menten1990}, this type of maser has been observed with interferometers in only three other star-forming regions: Cepheus~A \citep{Patel2007}, W49A \citep{Kramer2017}, and Source~I in Orion~KL \citep{Hirota2014a}.
S255~NIRS~3 is the fourth star-forming region for which the 321~GHz H$_{2}$O masers have been imaged. 
Figure \ref{fig-321GHz} shows the spectrum of the 321~GHz H$_{2}$O maser integrated over a 0.5\arcsec$\times$0.5\arcsec \ region. 
There are multiple spike-like velocity components in the range $-5$ to 25~km~s$^{-1}$ covering a slightly larger velocity range than that of the 22~GHz H$_{2}$O maser spectra (Fig. \ref{fig-spVERA}). 
We note that the broader feature from $-10$ to +20~km~s$^{-1}$ could be due to multiple HCOOCH$_{3}$ lines \citep{Hirota2014b}. 

The right panel of Fig. \ref{fig-mapALMA} shows the distribution of the 321~GHz H$_{2}$O maser features superposed on the moment~0 map of the 321~GHz H$_{2}$O maser line, and the continuum emission at 1~mm obtained from the ALMA Band-7 observation. 
The centroid positions of all features were determined via Gaussian fitting of the emission at the peak velocities. 
The moment~0 map shows three peaks: one at the continuum position, and one at each of the NE and SW lobes, which are also detected in the 22~GHz H$_{2}$O maser line. 
Therefore, the 321~GHz H$_{2}$O masers trace the outflow, as suggested for other HMYSOs \citep{Patel2007, Hirota2014a}. 
It is worth noting that there seems to be a velocity gradient in the 321~GHz H$_{2}$O maser spots close to the central continuum peak. 
The direction is perpendicular to the outflow axis, and it is consistent with those detected in other thermal molecular lines \citep{Zinchenko2015}. 
This gradient could trace a rotating disk/envelope or the base of the outflow \citep{Hirota2014a}. 
The 321~GHz maser detections are given in Table \ref{tab-321GHz}, where we list positions, velocities, and flux densities for the spots at the peak velocity in individual features.
Altogether, we identified 11 features at different positions and velocities. 
Among them, five features correspond to the positions of the 22~GHz H$_{2}$O maser features detected on May 14, 2017. 
The other six features, including the most redshifted and blueshifted spots, have no counterpart in the 22~GHz H$_{2}$O maser features, as in the case of Cepheus~A \citep{Patel2007}. 
The peak brightness temperature $T_{b}$ of the 321~GHz H$_{2}$O emission is close to 10$^{4}$~K for the brightest components (see Table \ref{tab-321GHz}). 
This value is to be taken as a lower limit if the emission is unresolved and clearly proves that we are observing maser emission in this water line, although we cannot rule out the possibility of thermal emission for the features with the lowest peak brightness temperatures ($T_{b}$=942~K; Table \ref{tab-321GHz}) and for the broad/weak spectral components (Fig. \ref{fig-321GHz}).

\begin{figure}[th]
\begin{center}
\includegraphics[width=9cm]{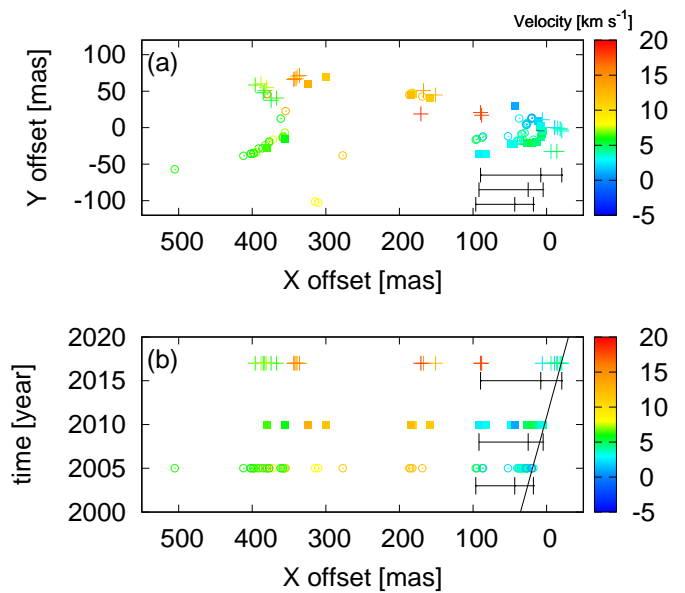}
\caption{(a) Distributions of the 22~GHz H$_{2}$O maser features obtained after rotating the plot in Fig. \ref{fig-mapVLBI} by 43.5~degrees. 
The abscissa is parallel to the outflow as indicated by a solid line in Fig. \ref{fig-mapVLBI}. 
Horizontal bars at the bottom right corner indicate the minimum, maximum, and mean positions of the SW lobe along the outflow axis for each epoch (2005, 2010, and 2017 from bottom to top). 
(b) Same as (a), where the displacement along the Y axis has been replaced by the observing epoch. 
The black solid line represents the movement of the head of the bow shock along the outflow axis. 
}
\label{fig-rotateVLBI}
\end{center}
\end{figure}

\begin{figure}[tbh]
\begin{center}
\includegraphics[width=9cm]{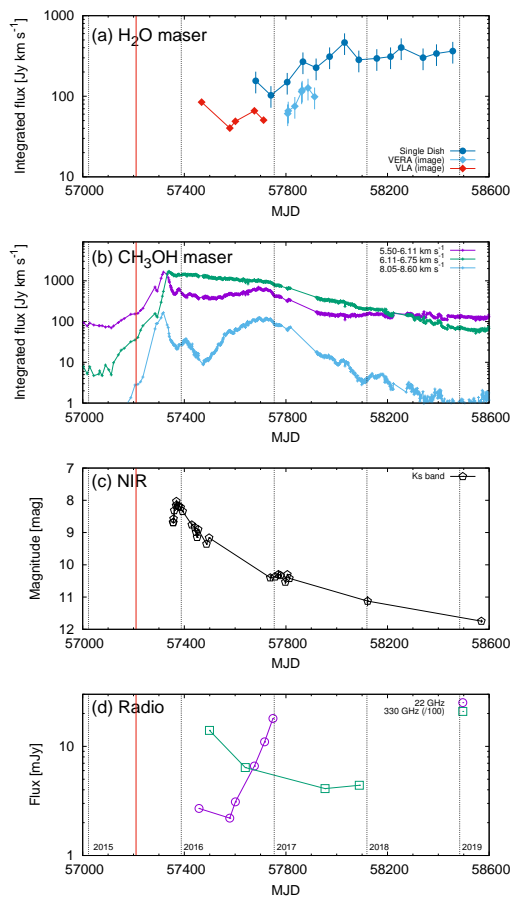}
\caption{Light curves of maser lines and continuum emissions. 
A red vertical line indicates the start of the 6.7~GHz methanol maser flare (July 07, 2015) reported by \citet{Fujisawa2015}. 
Dashed lines indicate the dates of January 01, from 2015 to 2019. 
(a) Flux densities of the H$_{2}$O masers integrated between $-15$ and $+25$~km~s$^{-1}$, from the VLA and VERA observations. 
The VERA results include the single-dish monitoring and the integrated spectra extracted from the VLBI image cubes. 
The VLA data were extracted from the image cubes, excluding the NW component. 
(b) Flux densities of representative 6.7~GHz methanol maser features observed with the Hitachi 32-m radio telescope \citep{Yonekura2016}, showing different behaviors among the light curves. 
(c) IR light curve at Ks band observed with the 1.5-m Kanata telescope in Hiroshima, Japan \citep{Uchiyama2020}. 
(d) Variation of the radio continuum emission at 22~GHz observed with the VLA \citep{Cesaroni2018} and ALMA Band 7 at $\sim$330~GHz \citep[][present paper]{Liu2018}. 
For the 330~GHz continuum emission on December 02, 2017 measured in the present study, we used the value obtained by integrating over a 5\arcsec aperture (420~mJy) rather than the result of the Gaussian fit (see discussion in Sect. \ref{subsec-cont}). }
\label{fig-variability}
\end{center}
\end{figure}

\section{Discussion}
\label{sec-discussion}

\subsection{Overall distribution of the 22~GHz H$_{2}$O masers}
\label{subsec-distribution}

The position angle of the NE-SW outflow traced by the 22~GHz H$_{2}$O maser features was derived to be 43.5$\pm$1.1~degrees (1$\sigma$ error) by fitting a line to the maser positions on the map. 
This is slightly different from that of the radio jet observed on December 27, 2016 \citep{Cesaroni2018}. 
On the other hand, the 22~GHz H$_{2}$O maser distribution coincides well with the 321~GHz H$_{2}$O masers.
The difference between the radio jet and maser orientations could be interpreted as the result of different outflow/jet directions in multiple/episodic ejection events;
the masers could be excited in the shock front away from the central protostar produced by a previous ejection episode. 
One could even speculate the mass accretion burst could have been responsible for changing the disk orientation. 

We find that the post-burst water maser positions and proper motions described in this work are basically the same as those in the pre-burst epochs \citep{Goddi2007, Burns2016}. 
The SW bow shock is persistent from the first VLBI observations in 2005 to the present observations from 2017 (Fig. \ref{fig-mapVLBI}). 
As found by \citet{Burns2016}, the shock front is moving toward the SW away from the central HMYSO. 
Figure \ref{fig-rotateVLBI} shows the distribution of the maser features rotated by 43.5~degrees from north to east, where the abscissa corresponds to the outflow axis fit on the image. 
The distribution of maser features in the SW lobe extends over 78~mas, 87~mas, and 110~mas along the outflow axis in 2005 \citep{Goddi2007}, 2010 \citep{Burns2016}, and 2017 (present study), respectively (horizontal bars in Fig. \ref{fig-rotateVLBI}). 
In the present study, there are redshifted features in the SW lobe that only appeared in 2017 at the trailing part of the lobe. 
If these two features are excluded, the distribution of the maser features in the SW lobe is smaller in 2017 than in previous epochs. 
On the other hand, the positions of the head of the bow shock measured as the minimum value along the outflow axis are moving significantly from the center. 
The position offsets between two consecutive epochs are 13~mas and 25~mas for 2005-2010 and 2010-2017, respectively. 
These offsets are considerably larger than the astrometric accuracy of VERA and the VLBA. 
The proper motion of the head of the bow shock is 3.26$\pm$0.27~mas~yr$^{-1}$ (1$\sigma$) toward the SW, which corresponds to 28~km~s$^{-1}$. 
This value is in good agreement with those of the maser features given in Table \ref{tab-feature}. 
If the bow shock is propagating from the continuum peak position without acceleration or deceleration, the dynamical timescale of the outflow traced by the H$_{2}$O masers is about 60~years. 
\citet{Burns2016} estimated a dynamical timescale of $<$130~years for the NE-SW outflow, based on the proper motion measurements of the H$_{2}$O maser features in two epochs between 2005 and 2010. 
We improve on their result using three epochs of data from 2005 to 2017, and our new value is consistent with the upper limit reported by them. 

\begin{figure}[tbh]
\begin{center}
\includegraphics[width=8cm]{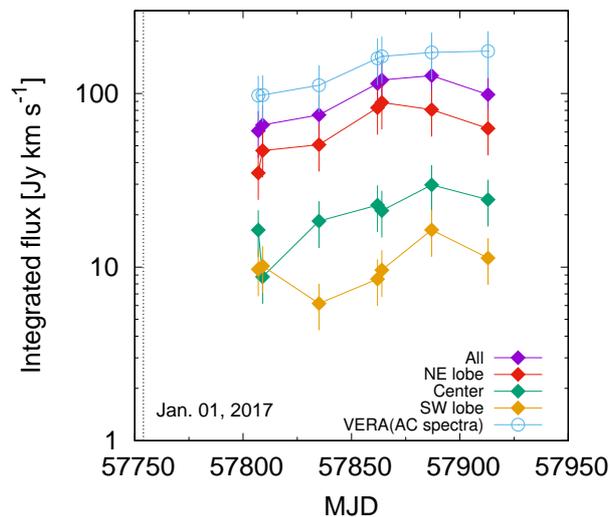}
\caption{Flux densities of the H$_{2}$O maser features in the whole region around NIRS~3, the NE lobe, central (continuum) position, and the SW lobe observed with VERA
(the corresponding spectra are shown in Fig. \ref{fig-ispecVERA}). 
The flux densities were computed by integrating the emission over the velocity range between $-15$ and $+25$~km~s$^{-1}$. 
The error bars represent the 30\% flux uncertainty. 
A dashed vertical line indicates January 01, 2017. }
\label{fig-fluxVERA}
\end{center}
\end{figure}

\begin{figure*}[th]
\begin{center}
\includegraphics[width=18cm]{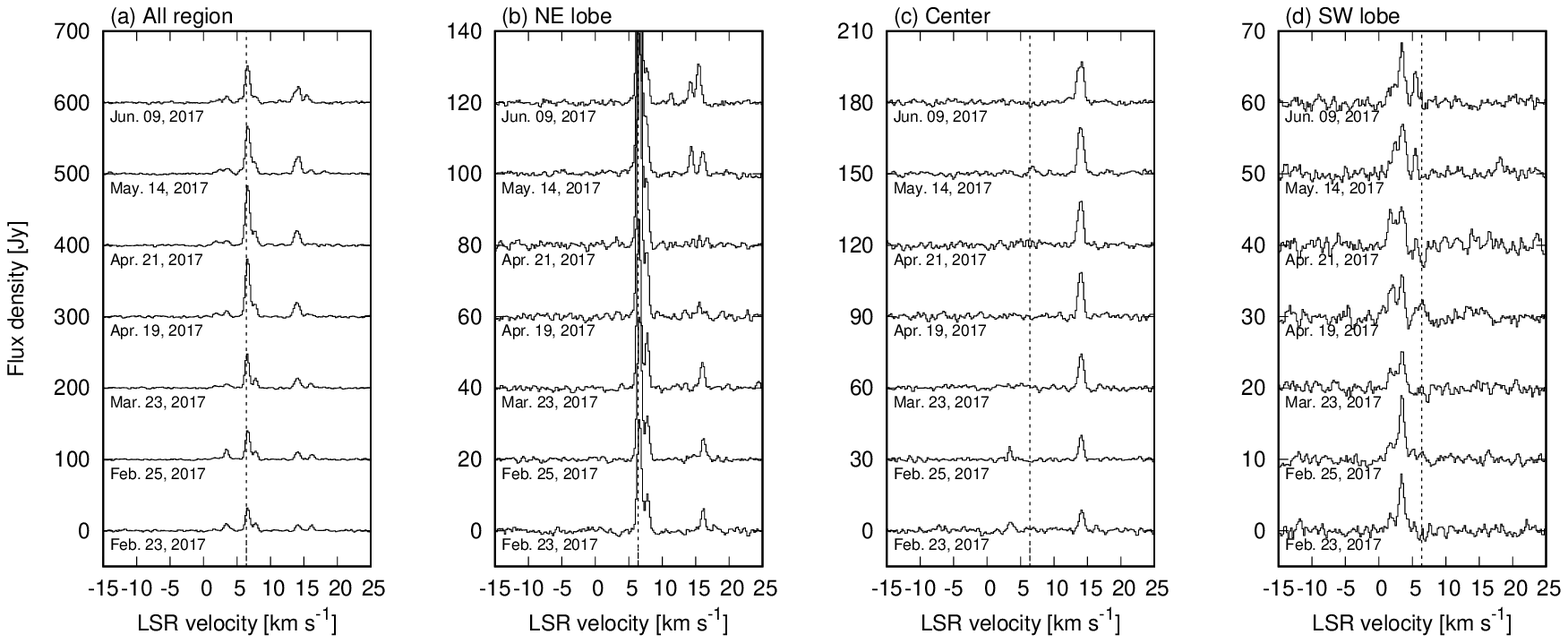}
\caption{Spectra of the H$_{2}$O maser features integrated over different regions in the image cubes from the VERA monitoring data: 
(a) the whole emission region around NIRS~3; (b) the NE lobe; (c) the central (continuum) position; and (d) the SW lobe. 
A vertical dashed line indicates the LSR velocity of 6.4~km~s$^{-1}$, corresponding to the brightest components. 
In panel (b), one can see the profiles of the weakest components at $\sim$15~km~s$^{-1}$, while the brightest features at 6.4~km~s$^{-1}$ are better seen in panel (a). }
\label{fig-ispecVERA}
\end{center}
\end{figure*}

\subsection{Maser flux variability}
\label{subsec-variability}

Flux variations are very common in H$_{2}$O maser sources, and S255~NIRS~3 is no exception to this rule \citep{Felli2007, Goddi2007}. 
This is also evident in previous VLBI observations in the pre-burst phase. 
For instance, \citet{Burns2016} reported that one of the redshifted features labeled J, located close to the continuum position, shows a steady increase from 4 to 82~Jy, while feature G in the SW blueshifted bow shock exhibits a gradual weakening in the observations with VERA from 2008 to 2010. 
The other maser features also show substantial variations \citep[see Fig. 1 in ][]{Burns2016}. 
In addition, the maximum flux densities of 15 and 20~Jy observed, respectively, in the NE and SW lobes by \citet{Goddi2007} are significantly different from those of the present study, although such a difference could be due to the different angular resolutions.

Flux variations of all maser features detected with VERA are shown in Appendix \ref{sec-appendix}. 
Although there is no overall trend in variability, about half (11) of the identified features increase their flux densities, including the brightest feature ID13. 

We plot the light curves of the H$_{2}$O masers in Fig. \ref{fig-variability}. 
For the single-dish monitoring data, we simply integrated them over the whole velocity range where emission is detected, namely from $-15$ to $+25$~km~s$^{-1}$. 
We note that the newly found NW component (see Fig. \ref{fig-spVLA}) could contribute to the total flux by less than 10\%,
if we assume that the 0~km~s$^{-1}$ spectral feature belongs to this NW component. 
Furthermore, the 0~km~s$^{-1}$ feature is not detected in the VERA monitoring (Fig. \ref{fig-spVERA}), and thus cannot affect our variability study of the masers in NIRS3.

The VERA single-dish monitoring suggests a gradual increase in the total flux density after the end of the VLBI monitoring period in 2017, but subsequently, during 2018, the total flux remains almost constant (see Fig.\ref{fig-variability}(a)). 
During VLBI monitoring in the first half of 2017, integrated spectra obtained from the VLBI image cubes follow a trend similar to that of the single-dish monitoring. 
These results suggest a presence of spatially extended emission components resolved out with VERA. We note that the total flux of both the extended and unresolved compact features increase with time.
The existence and variability of an extended maser component is further discussed in Sects. \ref{subsec-extend} and \ref{subsec-origin}.

In Fig. \ref{fig-fluxVERA}, we plot the integrated flux densities extracted for the NE and SW outflow lobes and from the central region close to the continuum peak, of which the spectra are shown in Fig. \ref{fig-ispecVERA}. 
The flux densities are obtained by integrating the VERA spectra over the entire velocity range in which emission is detected (i.e., from $-15$ to $+25$~km~s$^{-1}$), with 30\% uncertainties as discussed in Sect. \ref{subsec-spectra}. 
The NE lobe clearly dominates the total flux, whereas the central region and the SW lobe contribute to only about 30\% and 10\% of the total emission, respectively. 
This could imply that the interaction between the radio jet and the surrounding environment is occurring mostly in the NE lobe.
Significant flaring activity is detected in no regions, and in all cases a moderate flux increase, at most by a factor of 2, is seen during the VERA monitoring period. 
The auto-correlation spectra show the same trend. 

\begin{figure}[th]
\begin{center}
\includegraphics[width=7cm]{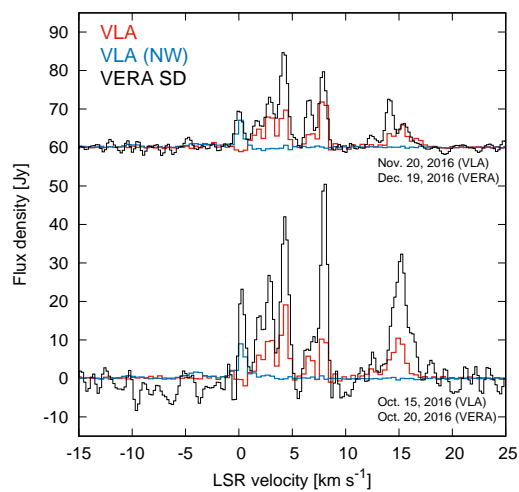}
\caption{Comparison between VERA single-dish and VLA spectra observed at almost the same time in October 2016 and November (VLA)/December(VERA) 2016. 
Black, red, and blue solid lines show, respectively, the total-power spectrum from the VERA single-dish monitoring, the integrated VLA spectrum of NIRS~3, and that of the newly found maser source NW of NIRS~3. }
\label{fig-spVERAVLA}
\end{center}
\end{figure}

Prior to the VERA monitoring, we observed the H$_{2}$O maser spectra with the VLA in 2016. 
When compared with the VERA results (Fig. \ref{fig-spVERA}), the spectral profiles observed with the VLA (Fig. \ref{fig-spVLA}) look significantly different, as we can see plotted in Fig. \ref{fig-spVERAVLA}. 
In the VLA spectra, only weak emission at 6.4~km~s$^{-1}$ is detected, whereas at this velocity one finds the brightest emission during the VERA observations. 
The 5~km~s$^{-1}$ feature becomes less prominent at the end of the VLA monitoring and it can hardly be seen in the VERA spectra, although there may be a corresponding feature slightly more blueshifted at a few km~s$^{-1}$. 
The most redshifted 15~km~s$^{-1}$ feature is gradually increasing its flux, as is clearly seen during the VERA monitoring and the total-power spectra especially. 

Figure \ref{fig-variability}(a) seems to show a flux increase of the H$_{2}$O masers between mid 2016 (July and August) and late 2016 (October and November). 
In the latter two epochs, the VLA was in a more extended configuration (A-configuration) than in the former (B-configuration), which rules out the possibility that this increase could be due to part of the flux being resolved out in the most extended configuration. 
However, the increase is not very prominent, and the flux appears to decrease between October and November. 
Furthermore, the VLA spectra in October and November are partly resolved out compared to the corresponding single-dish measurements (Fig. \ref{fig-spVERAVLA}), and the total flux density of the VLA C-configuration data on March 11, 2016 is almost at the same level as those observed with the VERA single-dish monitoring in late 2016. 
We conclude that the total flux has probably undergone some variation between March and November 2016, but a true persistent increase was only observed during 2017, as proven by our single-dish monitoring (blue line in Fig. \ref{fig-variability}(a)). 

\subsection{Extended H$_{2}$O maser emission}
\label{subsec-extend}

We already highlighted a difference between the total flux recovered by the VERA interferometer and the corresponding single-dish flux.
This finding hints at the existence of extended H$_2$O maser emission. 
Our VLA observations seem to confirm this idea, because the flux measured with the VLA decreases significantly as the array configuration becomes more and more extended (from C to B, to A), as one can see in Fig. \ref{fig-spVLA}. 
While such an effect could be the result of intrinsic variability of the maser emission, the comparison between the VLA spectrum obtained on October 15 and November 20, 2016 with the VERA single-dish spectra observed at almost the same time (on October 20 and December 19, 2016, see Fig. \ref{fig-spVERAVLA}) proves that 40-50\% of the maser emission is indeed resolved out by the A-array configuration of the VLA. This implies that such an extended emission must arise from a region of at least $\sim$1\arcsec\ (1780 au).
To our knowledge, this is the first time that water maser emission has been resolved out with the VLA in a star-forming region. This finding raises questions about the actual origin of the emission, as collisions in shock fronts appear inadequate to pump the 22 GHz line over such a large scale.
We discuss this issue and propose a possible explanation in Sect. \ref{subsec-origin}.

\subsection{Origin of the maser flux increase in relation to the accretion burst}
\label{subsec-origin}

While our observations have revealed no strong flare in the 22~GHz H$_{2}$O maser line, Fig. \ref{fig-variability}(a) clearly shows that a steady increase of the maser flux was present for almost a year, starting from the beginning of 2017. In the following, we attempt to explain the origin of such an increase.

In Fig. \ref{fig-variability}(c) and Fig. \ref{fig-variability}(d), we plot the total flux emitted by S255~NIRS~3 at NIR \citep{Uchiyama2020}, submillimeter \citep[][and present results]{Liu2018}, and radio wavelengths \citep{Cesaroni2018} on the same timescale used in Fig. \ref{fig-variability}(a) for the H$_2$O masers.
The comparison between the different light curves is very informative.
As noted by \citet{Cesaroni2018}, the time lag between the NIR and radio bursts indicates that the latter cannot be directly related to the former, due to the long time lag ($\ga$7~months) between the peak of the NIR burst and the beginning of the radio brightening.
\citet{Cesaroni2018} demonstrated that the increase of the centimeter radio flux can be due to the expansion of a thermal jet from NIRS~3. Vice versa, the submillimeter flux increase occurs prior to that at radio wavelengths, and the delay of $<$4 months with respect to the NIR peak is consistent with the heating of the dusty core enshrouding NIRS3 with IR photons.

It is suggested that the 22~GHz H$_{2}$O masers show anticorrelation with the CH$_{3}$OH masers. 
For instance, an intermediate-mass protostar in G107.298+5.639 exhibits alternating flares of the H$_{2}$O and CH$_{3}$OH masers with a period of 34.4 days, even though these two masers spatially coexist around the same protostar within 360~au \citep{Szymczak2016}. 
Although the detailed mechanism of the anticorrelation between the H$_{2}$O and CH$_{3}$OH masers in G107.298+5.639 is still unclear, it could be caused by the different physical conditions required for these masers. 
One possible explanation is that the increase of the NIR emission has opposing effects on the intensities of the H$_{2}$O and CH$_{3}$OH masers. 
The 22~GHz H$_{2}$O maser could be quenched by the NIR radiation, while the 6.7~GHz CH$_{3}$OH maser could be radiatively pumped by NIR emission, as suggested by \citet{Szymczak2016}. 
However, this is probably not the case for S255~NIRS~3 because the flare of the CH$_{3}$OH masers does not occur at the same time as the fading of the H$_{2}$O masers \citep{Moscadelli2017}, and these two different masers could reflect different physical conditions in different locations. 
In fact, the CH$_{3}$OH maser flux is not always anticorrelated with that of the H$_{2}$O maser, as shown in Fig. \ref{fig-variability}(a) and (b). 
Furthermore, it would be difficult to discuss any anticorrelation of these two masers in detail, such as the dimming of the H$_{2}$O maser at the CH$_{3}$OH flares, because of a lack of simultaneous observational data around the maximum of the CH$_{3}$OH maser in late 2015. 
Future simultaneous monitoring of the CH$_{3}$OH and H$_{2}$O masers will be key to revealing a possible anticorrelation of these two masers in the maser flare events. 

Before discussing the origin of the increase of the H$_{2}$O maser flux, it is important to remember that this is mostly due to features lying in the NE lobe (feature ID13 at 6.40~km~s$^{-1}$ and ID22 at 15.96~km~s$^{-1}$s) at a distance of a few 100~au from the central star (see Table \ref{tab-feature}, Fig. \ref{fig-feature13} and Fig.\ref{fig-feature22}).
This implies that the flux of these features could be affected by the expansion of the radio jet once the latter has reached the location of the NE masers at approximately 300~au from the star.
According to \citet{Cesaroni2018}, this should occur at the end of 2016, which is consistent with the beginning of the increase of the maser flux.
Despite the temporal coincidence between the two events, their relationship is not fully established, as one should explain why only a moderate and slow increase in the maser intensity is seen, instead of a strong flare.
Another problem is that the H$_{2}$O masers do not lie in front of the expanding radio jet imaged by \citet[][as plotted by purple contours in Fig. \ref{fig-mapVLBI}(a)]{Cesaroni2018}, as one would expect if such masers were excited through collisions in a bow shock.

\begin{figure*}[th]
\begin{center}
\includegraphics[width=7cm]{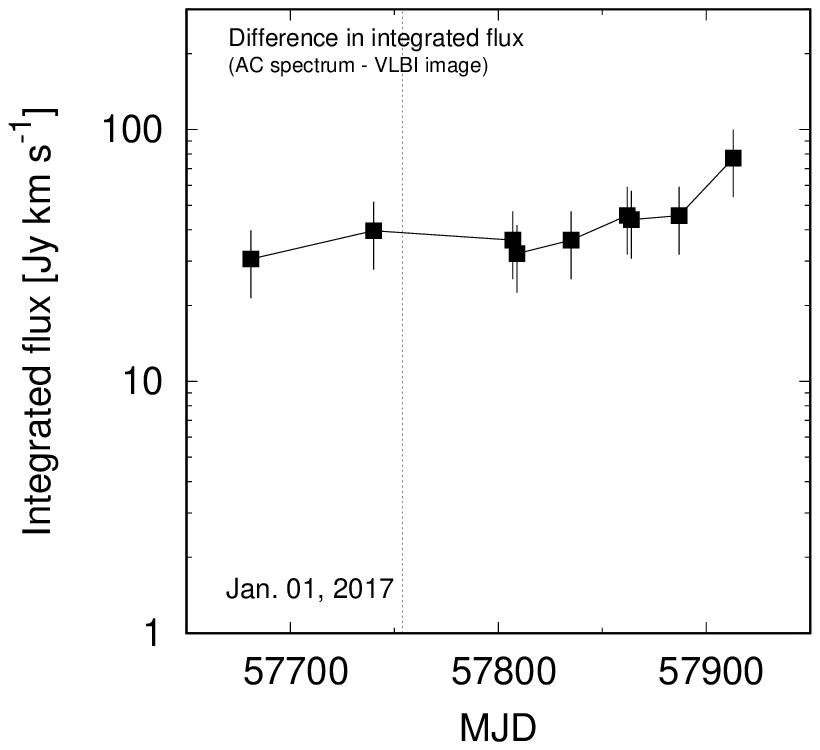}
\includegraphics[width=7cm]{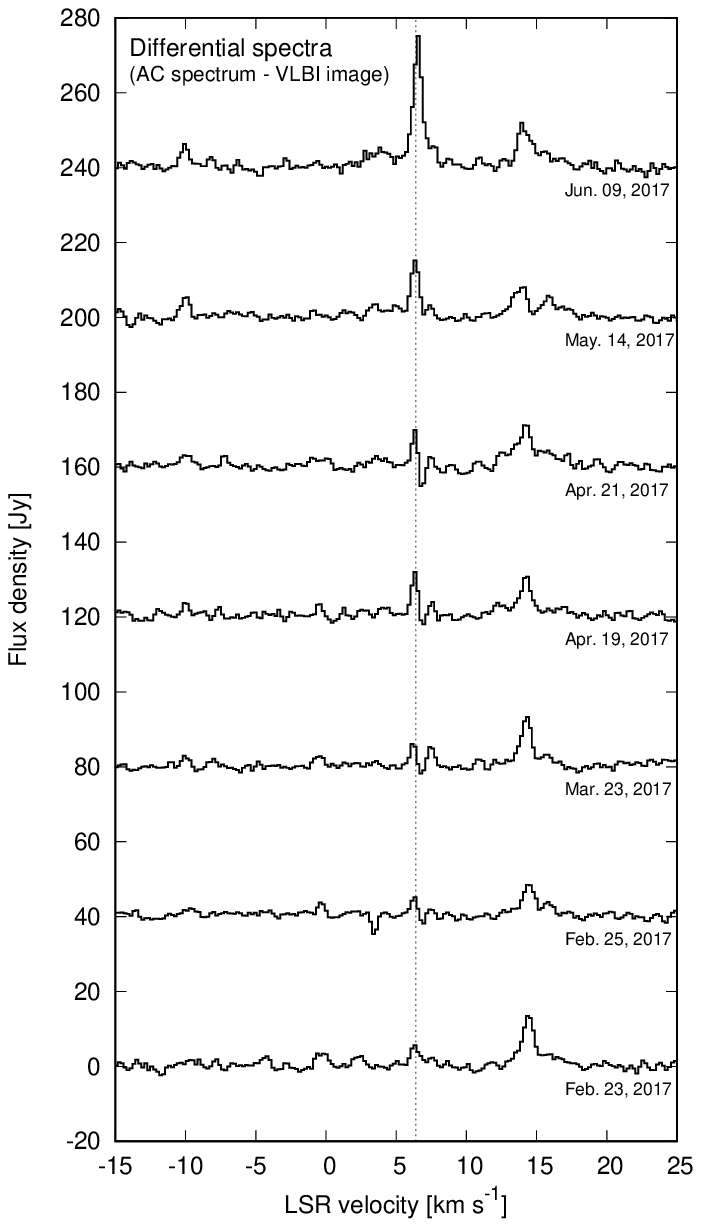}
\caption{
Difference between auto-correlation spectra and those from VLA/VERA images. 
Left and right panels show, respectively, flux densities integrated over all velocity components and differential spectra. 
In the left panel, the error bars correspond to 30\% uncertainty, and the dotted vertical line indicates January 01, 2017. 
In the right panel, the dotted vertical line marks the systemic velocity. }
\label{fig-diffsp}
\end{center}
\end{figure*}

The previous considerations apply to the compact maser features, but it is also interesting to find out what happens to the extended maser emission resolved out in our VERA and (to some extent) VLA observations.
To study the variation of this component, we computed the difference between the integrated flux of the autocorrelation spectrum and that of the spectrum extracted from the VERA image.
The result is plotted as a function of time in Fig. \ref{fig-diffsp} and clearly shows that the flux increase is also seen in the extended maser component.
On one hand, this result suggests that both the extended and compact maser emissions are excited through the same mechanism.
On the other hand, it makes it difficult to believe that the increase of maser intensity can be due to the radio jet impinging on the H$_{2}$O gas, because the jet is too collimated to trigger maser emission over an extended region of a few arcseconds or $\sim10^{4}$~au.

In conclusion, the lack of a true maser flare and the presence of extended maser emission undergoing the same increase as the compact maser features cast some doubt on the idea that the maser flux increase can be directly triggered by collisions in the radio jet.
While it is possible that the water maser variation observed in our monitoring is not related at all to the outburst observed in other tracers, we propose a solution of the conundrum in the wake of a similar event observed in another HMYSO.
\citet{Brogan2018} observed a strong maser flare associated with an accretion outburst in NGC6334I-MM1, with a delay of two years after the submillimeter burst.
They proposed that the maser flare is caused by the radiation of the outburst propagating along the outflow cavity and terminating at the bow shock described by the masers.
A similar scenario might apply to our case, where the radio jet could be piercing the molecular core, thus opening a free path along which the photons of the (already declining) IR outburst could escape and reach the shock where the masers are excited.
In this scenario, the interplay between the expanding jet and the declining IR emission (see Fig. \ref{fig-mapVLBI}) could justify the lack of a sudden and strong increase of the maser intensity.
At the same time, maser excitation through radiation could also explain the presence of maser emission over an extended region.
A possibility of radiative excitation of the H$_{2}$O masers is supported theoretically by the model of \citet{Gray2016}, in which the 22~GHz H$_{2}$O maser is both collisionally and radiatively pumped. 

\subsection{Physical properties of the shocked region}
\label{subsec-property}

The physical properties of the H$_{2}$O maser-emitting region can be estimated from the ratio between the luminosity of the 22~GHz maser and that of the 321~GHz masers, $R\equiv L_{22}/L_{321}$,
for maser features with the same position and velocity \citep{Neufeld1990,Patel2007}. 
Assuming that the beaming angle is the same for both maser transitions, $R$ is also equal to the flux density ratio, $R=F_{22}/F_{321}$. 
Table \ref{tab-321GHz} compares the positions, peak velocities, and peak flux densities of the five features with both 321~GHz and 22~GHz H$_{2}$O maser emission.
The variability of the 22~GHz H$_{2}$O masers was established on the basis of Figs. \ref{fig-feature1}-\ref{fig-feature0}. 
From these, we estimate a maximum uncertainty on the flux density, due to variability, of about a factor 2. 
We note that the faint features ID23 and ID25 were detected only in one epoch, and, consequently, for these two features we can only derive an upper limit for the ratio $R$.
The ratio $R$ ranges from 0.5 to 14.2 for the five features, where the lowest value is found for the redshifted ($\sim$17~km~s$^{-1}$) feature (J) located close to the continuum peak.
The features in the blueshifted SW lobe (D) and the redshifted feature located to the SW of the continuum peak (H) have a ratio of $\sim$2. 
On the other hand, the other two features in the NE lobe (E) and closer to the continuum peak (F) have the highest value of $R\sim$14.

According to Fig. 3 of \citet{Neufeld1990}, features with low $R$ values have higher temperatures of $>$1000~K. 
This implies that the continuum peak and SW bow shock should have higher temperatures ($>$400~K) than that of the NE lobe.
More recent theoretical models suggest that the 321~GHz H$_{2}$O masers tend to be excited most efficiently at a hydrogen density of $\sim10^{8}$~cm$^{-3}$, which is slightly higher than that needed to excite the 22~GHz masers \citep[Fig. 8 in ][]{Richards2014}. 
Similar physical conditions were predicted by \citet{Gray2016}, in which the 22~GHz H$_{2}$O maser is also pumped in a lower H$_{2}$O density region of $\le$10$^{4}$~cm$^{-3}$. 

The long-term VLBI monitoring of more than a decade suggests that the SW maser features are more stable than the others.
This behavior is the opposite to that of the radio jet, where the structure of the NE free-free knot is more stable than that of the SW knot \citep[see Fig. 1 of][]{Cesaroni2018}. 
We speculate that the inhomogeneous density structure of the envelope and/or outflow cavity might affect the shape of the smaller scale outflow/jet.
The higher density in the SW direction could produce a stronger maser emission at the bow shock. 
This scenario seems consistent with the lower values of $R$. 
In contrast, the outflow cavity in the NE lobe could favor a more prominent radio jet structure due to the lower density envelope or outflow cavity, possibly created by a previous outburst.
Although the NE masers are also stable, the main difference with respect to the SW lobe consists of more scattered positions and line-of-sight velocities. 
This could be due to the ambient medium in the NE lobe being less homogeneous, although not necessarily less dense.

There are several 321~GHz H$_{2}$O maser features not associated with the 22~GHz masers, with $R$ values close to zero ($<$1).
It has been reported that the majority of the 321~GHz maser features in Cepheus~A do not coexist with 22~GHz features \citep{Patel2007}. 
In the case of the red supergiant VY~CMa, long-baseline observations with ALMA revealed that the 321~GHz and 22~GHz masers do not always overlap \citep{Richards2014}. 
In these cases, the discussion based on the emissivity ratio $R$ \citep{Neufeld1990} cannot be applied.
We cannot rule out the possibility that in our case the non-detection of the 22~GHz H$_{2}$O masers and the low value of $R$ are due to the variability of the 22 GHz masers and/or the higher spatial resolution of the VLBI data compared to the ALMA images, which could resolve the 22~GHz H$_{2}$O maser features. 
In fact, possible signatures of velocity components other than 3~km~s$^{-1}$, 6~km~s$^{-1}$, 13~km~s$^{-1}$, and 18~km~s$^{-1}$ appear in the single-dish spectra from October 2017 to February 2018, showing significant variation (Fig. \ref{fig-spSD}).
Further coordinated observations with ALMA and VLBI (and the VLA as well) are needed to address this issue. 

\section{Summary}
\label{sec-summary}

We carried out monitoring observations of the 22~GHz H$_{2}$O masers with both single-dish telescopes (VERA 20-m antennas) and interferometers (VLA and VERA) after the mass accretion burst event in the high-mass protostar S255~NIRS~3. 
The H$_{2}$O maser flux density shows a gradually increasing trend in 2017, about two years after the outburst in mid 2015, and it becomes almost stable in 2018. 
The flux increase is mainly seen in the maser features associated with the NE outflow lobe or the continuum source. 

We found that a part of the H$_2$O maser emission is resolved out even with the A-configuration of the VLA. 
To our knowledge, this is the first time that extended ($>1700$ au) maser emission has been detected in a star-forming region.
Both the compact features and the extended component show a similar flux variation over time, suggesting a common origin and excitation mechanism, which we propose to be radiative excitation due to the combined effect of the IR outburst and the expanding jet.
 
The H$_{2}$O masers associated with the SW bow shocks seem to be unchanged compared with the previous VLBI monitoring \citep{Goddi2007, Burns2016}. 
The measured proper motions of the H$_{2}$O masers in this SW bow shock suggests a dynamical timescale of 60~years, which is much longer than the current outburst. 

In addition, we conducted followup observations of the submillimeter continuum and the 321~GHz H$_{2}$O masers with ALMA at Band~7. 
The outflow structure is also traced by the 321~GHz H$_{2}$O masers and is slightly tilted compared with the radio jet powered by the mass accretion burst \citep{Cesaroni2018}. 
The ALMA Band-7 continuum does not show significant variation compared with the previous observations \citep{Liu2018}. 

Considering all the above characteristics, we suggest that the gradual increase of the 22~GHz H$_{2}$O maser flux in S255~NIRS~3 could be caused by the expanding radio jet. 
The change in the flux density seems to be more moderate compared with another case of mass accretion burst events in NGC6334I-MM1 \citep{Brogan2018}.
As already suggested \citep[e.g.,][]{Burns2020}, mass accretion bursts in high-mass star-forming regions appear to be quite different from each other.
Further surveying and monitoring of these kinds of sources/events, such as the M2O program, are needed to shed light on the mass accretion processes in high-mass star-formation.

\bigskip

\begin{acknowledgement}

This paper makes use of the following ALMA data: ADS/JAO.ALMA\#2017.1.00178.S. 
ALMA is a partnership of ESO (representing its member states), NSF (USA) and NINS (Japan), 
together with NRC (Canada), NSC and ASIAA (Taiwan), and KASI (Republic of Korea), 
in cooperation with the Republic of Chile. 
The Joint ALMA Observatory is operated by ESO, AUI/NRAO and NAOJ. 
TH is financially supported by the MEXT/JSPS KAKENHI Grant Number 17K05398. 
KS is financially supported by the MEXT/JSPS KAKENHI Grant Number 19K03921. 
Data analysis were in part carried out on common use data analysis computer system at the Astronomy Data Center, ADC, of NAOJ. 
\end{acknowledgement}

\clearpage
\appendix

\section{Proper motion measurements and fitting results}
\label{sec-appendix}

First, we measured the absolute position and proper motion of one of the selected maser spots in the feature ID19 at $\sim$14~km~s$^{-1}$ (Table \ref{tab-feature}), which had stable spatial and velocity structures during our observation period. 
We could obtain phase-referenced images for five epochs from seven observing sessions (February 23, February 25, April 19, May 14, and June 09, 2017). 
The derived absolute coordinates of this maser spot are 
\begin{eqnarray*}
\mbox{RA(J2000)} &=& 06\mbox{h}12\mbox{m}54.0114935\pm 0.0000007\mbox{s} \\
\mbox{Decl(J2000)} &=& +17^{\circ}59\arcmin23.103476\pm 0.000011\arcsec
\end{eqnarray*}
for the first epoch (February 23, 2017), in which the error bars are formal uncertainties in the Gaussian fitting, and hence do not include systematics due to calibration errors. 
We compared the absolute positions of this maser spot in the first and second epochs, February 23 and 25, 2017, which are separated by only a two-day interval, and we found that the position difference between these two epochs is 0.03~mas. 
We regard this position difference as the typical uncertainty in the absolute positional measurements, which
is taken into account in the absolute proper motion fitting as a systematic error. 

We used the results from all five epochs in which we succeeded in phase-referencing astrometry to obtain the absolute proper motion of this spot by fitting the position movement as a function of time with the fixed annual parallax value of 0.563~mas \citep[1.78~kpc; ][]{Burns2016}. 
The best-fit absolute proper motion of this spot in the feature ID19 is $-1.08\pm0.13$~mas~yr$^{-1}$ in right ascension and $-0.42\pm0.13$~mas~yr$^{-1}$ in declination, with a post-fit residual of 0.04~mas in both coordinates. 

Next, we performed fringe fitting and self-calibration for the strongest maser spot in the feature ID13 at $\sim$6~km~s$^{-1}$ (see Table \ref{tab-feature}) to achieve a better phase calibration than that obtained in phase-referencing mode. 
In this procedure, the absolute position information was lost and the reference maser spot in the feature ID13 was placed at the (0,0) position, which is the phase-tracking center. 
Thus, the absolute position of the ID13 feature was registered by measuring its position offset from the spot in feature ID19 with absolute position known from the phase-referencing astrometry measurements. 

We only measured the proper motions of those features that were detected in more than three epochs. 
The position of a feature at each epoch was determined by the intensity-weighted average of all maser spots in the feature. 
As a result, we determined the proper motions for 15 features among the 25 identified, as shown in Table \ref{tab-feature}. 
The average post-fit residual of the proper motion fitting for each feature is 0.12~mas (0.14~mas in right ascension and 0.08~mas in declination). 
The residuals are larger than those of the absolute position measurement for the maser spot in the ID19 feature (0.04~mas). 
This is most likely due to structural changes in the maser features, which are sometimes seen in VLBI astrometry \citep[e.g.,][]{VERA2020}. 

In order to estimate the systemic motion of the outflow with respect to the barycenter, we calculated the average proper motions of the maser features associated with the NE and SW lobe of the outflow. 
As done in \citet{Burns2016}, we excluded the redshifted maser features located between the NE and SW lobes because they could not trace the NE-SW outflow. 
If the outflow motion is symmetric with respect to the barycenter, these motions should point toward the opposite direction at the same velocity. 
Finally, by subtracting the proper motion of the barycenter, we were able to obtain the proper motion of each maser feature with respect to the barycenter of the outflow ejected from S255~NIRS~3, as listed in Table \ref{tab-feature}. 
Due to the short monitoring period of only three~months, some of the features have large uncertainties. 
However, we could see systematic proper motions in the outflow, as shown in Fig. \ref{fig-mapVLBI}. 
The less ordered proper motions in the NE lobe could be due to interaction with the burst-regenerated radio jet (see Fig. \ref{fig-mapALMA}(a)).

The absolute proper motion of the barycenter was calculated from the absolute proper motion of the maser spot in the ID19 feature, and that relative to the barycenter
and is equal to $-0.03\pm0.40$~mas~yr$^{-1}$ in right ascension and $-0.29\pm0.30$~mas~yr$^{-1}$ in declination. 
This means that the absolute proper motion of the barycenter is negligible. 
The motion is consistent with those obtained by \citet{Rygl2010} ($-0.14\pm0.54$~mas~yr$^{-1}$, $-0.84\pm1.76$~mas~yr$^{-1}$) and \citet{Burns2016} ($-0.13\pm0.20$~mas~yr$^{-1}$, $-0.06\pm0.27$~mas~yr$^{-1}$). 
Therefore, we do not correct for the absolute positional offset due to the systemic motion with respect to the absolute coordinate frame when we compare our present results to those of previous works \citep{Goddi2007, Burns2016}.

\begin{figure*}[hbt]
\begin{center}
\includegraphics[width=5cm]{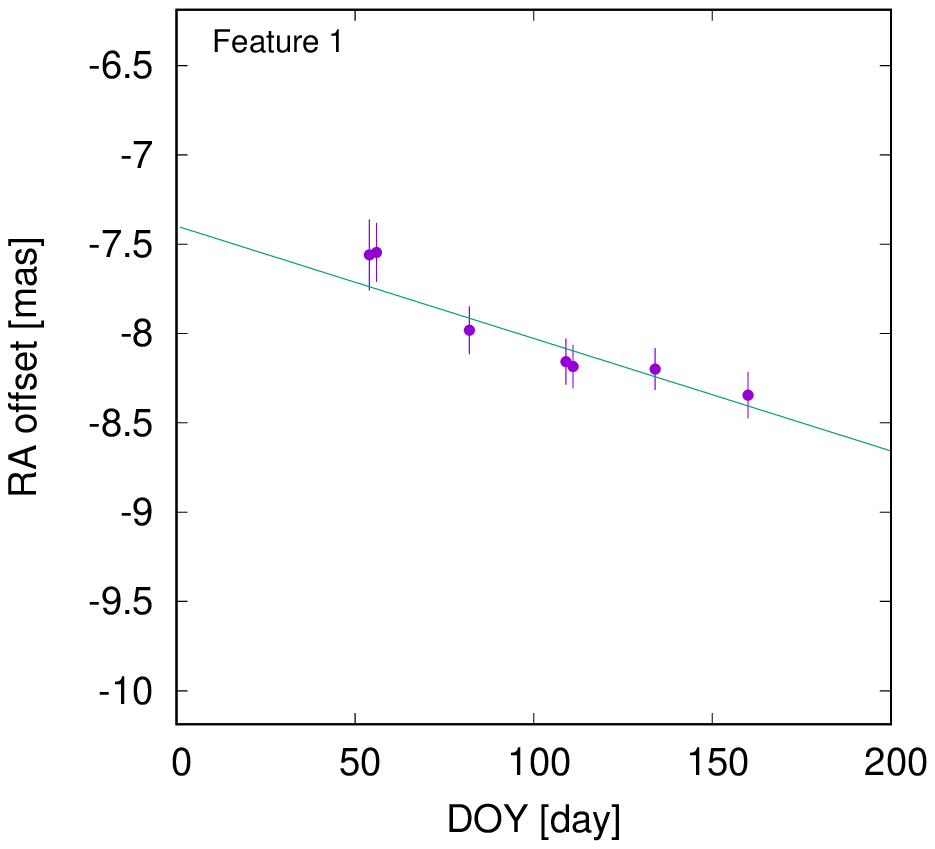}
\includegraphics[width=5cm]{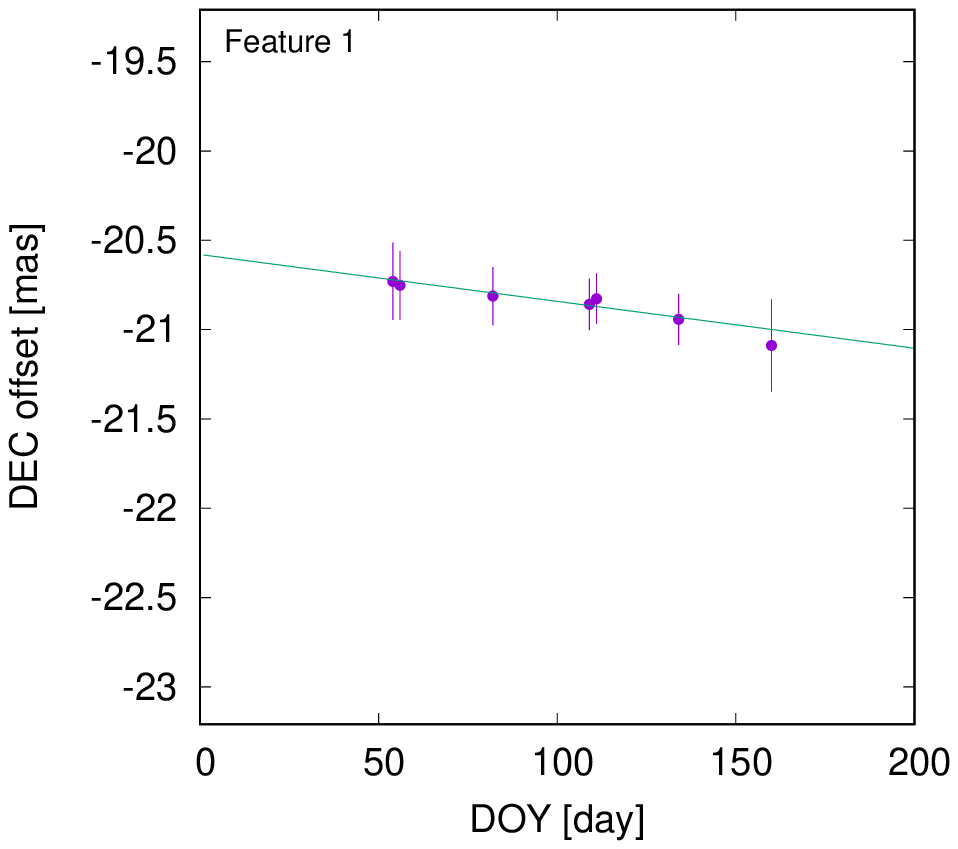}
\includegraphics[width=5cm]{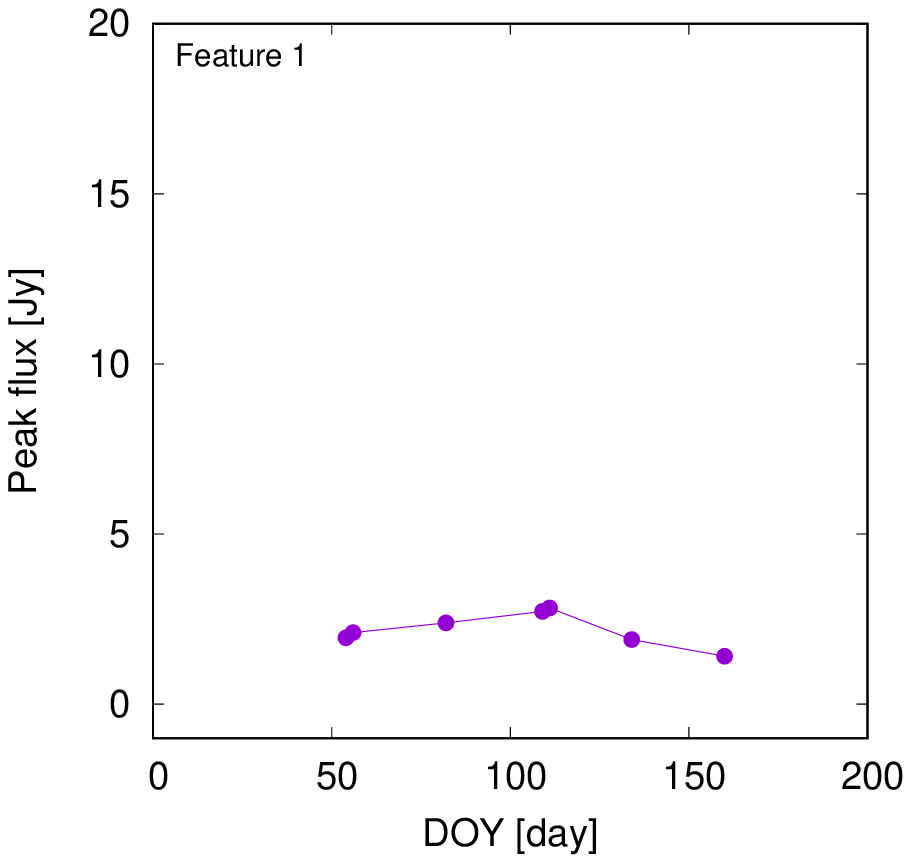}
\caption{Positions in right ascension (left) and declination (middle), and peak flux density (right) of maser feature 1 as a function of time.}
\label{fig-feature1}
\end{center}
\end{figure*}

\begin{figure*}[th]
\begin{center}
\includegraphics[width=5cm]{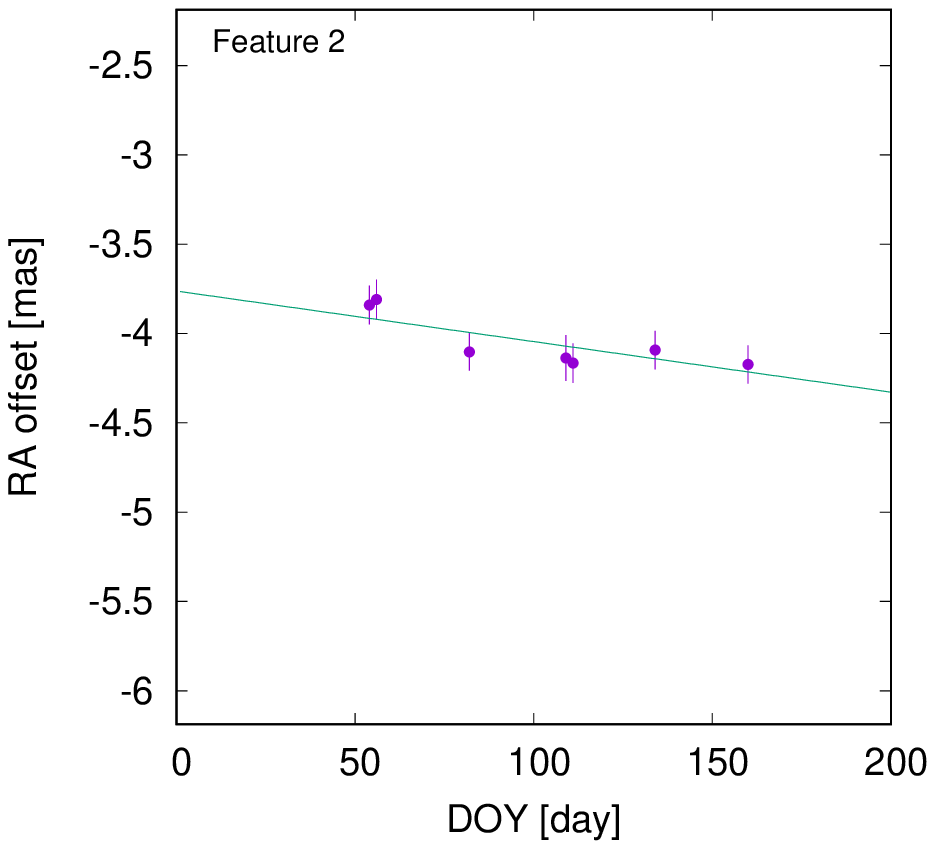}
\includegraphics[width=5cm]{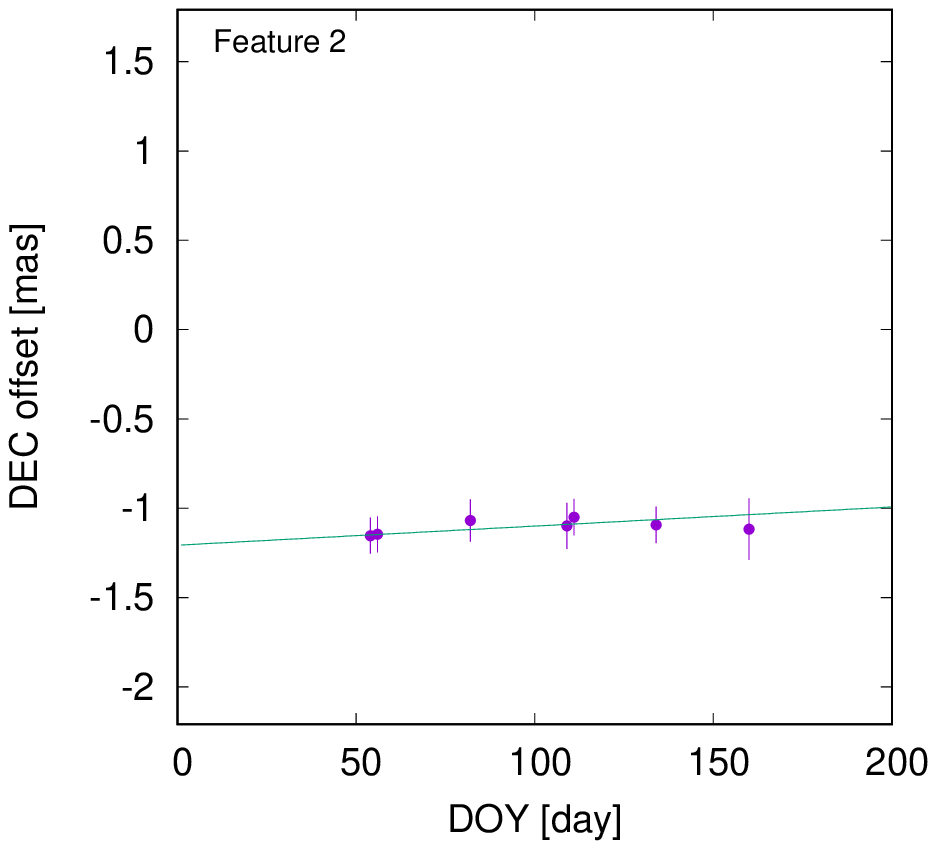}
\includegraphics[width=5cm]{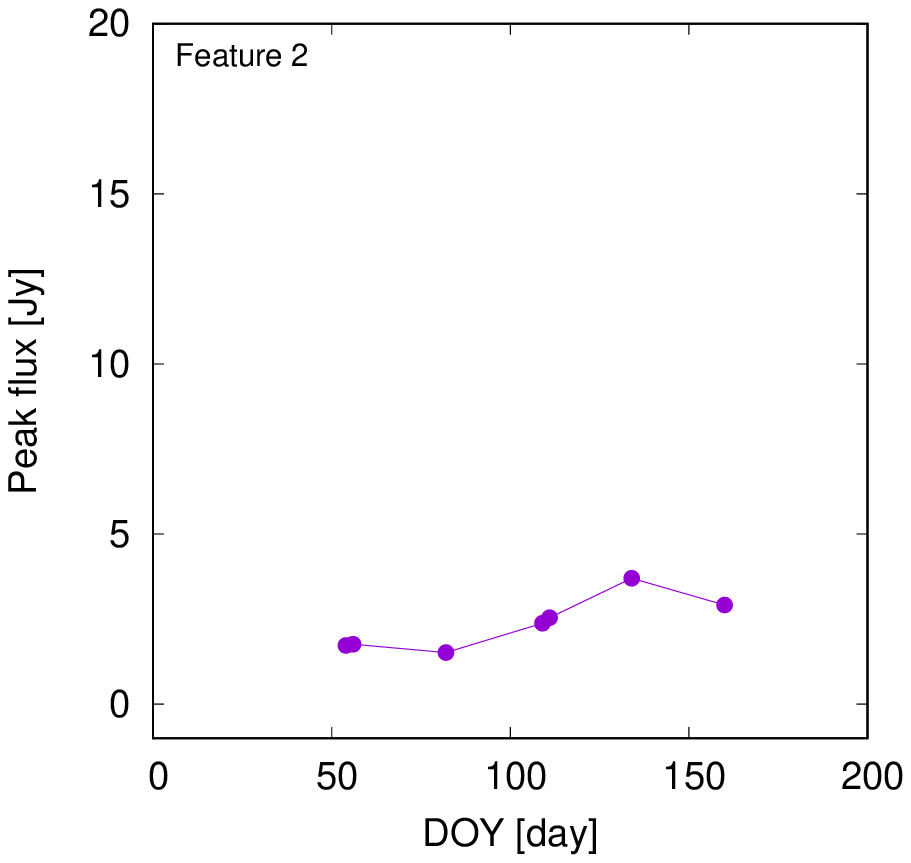}
\caption{Same as Fig. \ref{fig-feature1}, but for feature 2. }
\label{fig-feature2}
\end{center}
\end{figure*}

\begin{figure*}[th]
\begin{center}
\includegraphics[width=5cm]{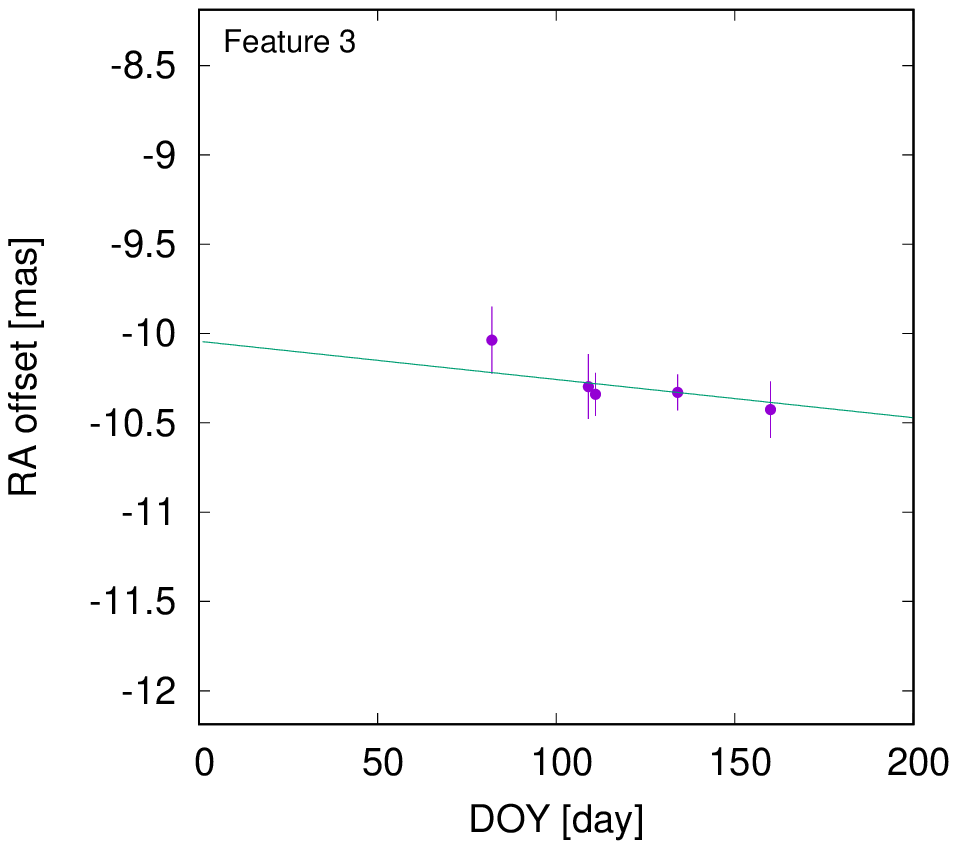}
\includegraphics[width=5cm]{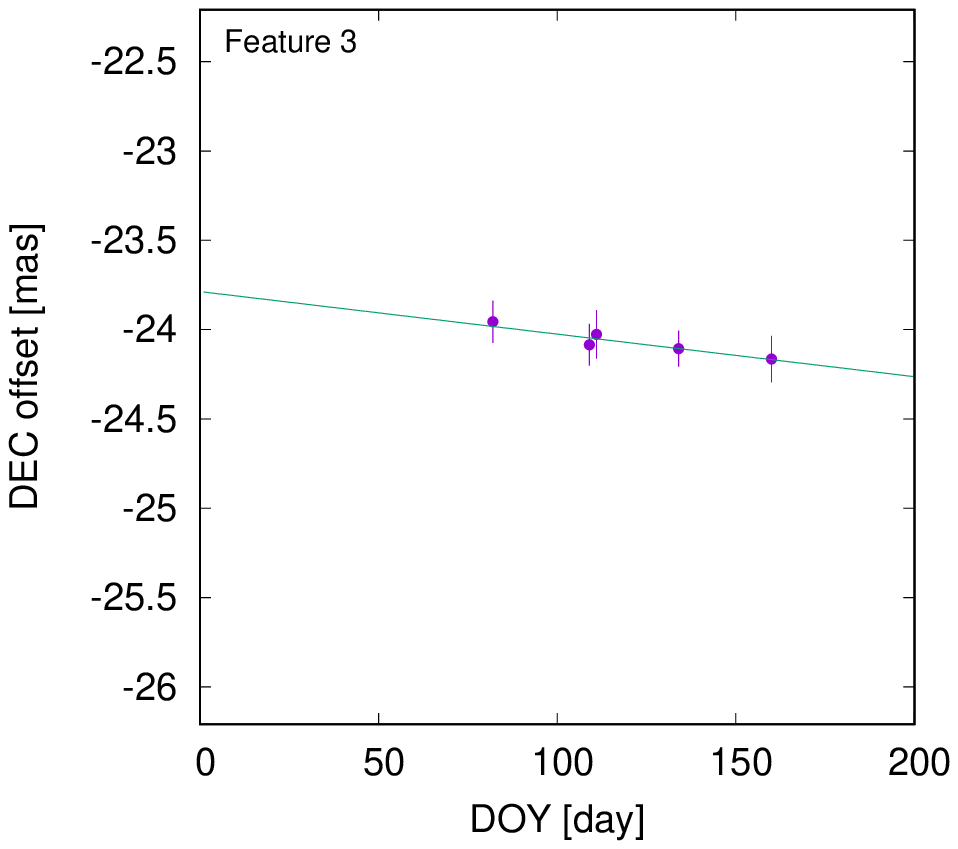}
\includegraphics[width=5cm]{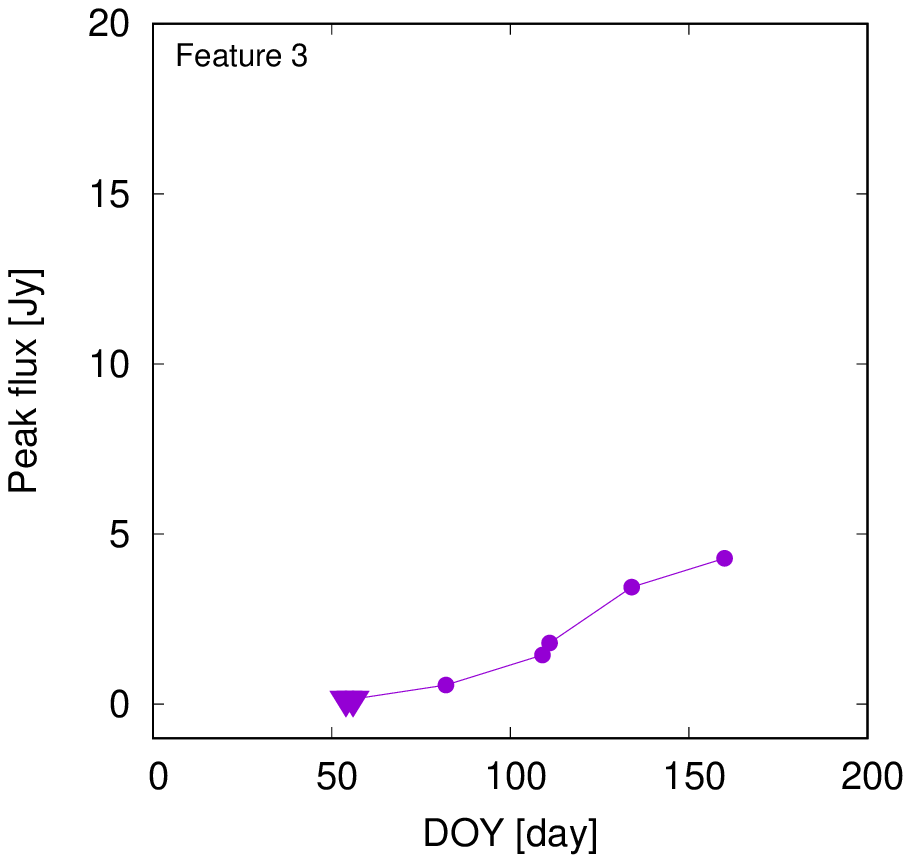}
\caption{Same as Fig. \ref{fig-feature1}, but for feature 3. 
Triangles in the right panel represent the upper limit of the flux densities. }
\label{fig-feature3}
\end{center}
\end{figure*}

\begin{figure*}[th]
\begin{center}
\includegraphics[width=5cm]{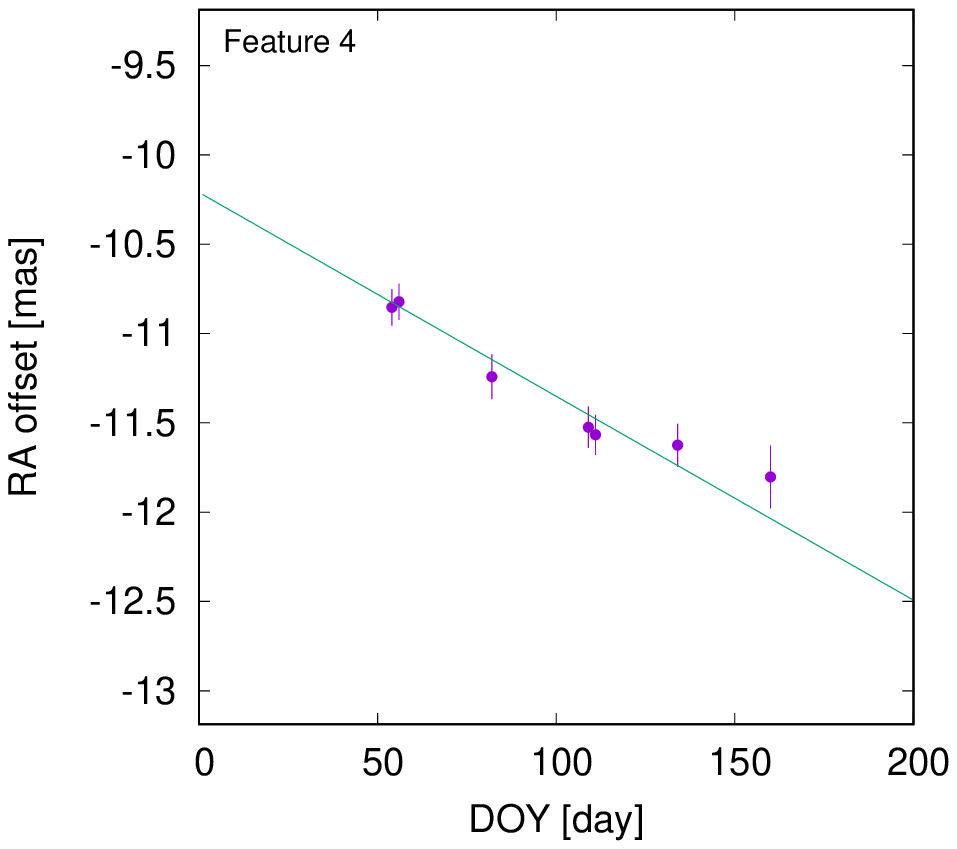}
\includegraphics[width=5cm]{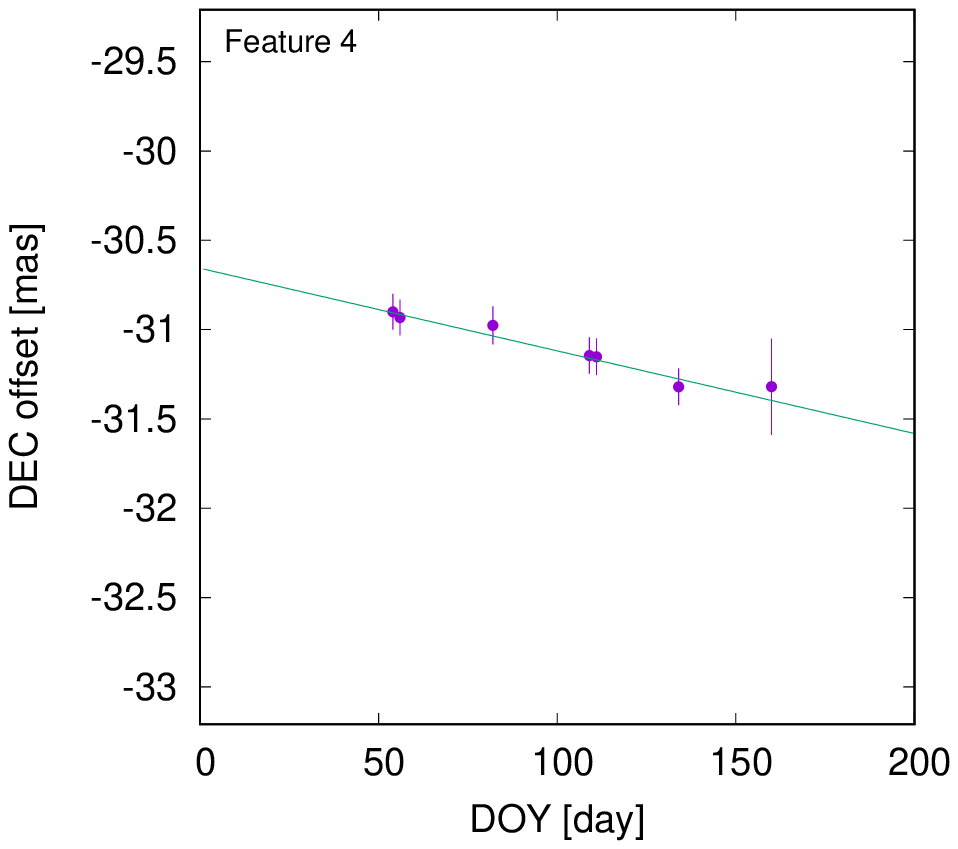}
\includegraphics[width=5cm]{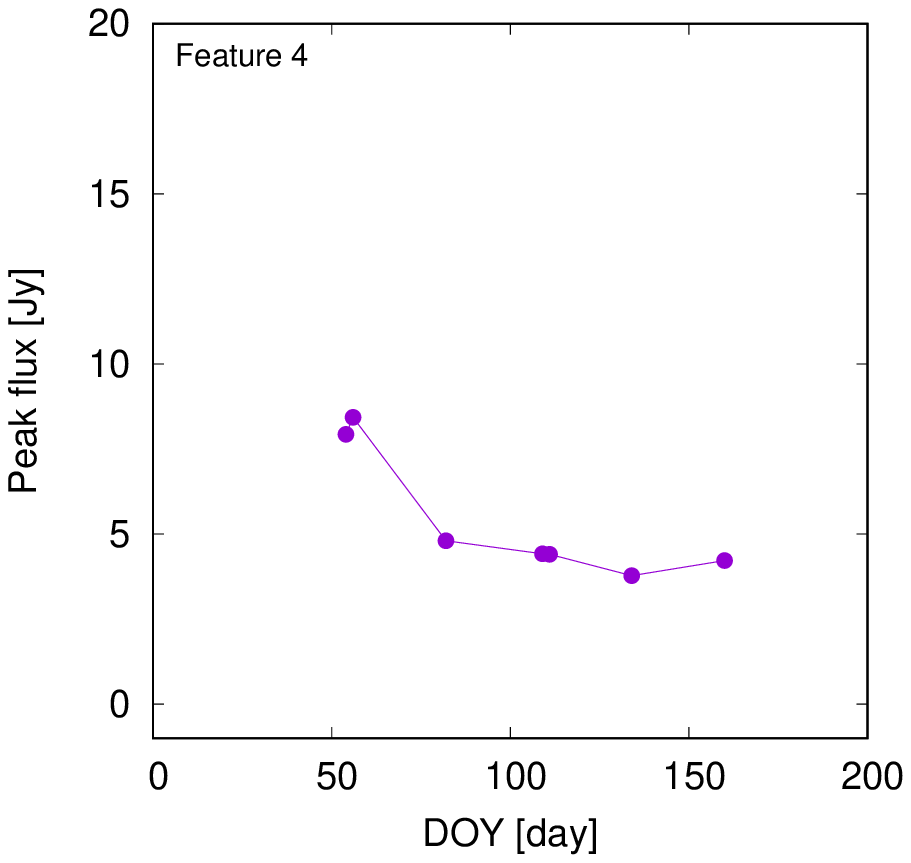}
\caption{Same as Fig. \ref{fig-feature1}, but for feature 4. }
\label{fig-feature4}
\end{center}
\end{figure*}

\begin{figure*}[th]
\begin{center}
\includegraphics[width=5cm]{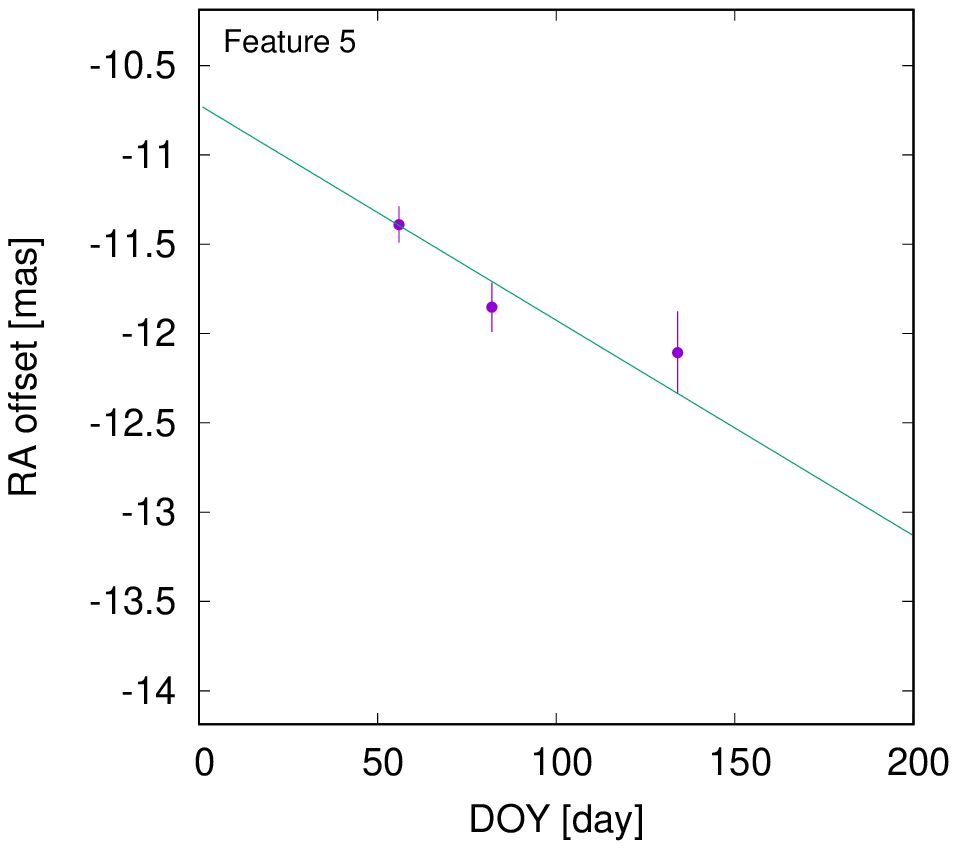}
\includegraphics[width=5cm]{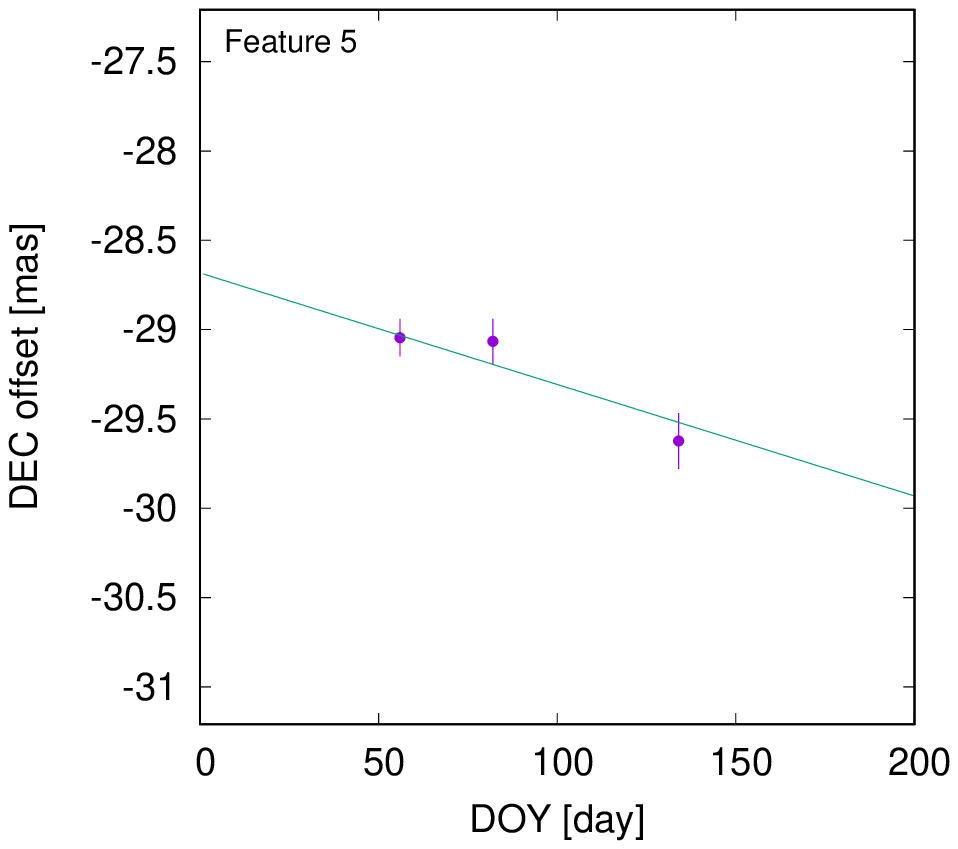}
\includegraphics[width=5cm]{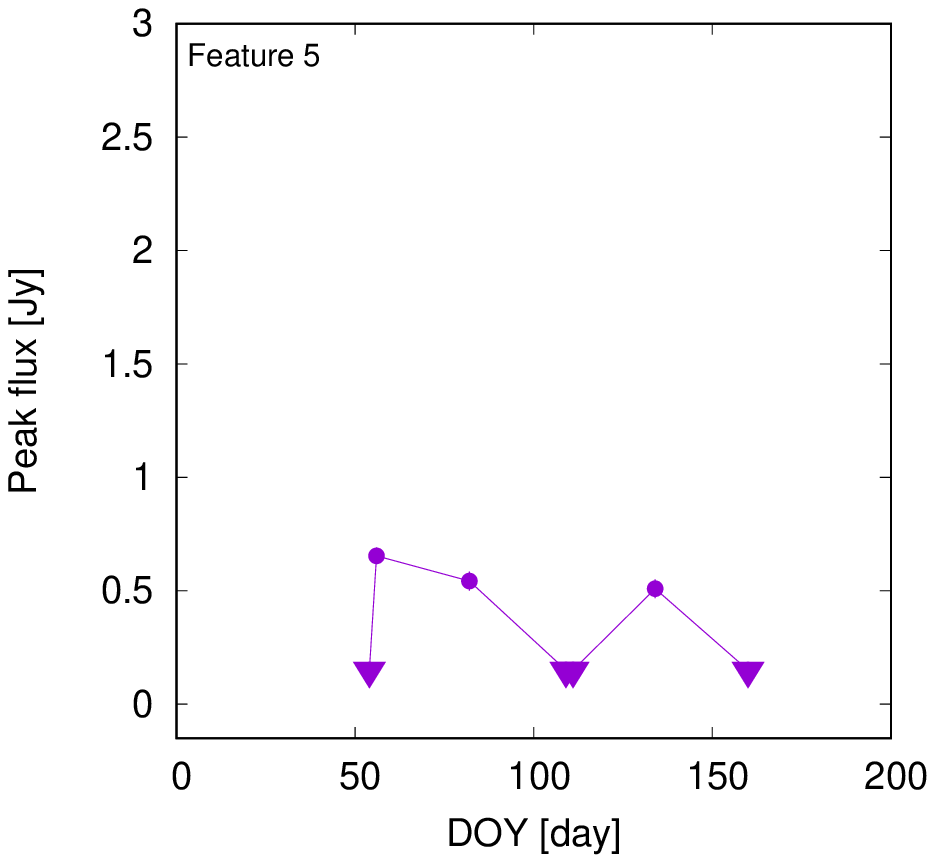}
\caption{Same as Fig. \ref{fig-feature1}, but for feature 5. 
Triangles in the right panel represent the upper limit of the flux densities. }
\label{fig-feature5}
\end{center}
\end{figure*}

\begin{figure*}[th]
\begin{center}
\includegraphics[width=5cm]{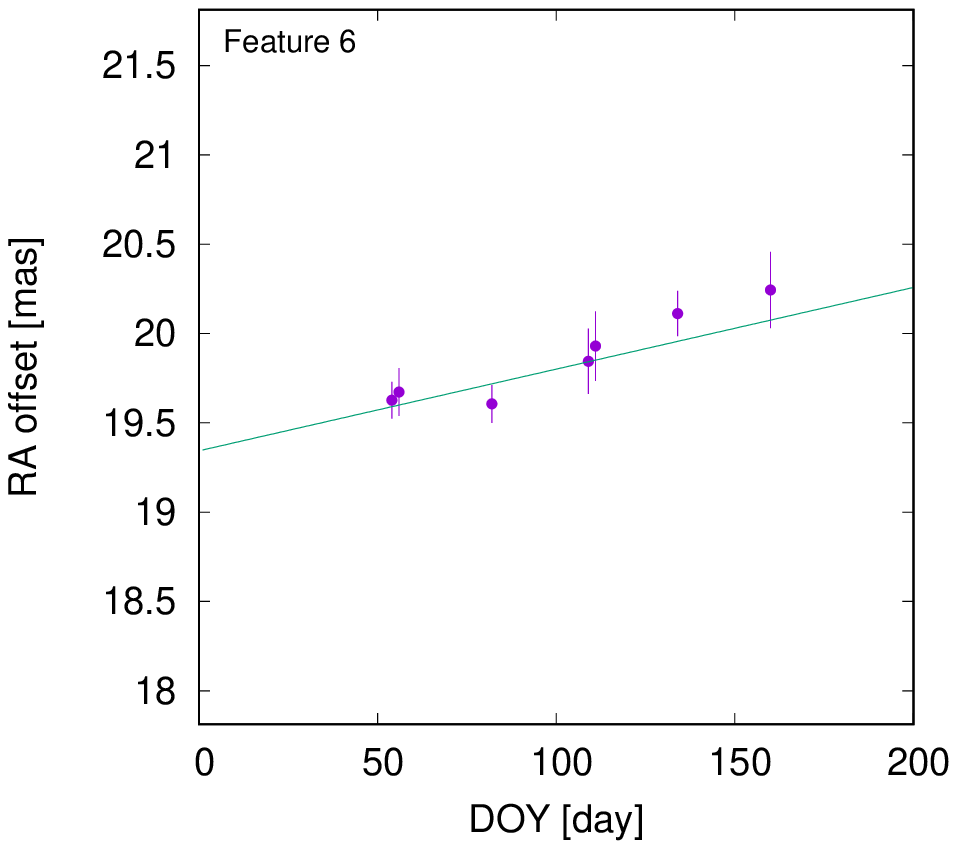}
\includegraphics[width=5cm]{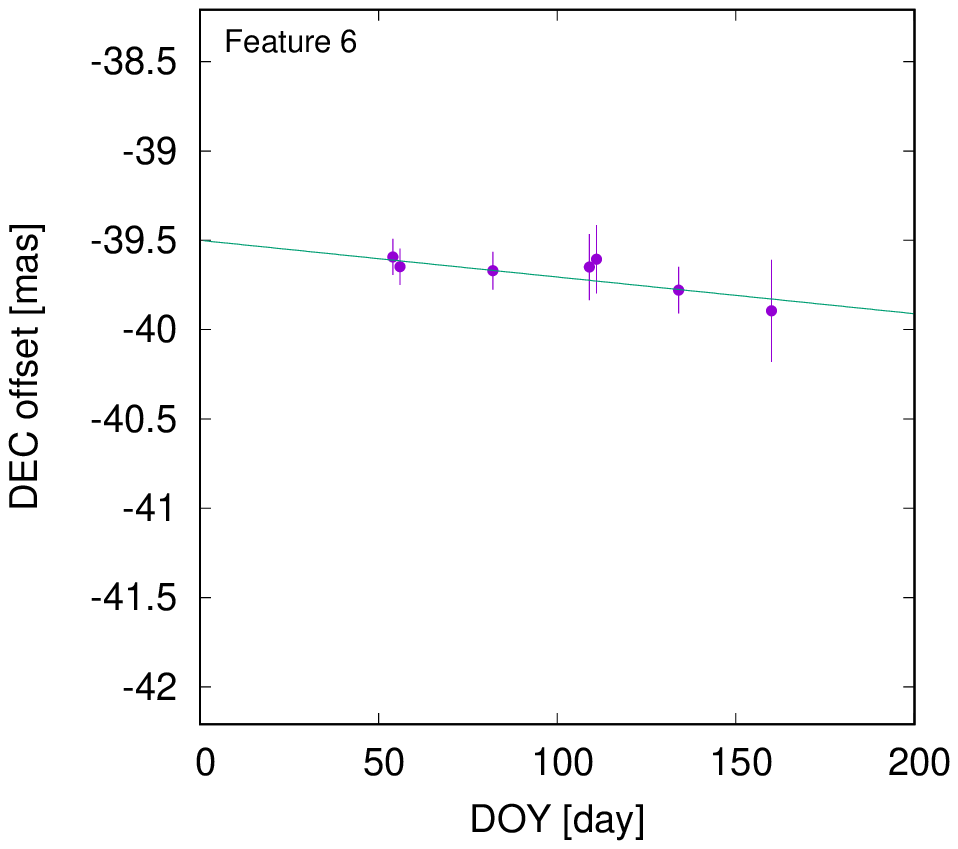}
\includegraphics[width=5cm]{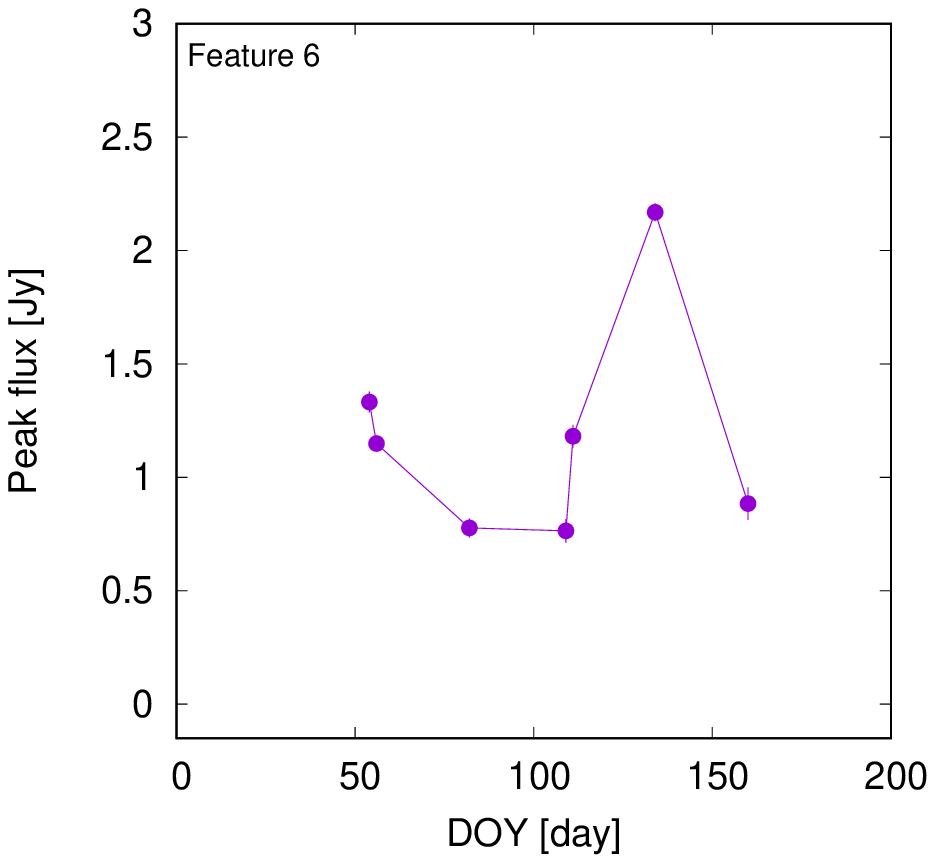}
\caption{Same as Fig. \ref{fig-feature1}, but for feature 6. }
\label{fig-feature6}
\end{center}
\end{figure*}

\begin{figure*}[th]
\begin{center}
\includegraphics[width=5cm]{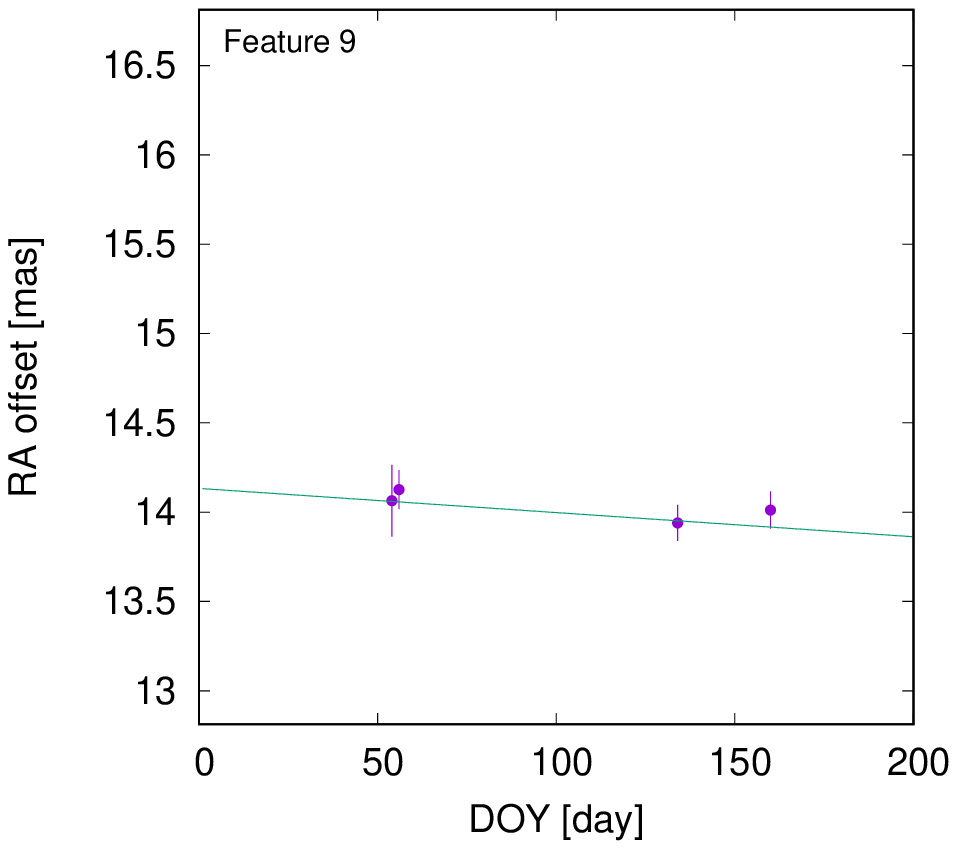}
\includegraphics[width=5cm]{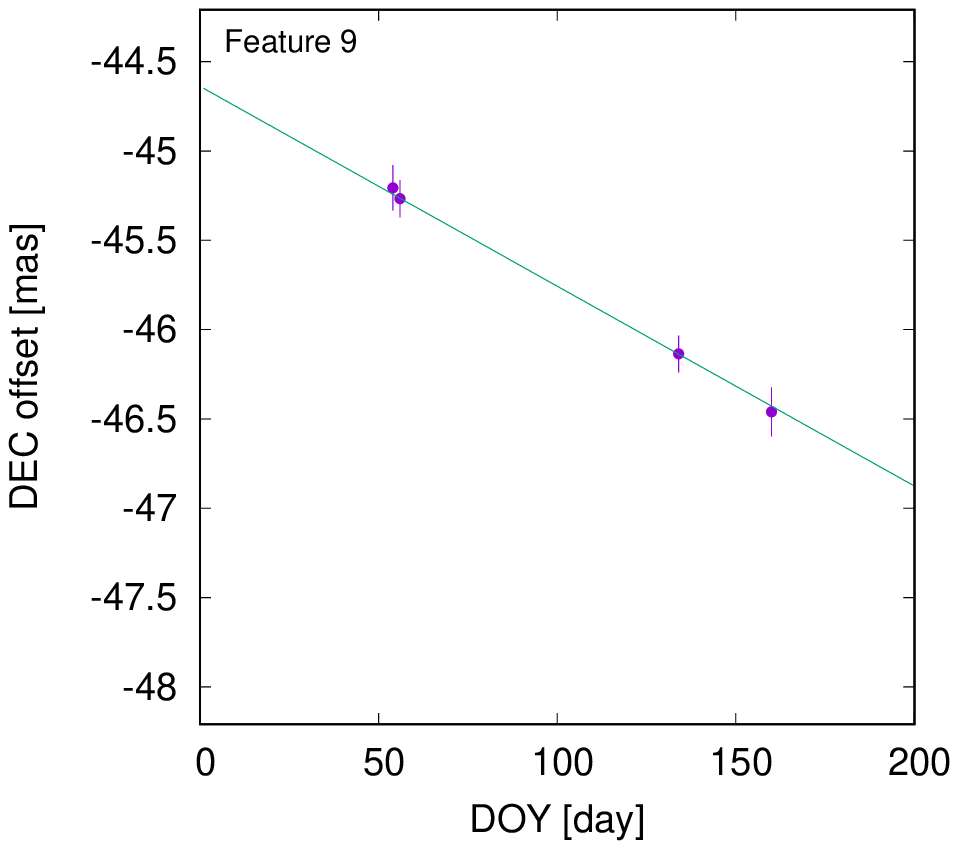}
\includegraphics[width=5cm]{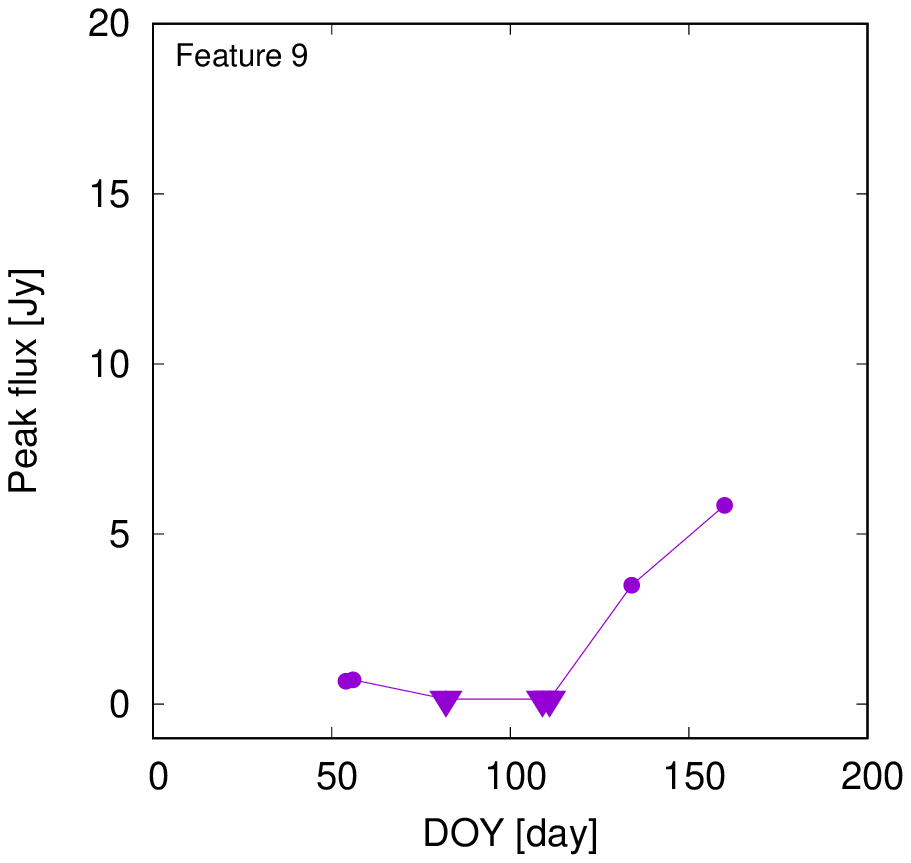}
\caption{Same as Fig. \ref{fig-feature1}, but for feature 9. 
Triangles in the right panel represent the upper limit of the flux densities. }
\label{fig-feature9}
\end{center}
\end{figure*}

\begin{figure*}[th]
\begin{center}
\includegraphics[width=5cm]{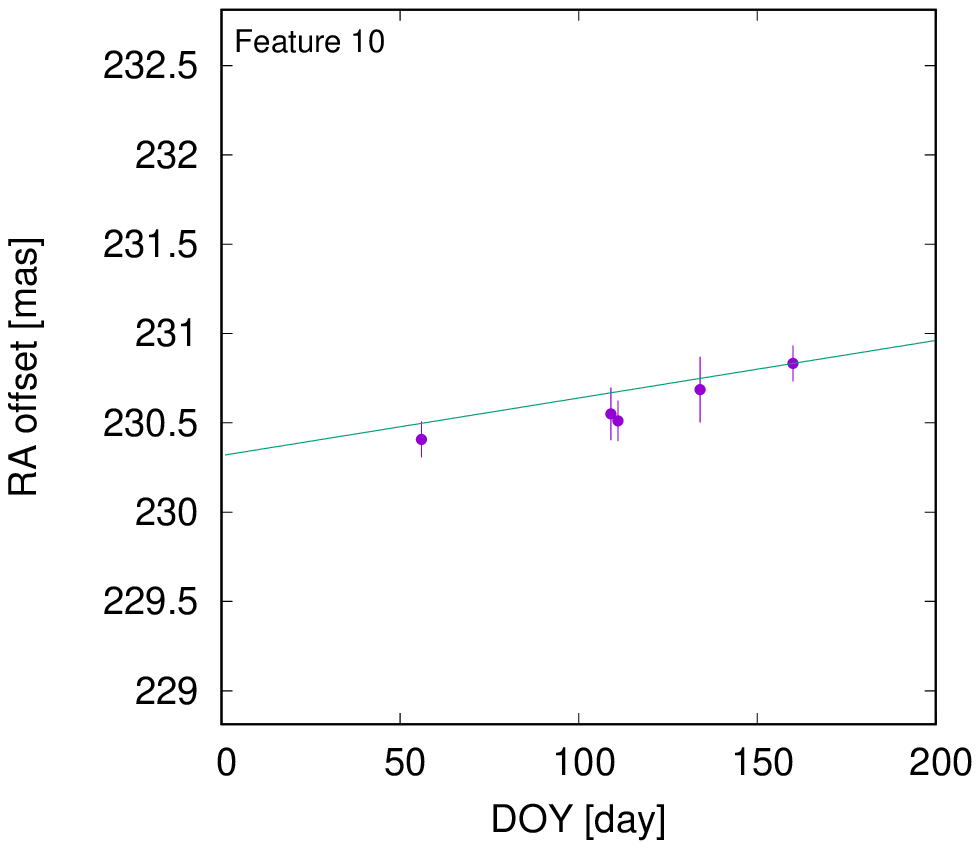}
\includegraphics[width=5cm]{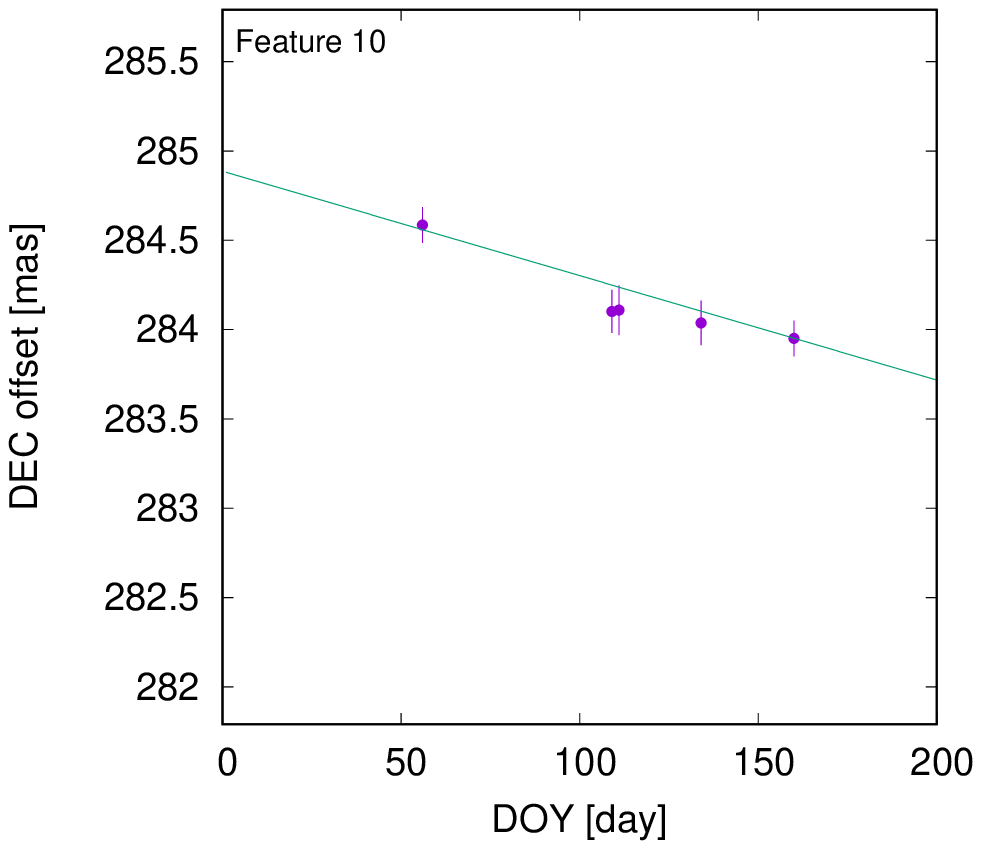}
\includegraphics[width=5cm]{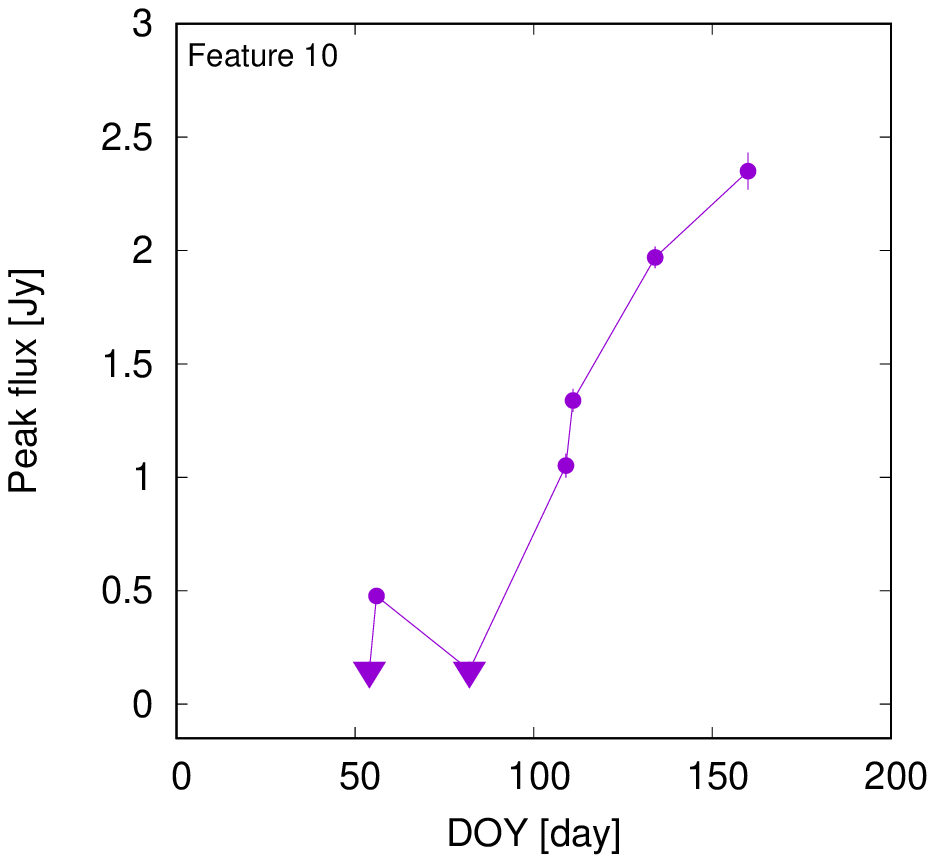}
\caption{Same as Fig. \ref{fig-feature1}, but for feature 10. 
Triangles in the right panel represent the upper limit of the flux densities. }
\label{fig-feature10}
\end{center}
\end{figure*}

\begin{figure*}[th]
\begin{center}
\includegraphics[width=5cm]{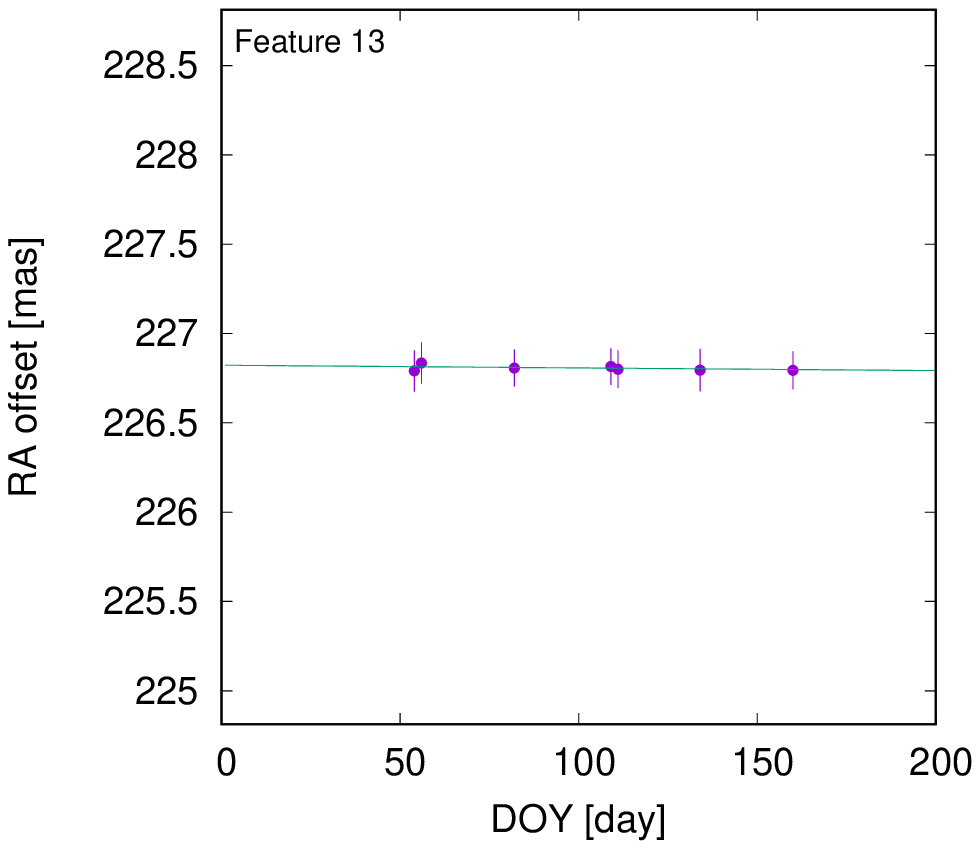}
\includegraphics[width=5cm]{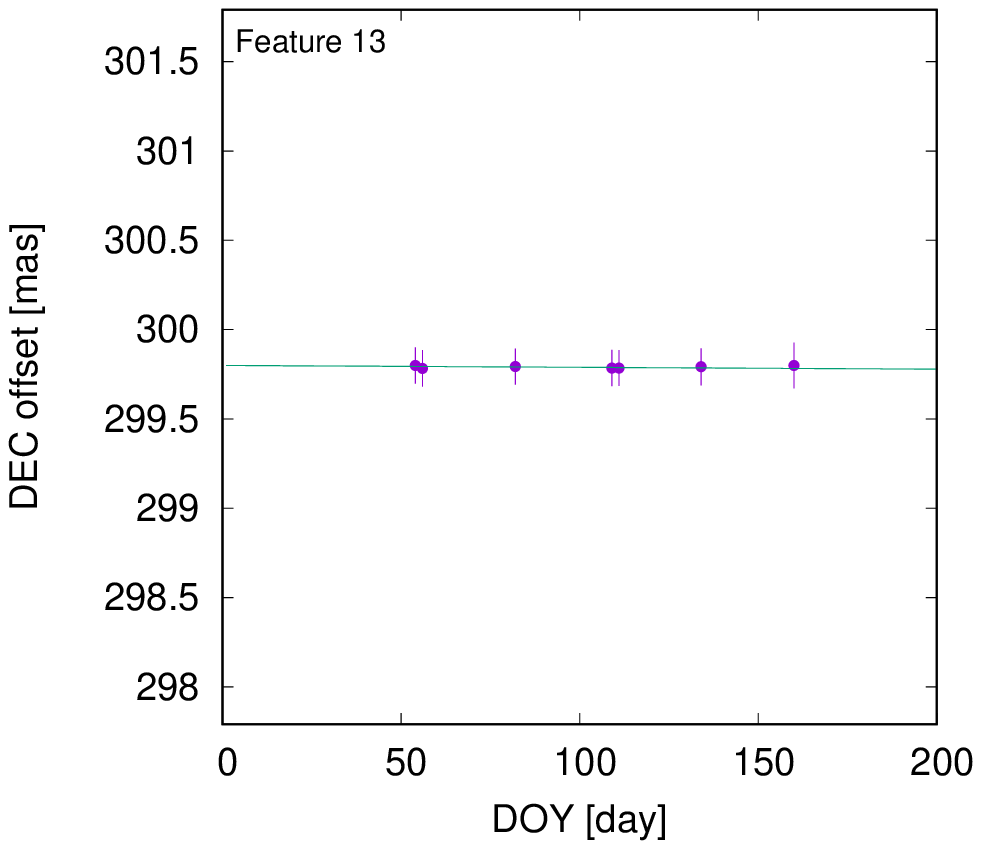}
\includegraphics[width=5cm]{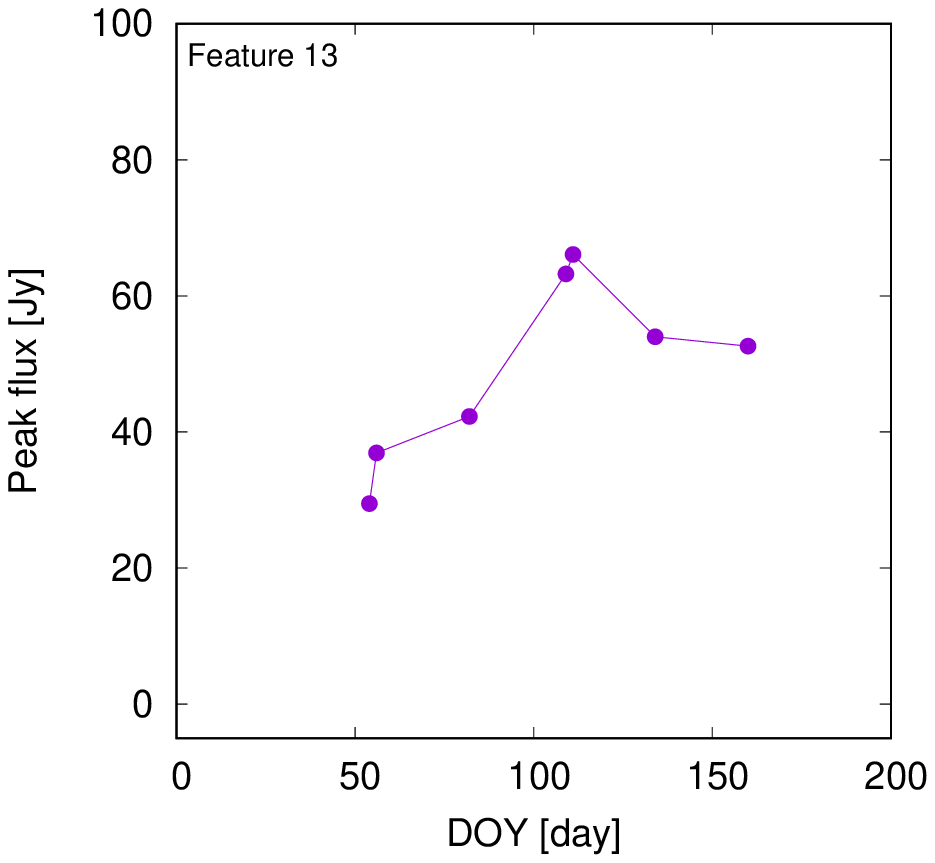}
\caption{Same as Fig. \ref{fig-feature1}, but for feature 13. }
\label{fig-feature13}
\end{center}
\end{figure*}

\begin{figure*}[th]
\begin{center}
\includegraphics[width=5cm]{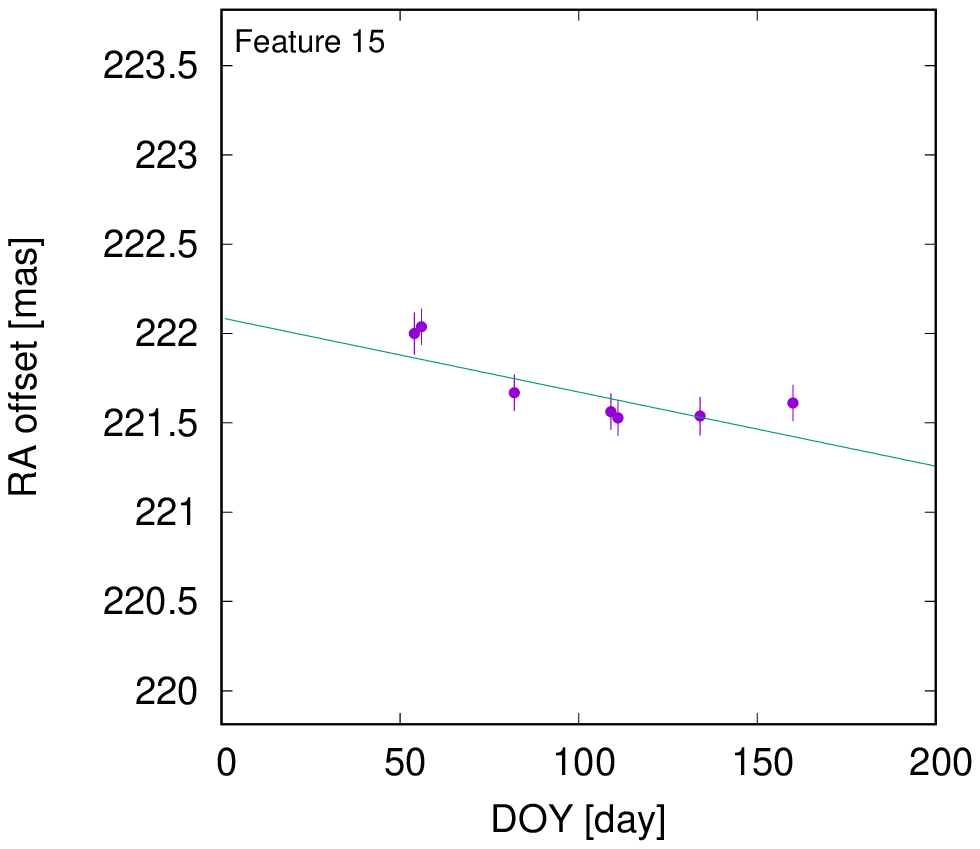}
\includegraphics[width=5cm]{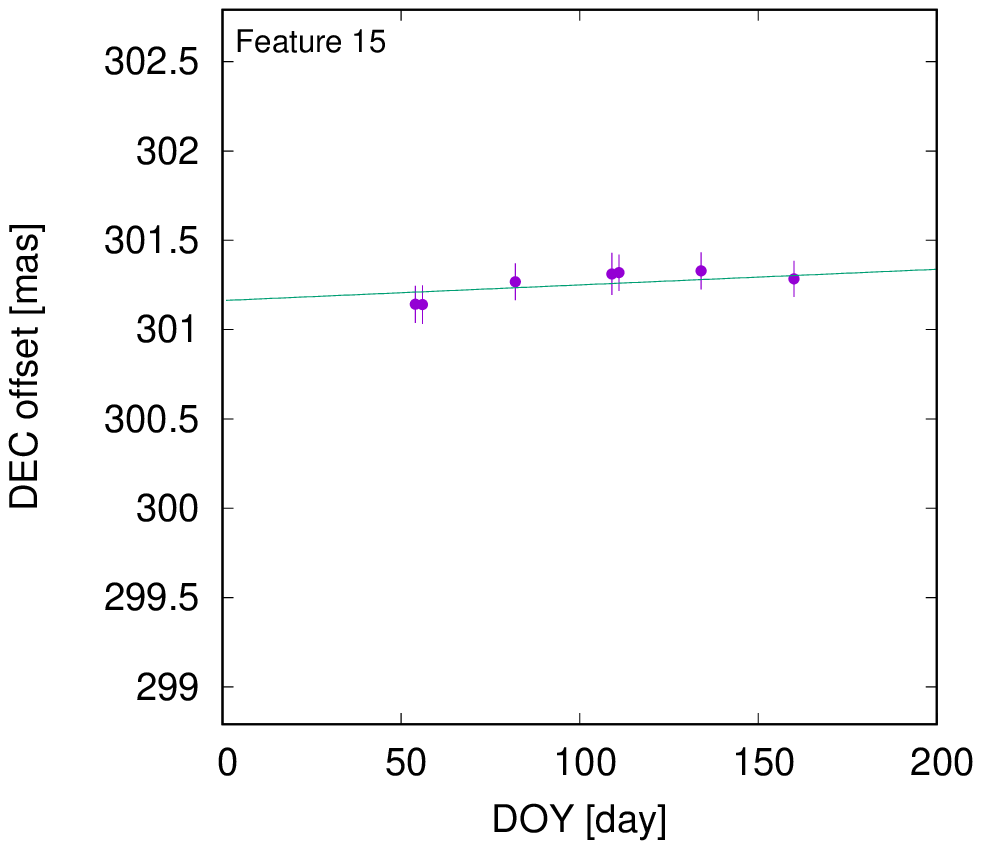}
\includegraphics[width=5cm]{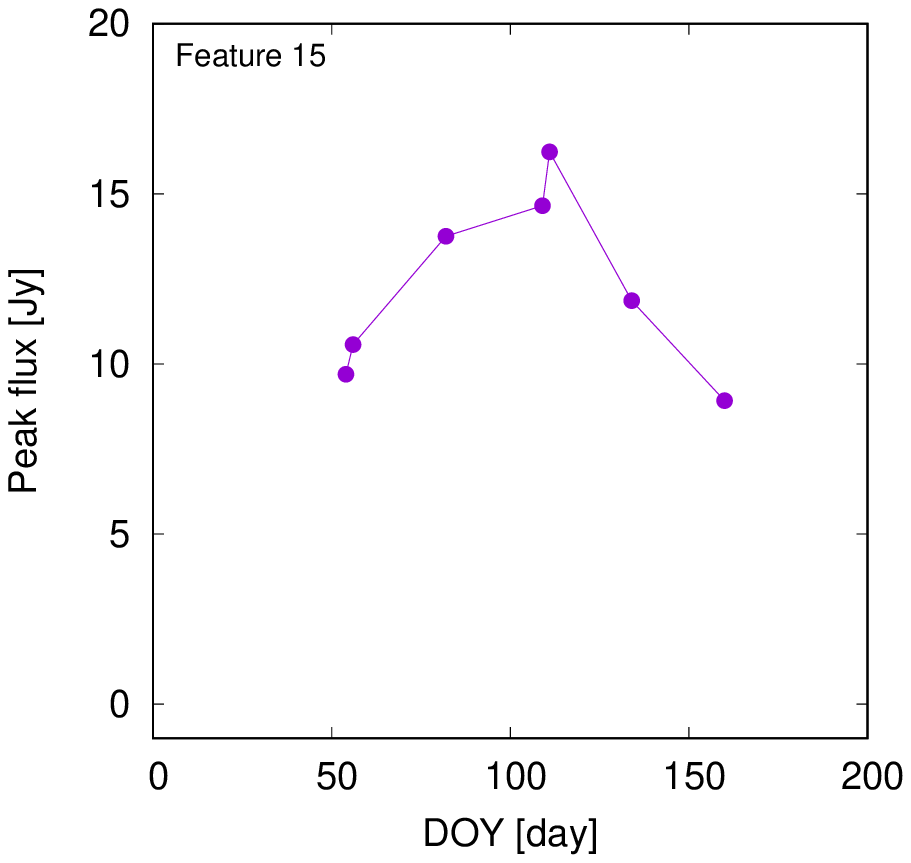}
\caption{Same as Fig. \ref{fig-feature1}, but for feature 15. }
\label{fig-feature15}
\end{center}
\end{figure*}

\begin{figure*}[th]
\begin{center}
\includegraphics[width=5cm]{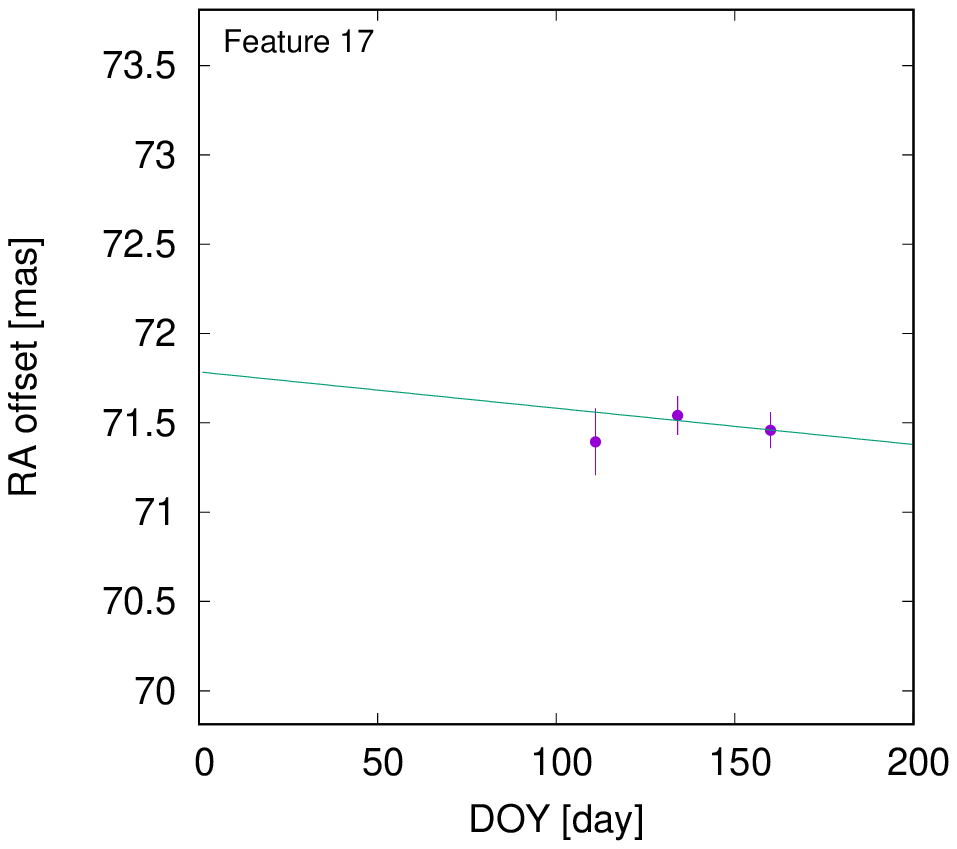}
\includegraphics[width=5cm]{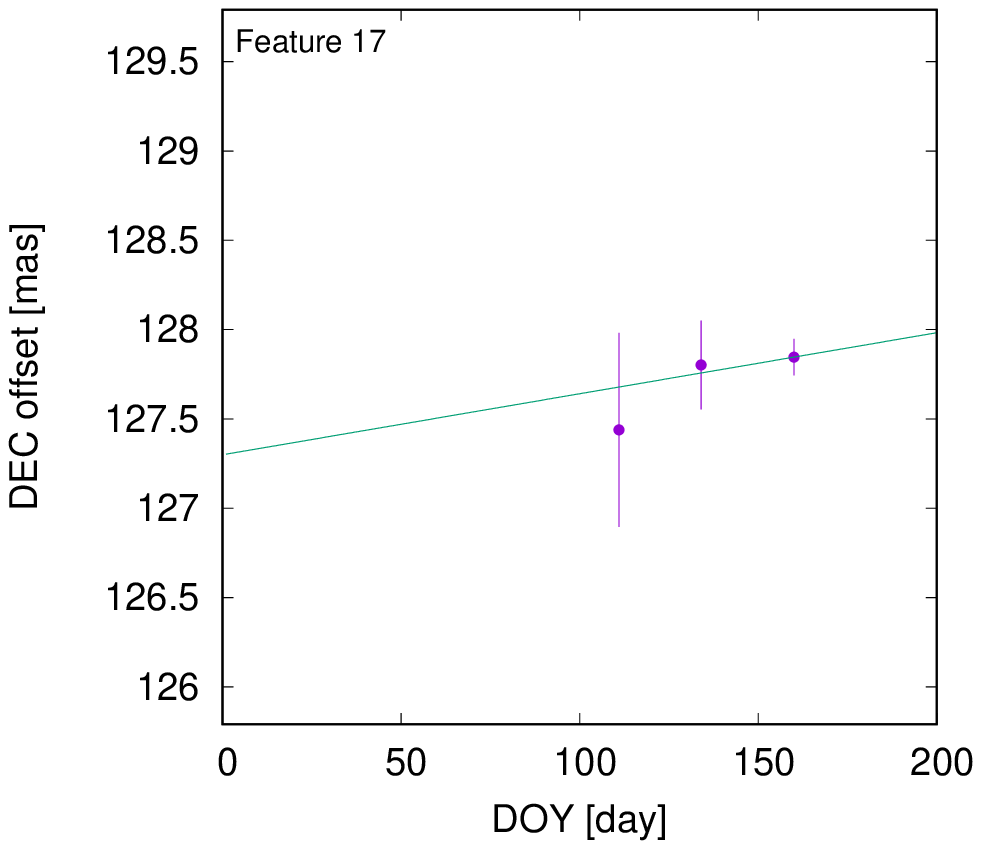}
\includegraphics[width=5cm]{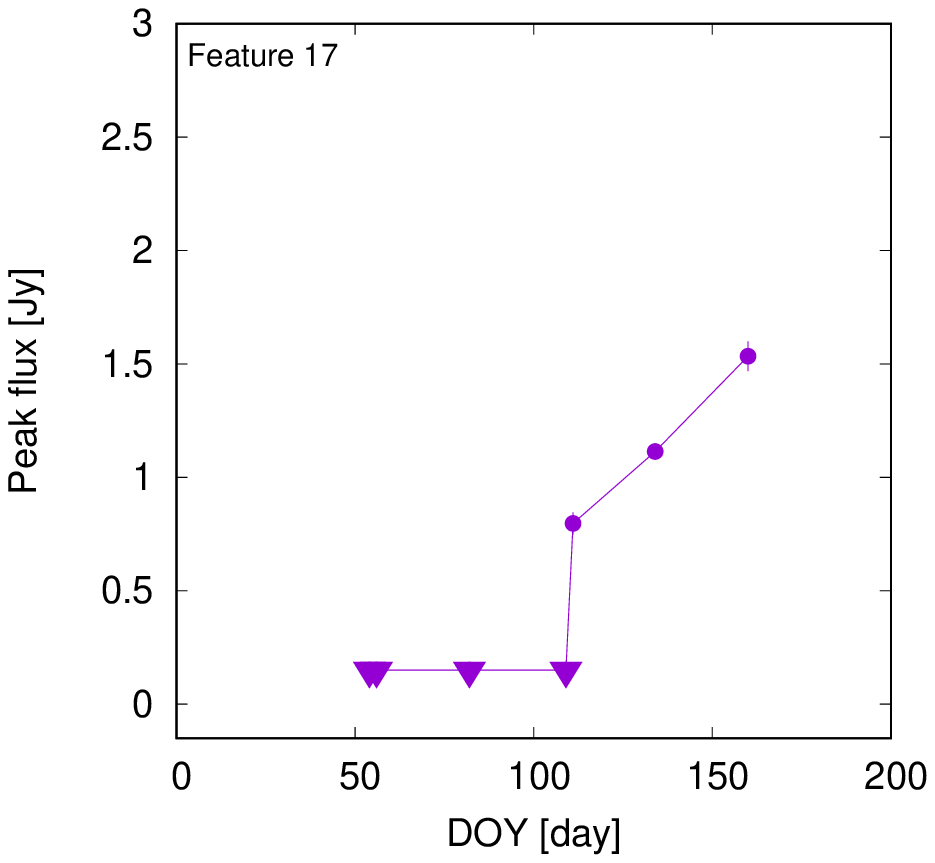}
\caption{Same as Fig. \ref{fig-feature1}, but for feature 17. 
Triangles in the right panel represent the upper limit of the flux densities. }
\label{fig-feature17}
\end{center}
\end{figure*}

\begin{figure*}[th]
\begin{center}
\includegraphics[width=5cm]{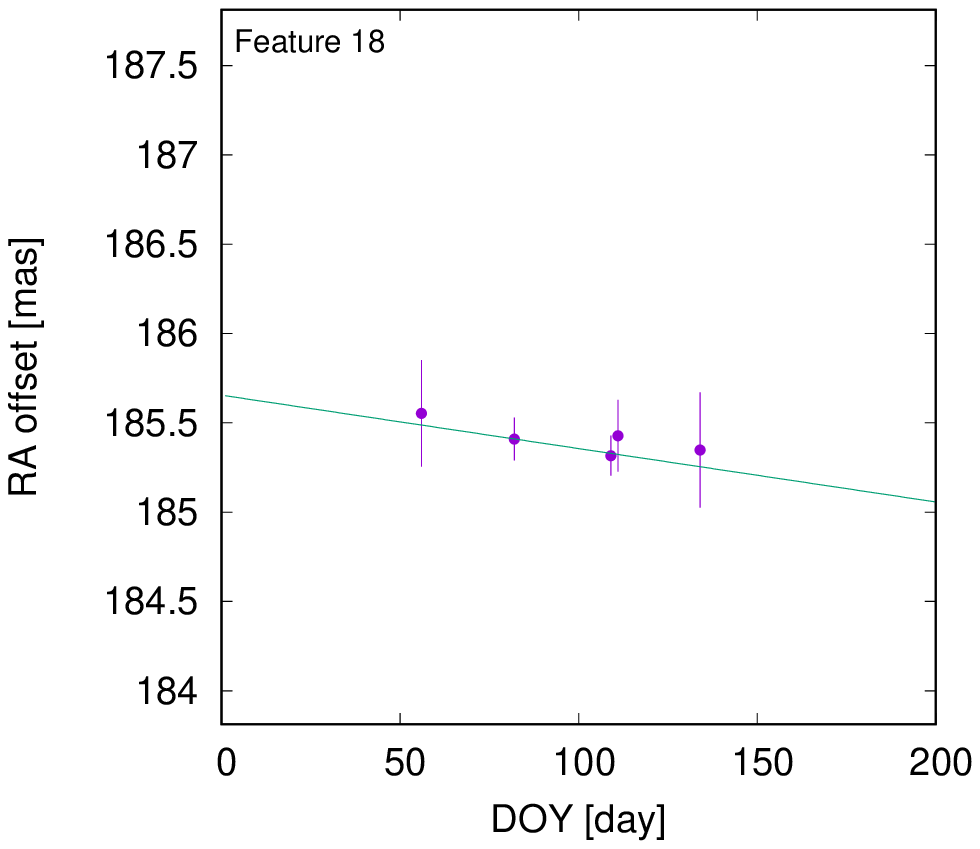}
\includegraphics[width=5cm]{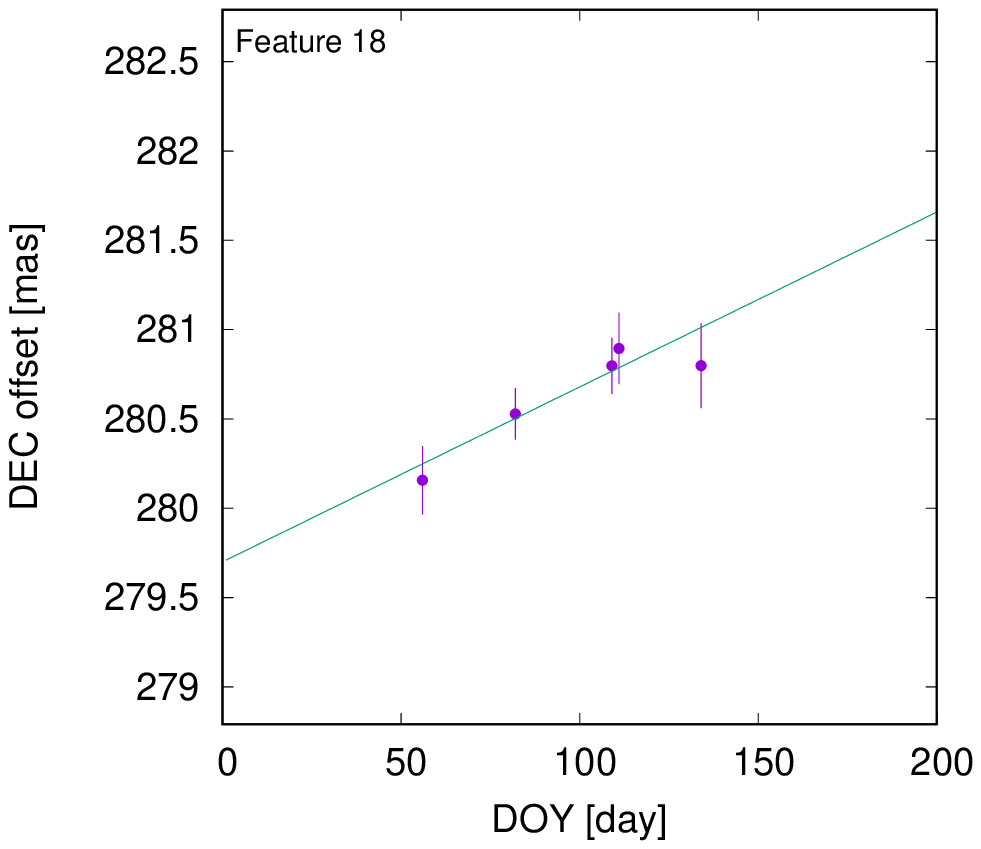}
\includegraphics[width=5cm]{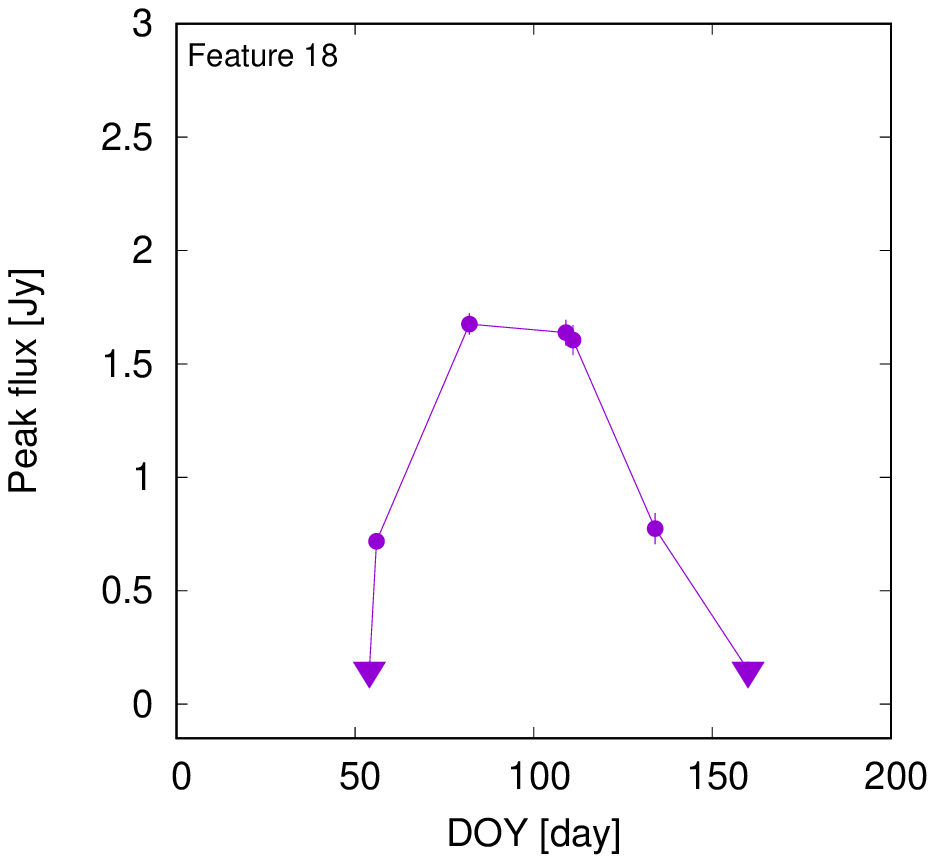}
\caption{Same as Fig. \ref{fig-feature1}, but for feature 18. 
Triangles in the right panel represent the upper limit of the flux densities. }
\label{fig-feature18}
\end{center}
\end{figure*}

\begin{figure*}[th]
\begin{center}
\includegraphics[width=5cm]{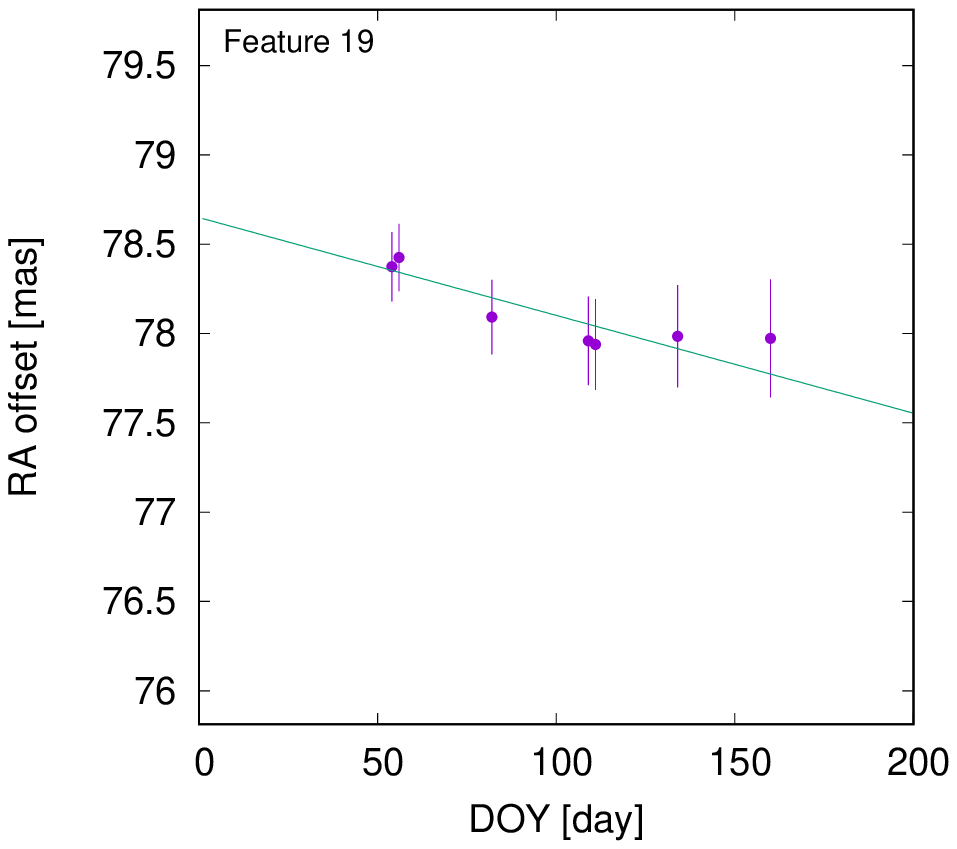}
\includegraphics[width=5cm]{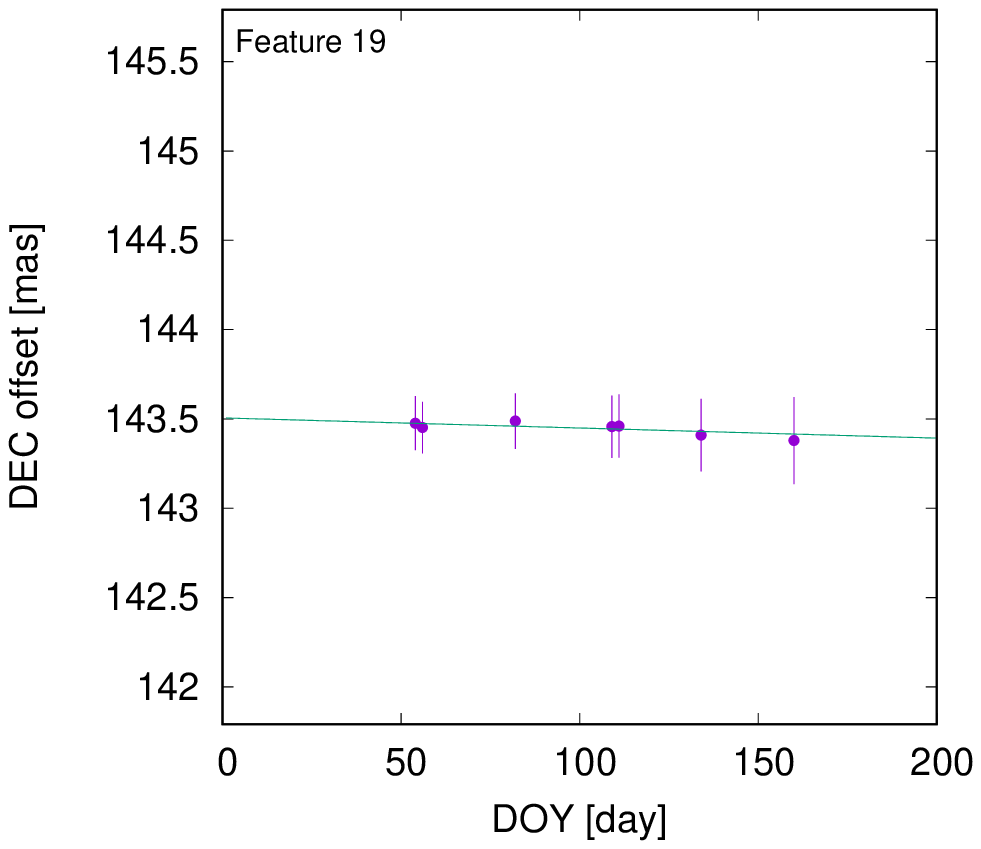}
\includegraphics[width=5cm]{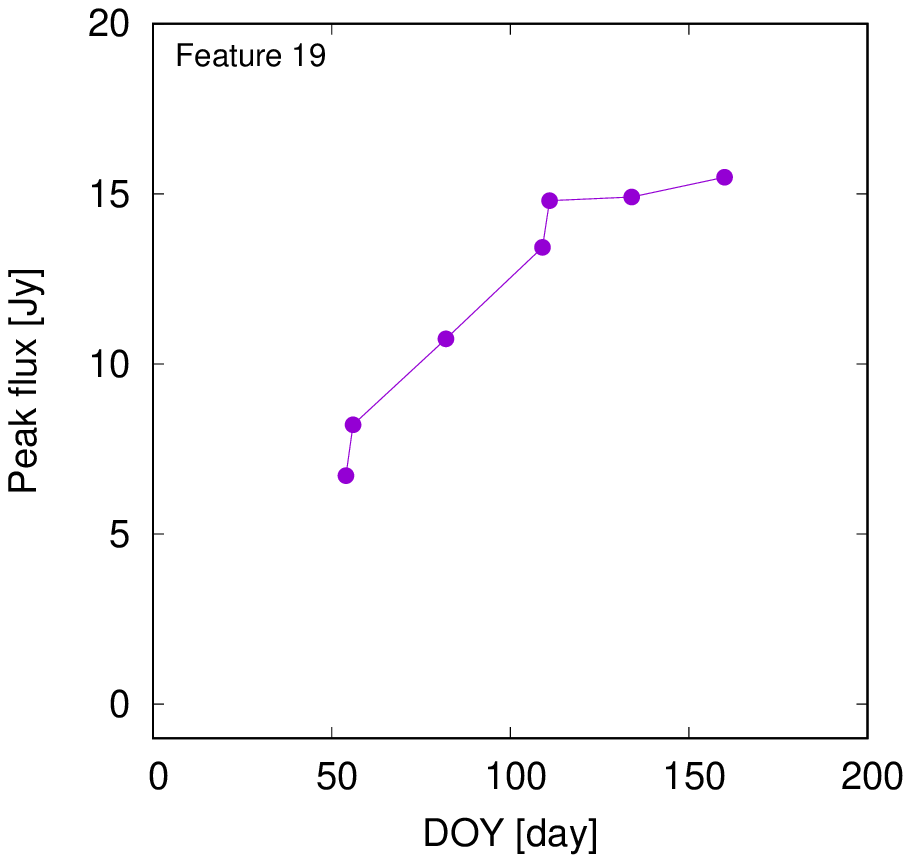}
\caption{Same as Fig. \ref{fig-feature1}, but for feature 19. }
\label{fig-feature19}
\end{center}
\end{figure*}

\begin{figure*}[th]
\begin{center}
\includegraphics[width=5cm]{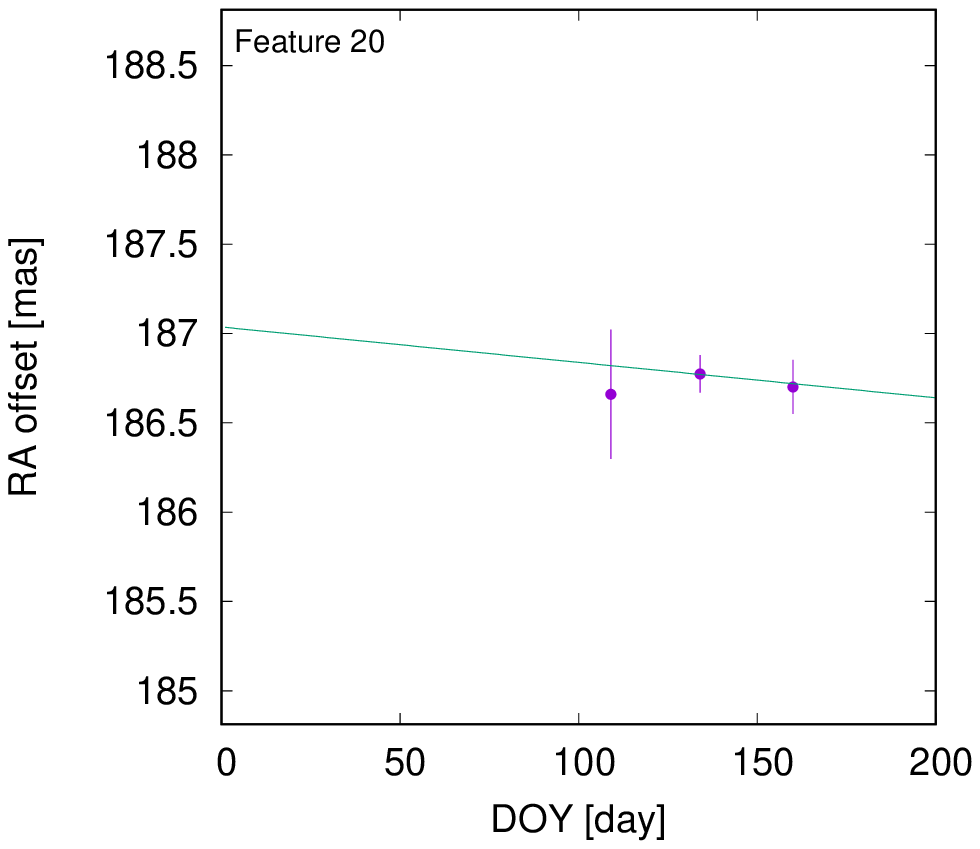}
\includegraphics[width=5cm]{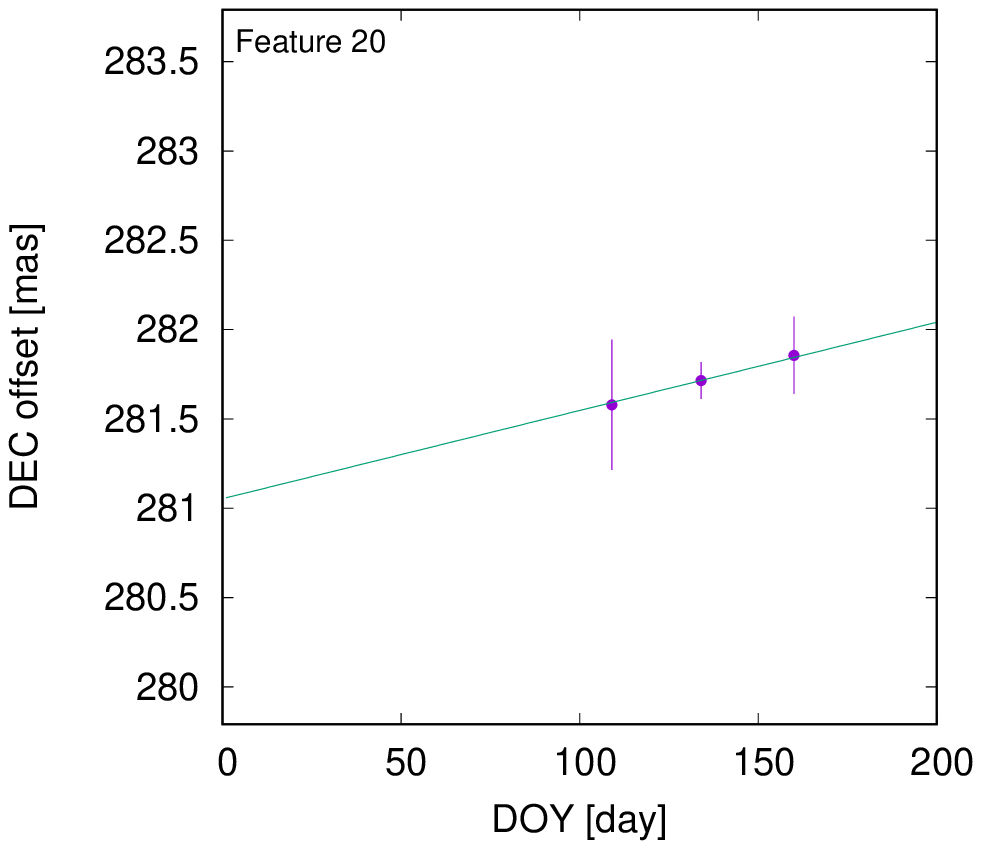}
\includegraphics[width=5cm]{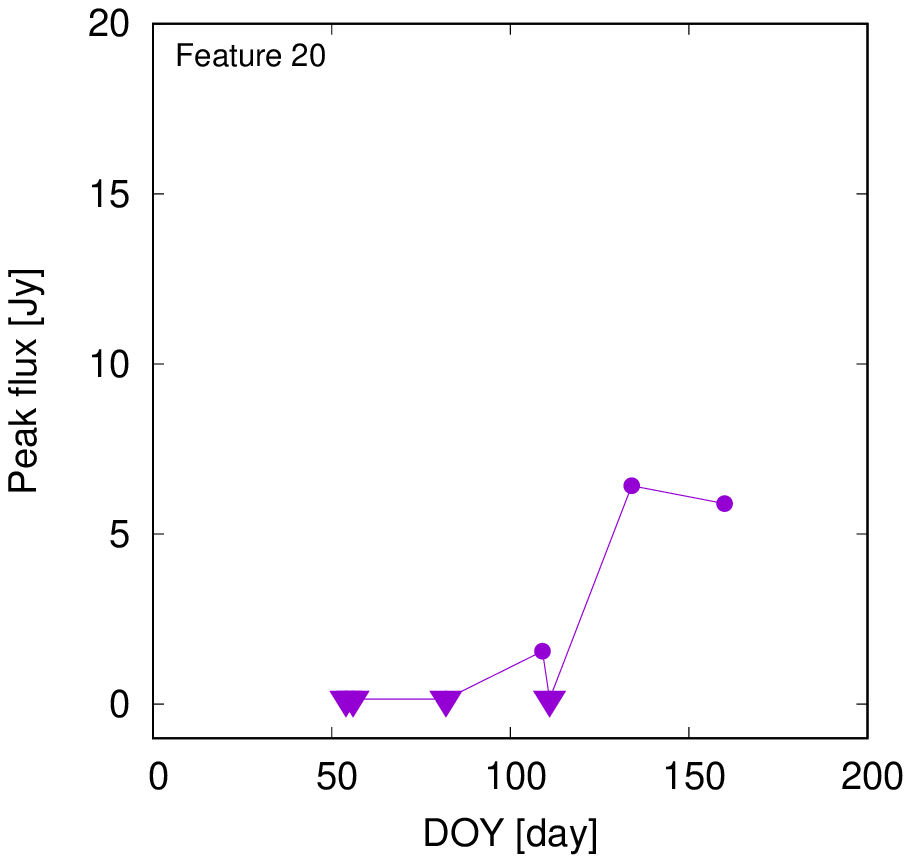}
\caption{Same as Fig. \ref{fig-feature1}, but for feature 20. 
Triangles in the right panel represent the upper limit of the flux densities. }
\label{fig-feature20}
\end{center}
\end{figure*}

\begin{figure*}[th]
\begin{center}
\includegraphics[width=5cm]{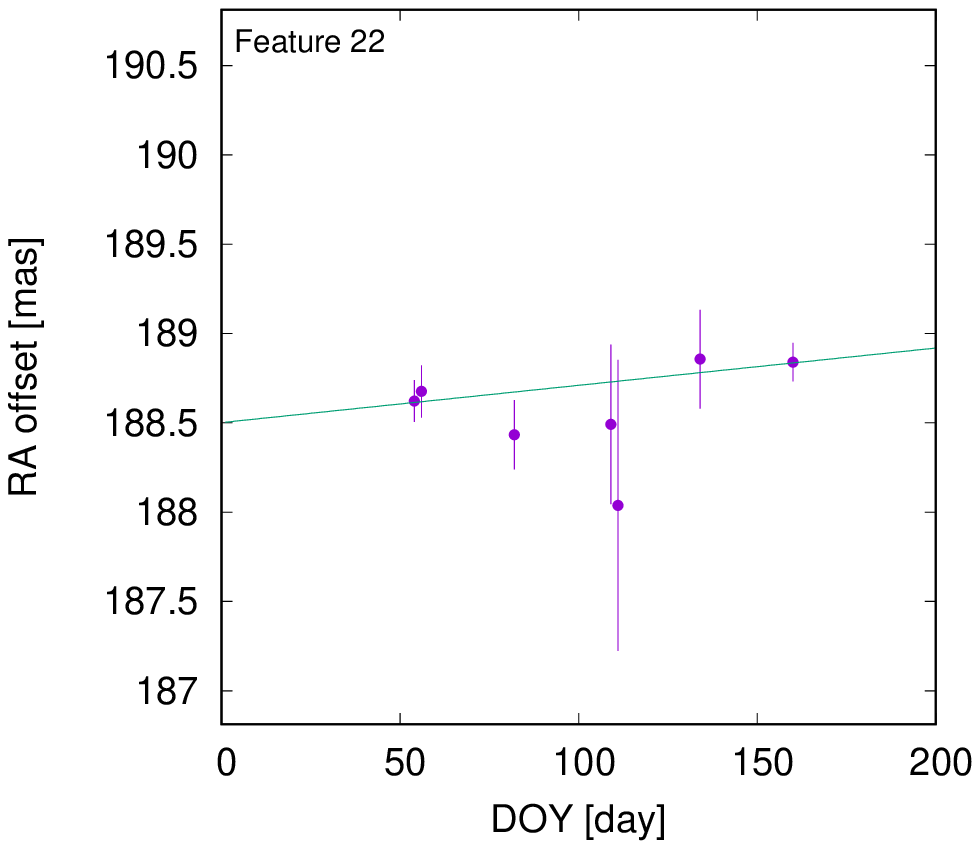}
\includegraphics[width=5cm]{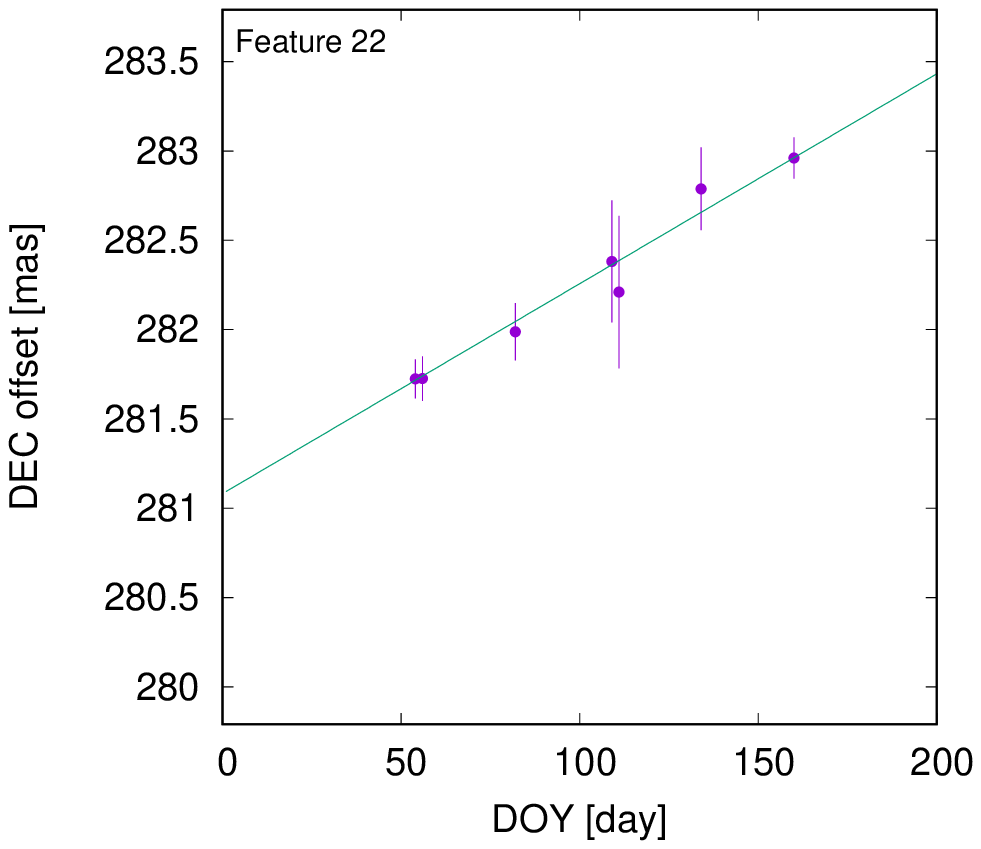}
\includegraphics[width=5cm]{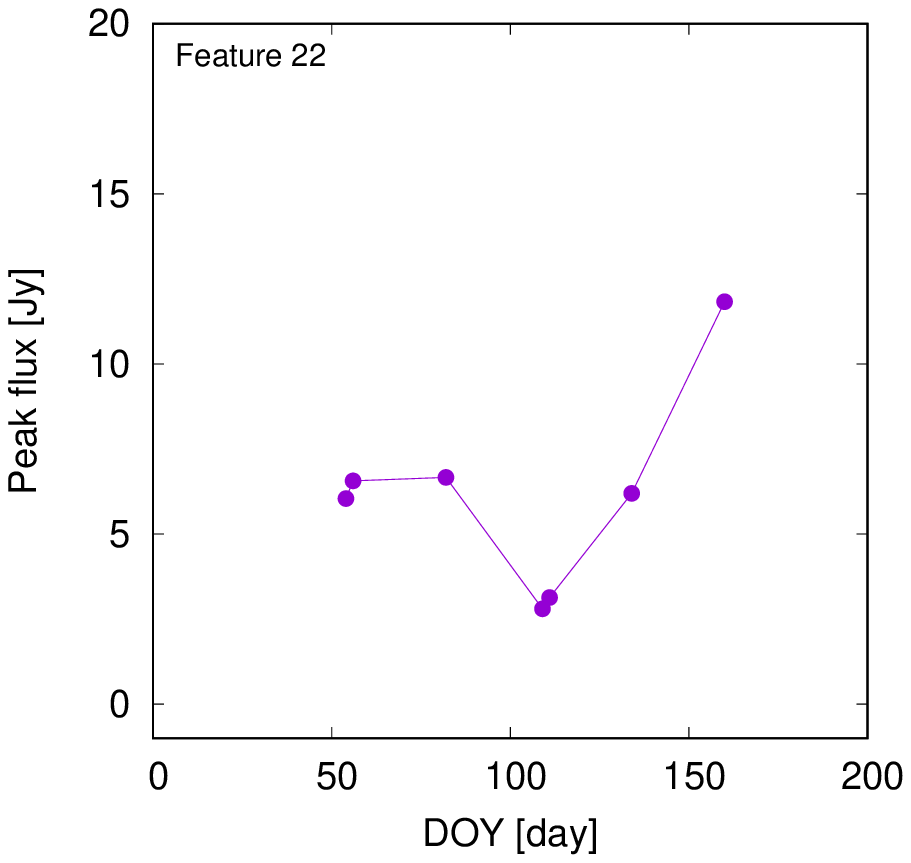}
\caption{Same as Fig. \ref{fig-feature1}, but for feature 22. }
\label{fig-feature22}
\end{center}
\end{figure*}

\begin{figure*}[th]
\begin{center}
\includegraphics[width=5cm]{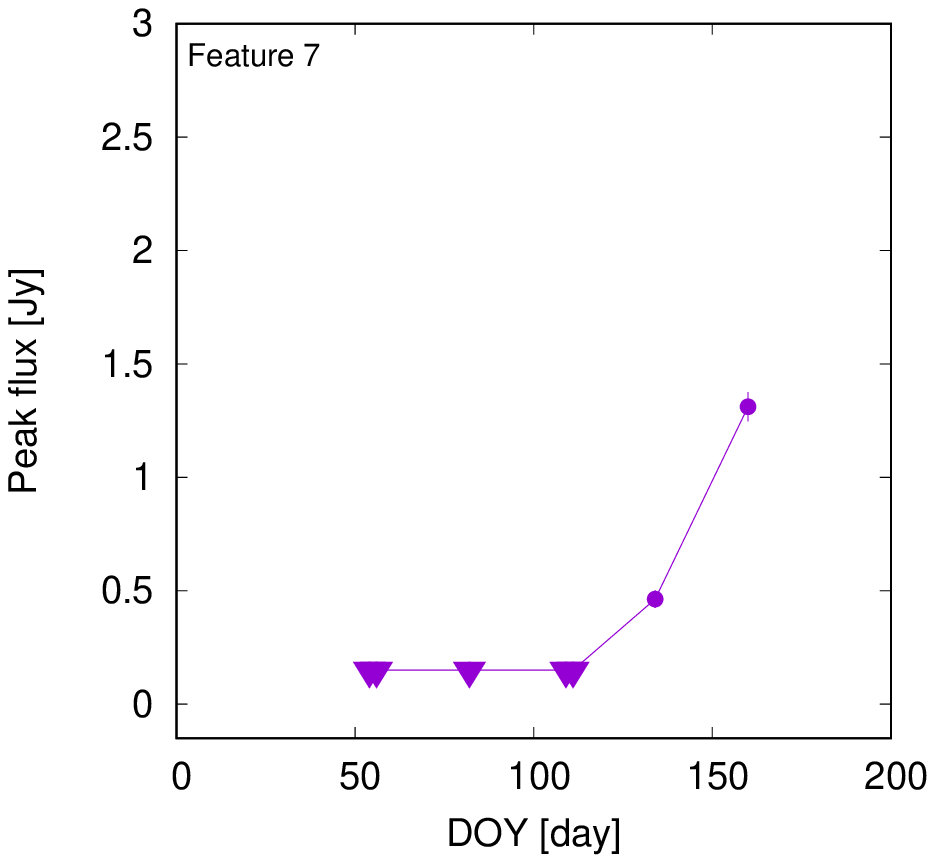}
\includegraphics[width=5cm]{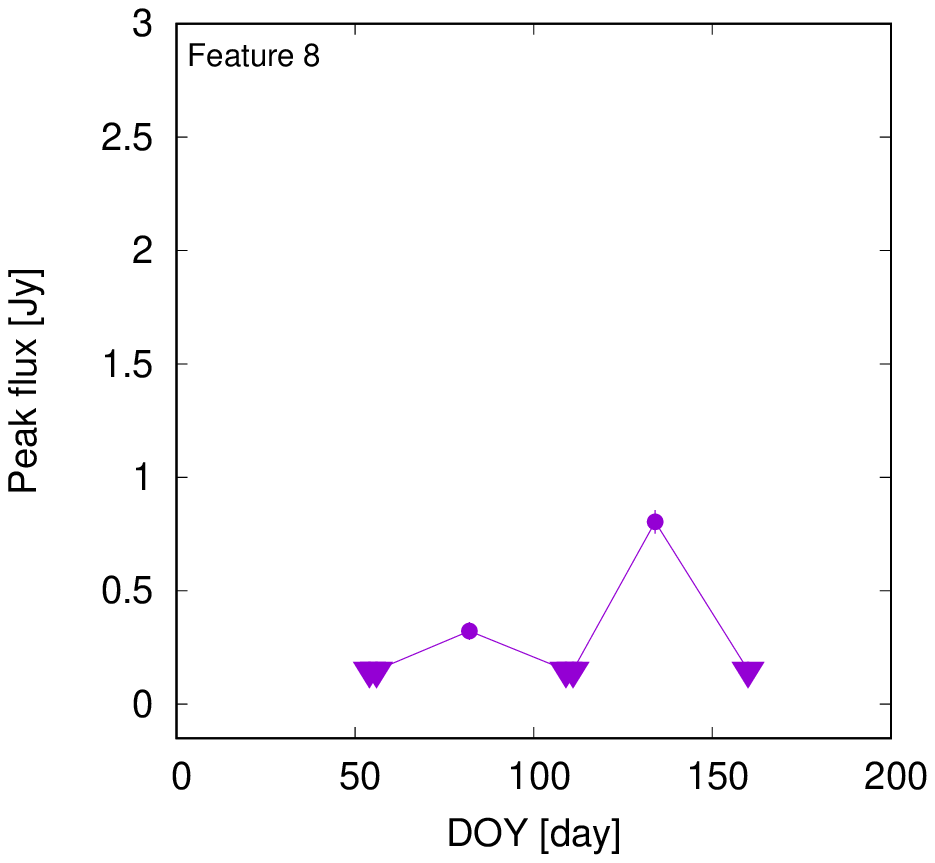}
\includegraphics[width=5cm]{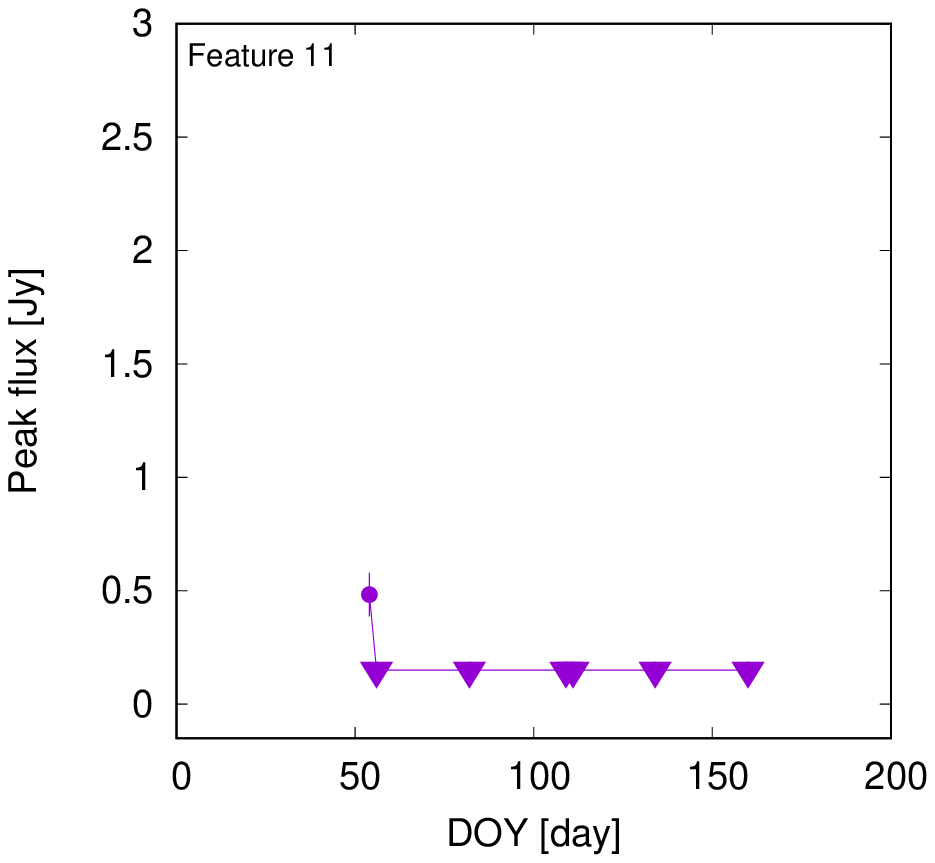}
\includegraphics[width=5cm]{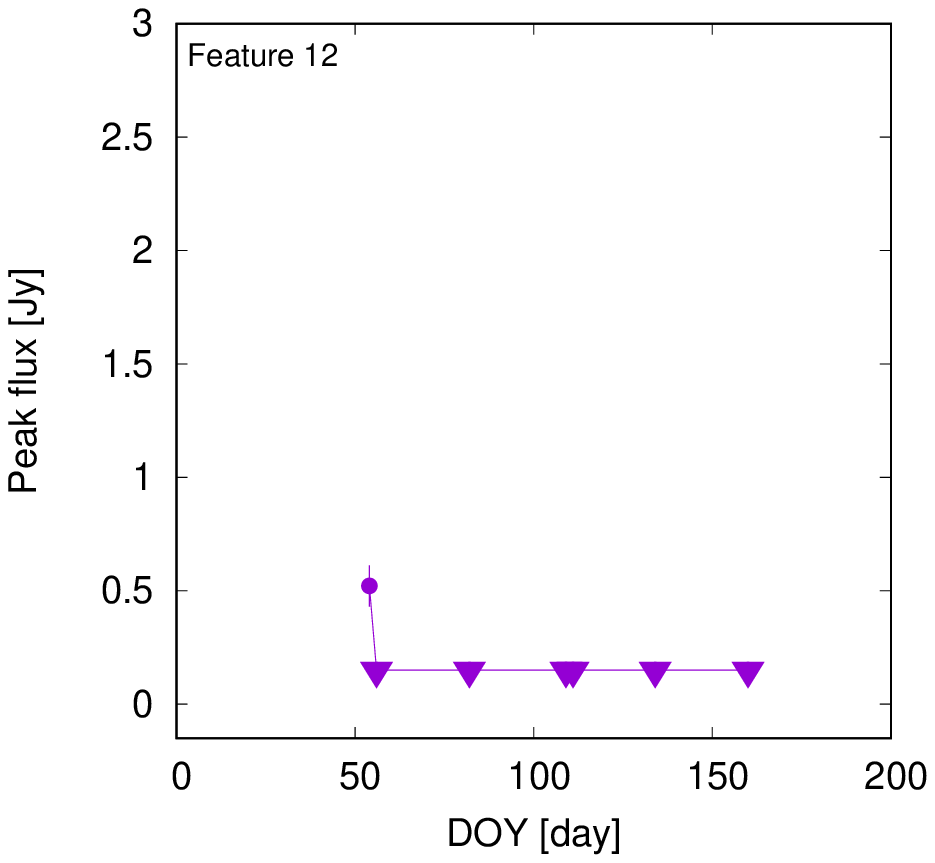}
\includegraphics[width=5cm]{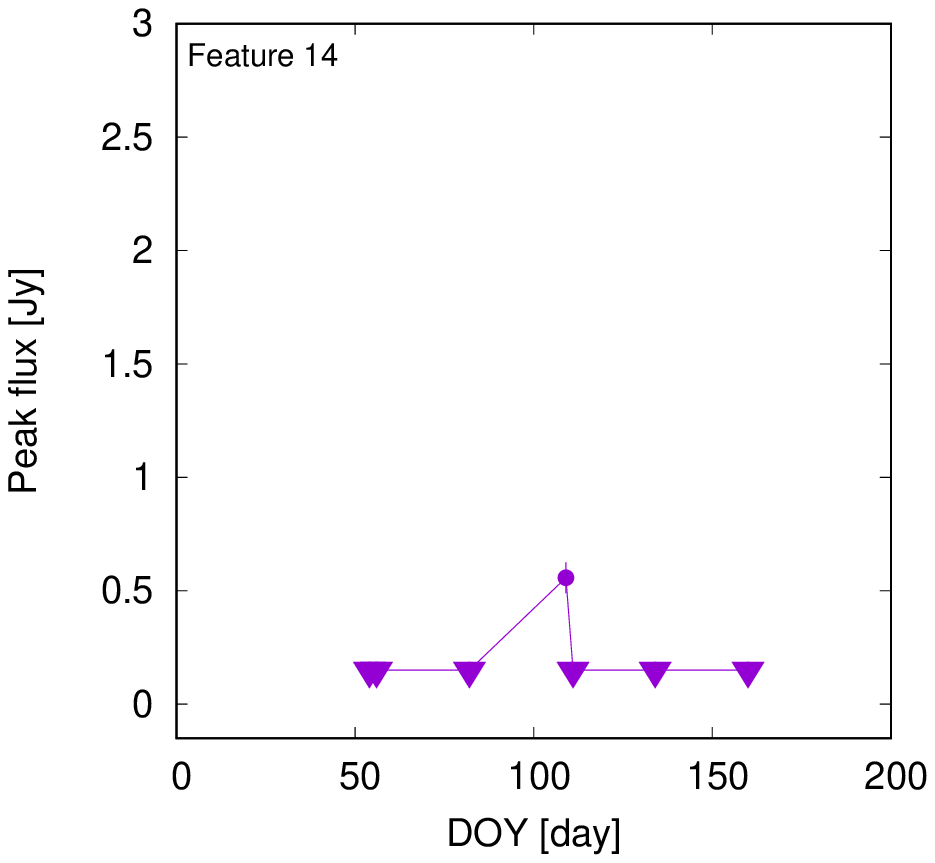}
\includegraphics[width=5cm]{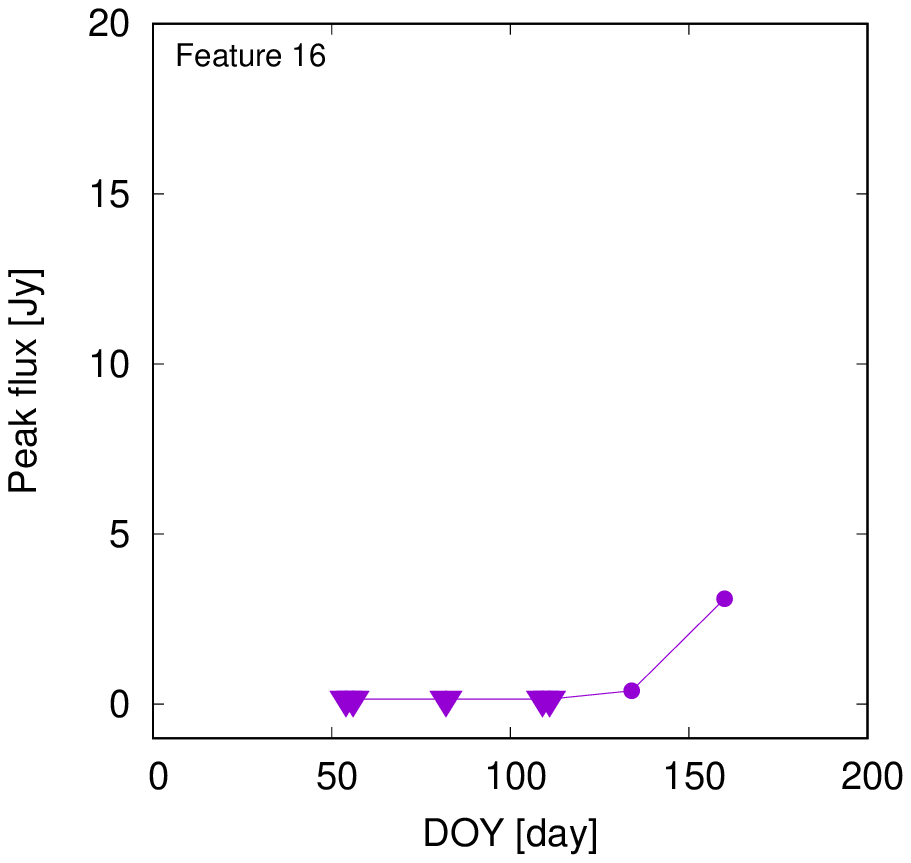}
\includegraphics[width=5cm]{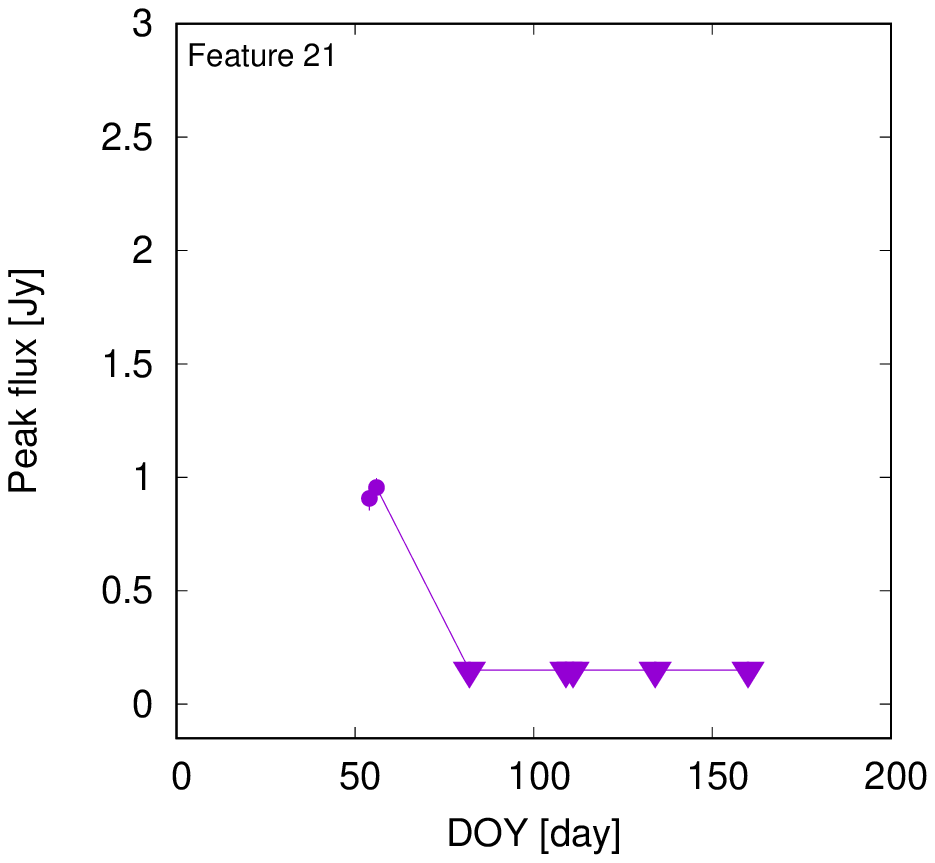}
\includegraphics[width=5cm]{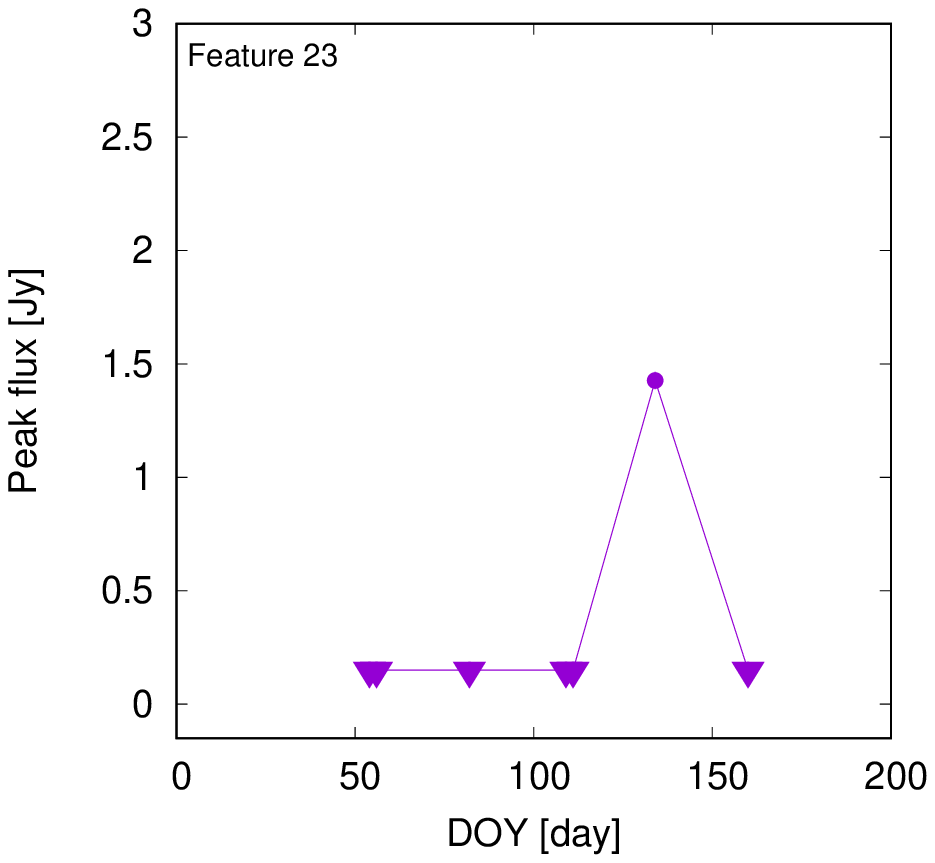}
\includegraphics[width=5cm]{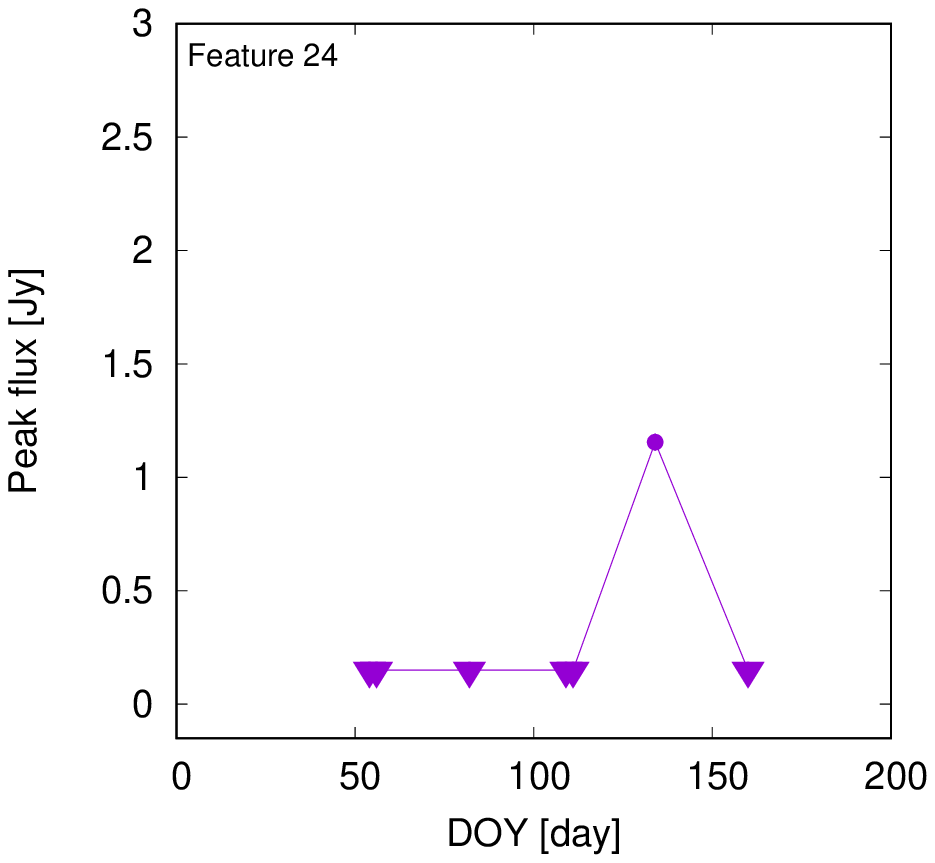}
\includegraphics[width=5cm]{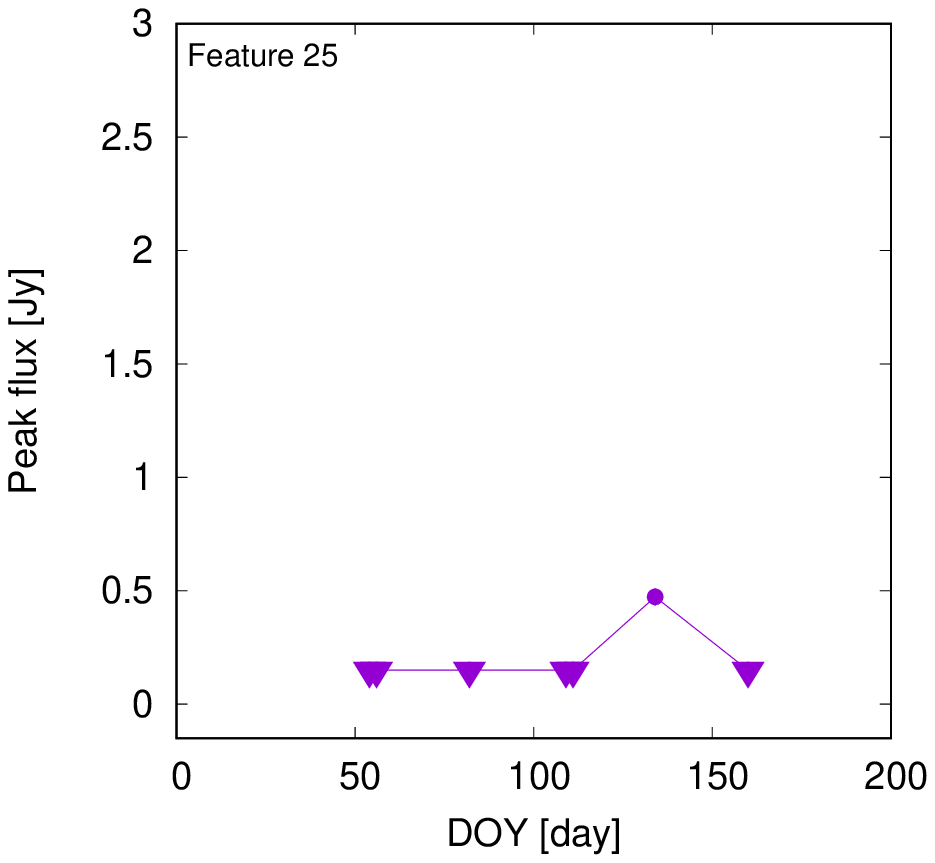}
\caption{Flux density variation for the features detected only in 1 or 2 VLBI epochs. 
Triangles in each panel represent the upper limit of the flux densities. 
}
\label{fig-feature0}
\end{center}
\end{figure*}

The proper motion fitting of all maser features are presented in Figs. \ref{fig-feature1}-\ref{fig-feature22}. 
The positions are measured with respect to that of the ID13 feature (not with respect to the barycenter or the absolute frame). 
The error bars are estimated from the Gaussian fitting errors, and the systematic errors of 100~mas summed in quadrature. 
In the same figures, we also plot the flux densities of the peak channels for each feature. 
In Fig. \ref{fig-feature0}, we plot the flux densities of those features that were detected in fewer than three epochs, which are insufficient for a proper motion measurement.

\end{document}